\documentclass[journal,twoside,web]{ieeecolor}
\usepackage{tmi}

\usepackage{siunitx}

\newcommand{\R}{\mathbb{R}}

\newcommand{\bk}{\mathbf{k}}
\newcommand{\bc}{\mathbf{c}}

\newcommand{\bK}{\mathbf{K}}
\newcommand{\bH}{\mathbf{H}}
\newcommand{\bv}{\mathbf{v}}

\newcommand{\bw}{\mathbf{w}}

\newcommand{\brho}{\boldsymbol{\rho}}

\newcommand{\N}{\mathbb{N}}
\usepackage{hyperref}
\usepackage[ruled,vlined]{algorithm2e}
\usepackage{hyperref}
\usepackage{stfloats}
\usepackage{verbatim}
\usepackage{amsmath}
\usepackage{xspace}
\usepackage{bbm}
\usepackage{xr}
\usepackage{stackengine}
\usepackage{array}
\usepackage{makecell}
\usepackage{mathpazo} % Use the Palatino font by default
\usepackage{pgfplots}
\usepackage{setspace}
\usepackage{multirow}
\usepackage{graphicx}
\usepackage{grffile}
\usepackage{tabularx}
\usepackage{cite}
\usepackage{url}
\usepackage{array}
\usepackage{booktabs}
\usepackage[normalem]{ulem}
\usepackage{subcaption}
\usepackage{lipsum}
\usepackage{colortbl}
\usepackage{multicol}
\usepackage{xcolor, mdframed}
\usepackage{tikz}
\usepackage{textcomp} %to add \textmu
\usepackage{pgfplots}
\DeclareMathAlphabet{\mathb}{OML}{cmm}{b}{it}
\newcommand{\Bw}[1]{\color{white} \bf #1}
\def\bigcro#1{\bigl[#1\bigr]}

\newcommand{\by}{\mathbf{y}}
\newcommand{\bY}{\mathbf{Y}}

\newcommand{\revcomment}[2]{#1}
\newcommand{\multirevcomment}[2]{#1}
\def\rem#1{}

\usepackage{varioref}

\begin{document}
\title{Optimizing full 3D SPARKLING trajectories for high-resolution $T_2$*-weighted Magnetic Resonance Imaging}

\author{Chaithya GR,
				Pierre~Weiss,
				Guillaume Daval-Fr\'erot,
		        Aur\'elien Massire,
				Alexandre Vignaud
		        and~Philippe~Ciuciu~\IEEEmembership{Senior Member, IEEE}
		        {\thanks{Chaithya G R, G. Daval-Fr\'erot and P. Ciuciu are with CEA, NeuroSpin, F-91191 Gif-sur-Yvette, cedex, France and Inria, Parietal, Universit\'e Paris-Saclay, F-91120 Palaiseau, France, whereas A. Vignaud is only affiliated with CEA, NeuroSpin~(Email: philippe.ciuciu@cea.fr). Pierre Weiss is with CNRS, IMT~(UMR 5219) and ITAV~(USR 3505), F-31106, Toulouse. Aur\'elien Massire and G. Daval-Fr\'erot are with Siemens Healthcare SAS, F-93527 Saint-Denis, cedex, France.}}}
\maketitle

 \IEEEaftertitletext{\vspace{-3cm}}
 \begin{abstract}
The Spreading  Projection  Algorithm  for  Rapid  K-space  samplING, or SPARKLING, is an optimization-driven method that has been recently introduced for accelerated 2D $T_2$*-w MRI using compressed sensing. It has then been extended to address 3D imaging using either stacks of 2D sampling patterns or a local 3D strategy that optimizes a single sampling trajectory at a time. 2D SPARKLING actually performs variable density sampling~(VDS) along a prescribed target density while maximizing sampling efficiency and meeting the gradient-based hardware constraints. However, 3D SPARKLING has remained limited in terms of acceleration factors along the third dimension if one wants to preserve a peaky point spread function~(PSF) and thus good image quality.
In this paper, in order to achieve higher acceleration factors in 3D imaging while preserving image quality, we propose a new efficient algorithm that performs \revcomment{optimization on full}{r:title} 3D SPARKLING. The proposed implementation based on fast multipole methods~(FMM) allows us to design sampling patterns with up to $10^7$ k-space samples, thus opening the door to 3D VDS. We compare multi-CPU and GPU implementations and demonstrate that the latter is optimal for 3D imaging in the high-resolution acquisition regime~(600\textmu m isotropic). Finally, we show that this novel \revcomment{optimization for full}{r:title} 3D SPARKLING outperforms stacking strategies or \revcomment{3D twisted projection imaging}{r:SOTA} through retrospective and prospective studies on NIST phantom and in vivo brain scans at 3 Tesla. Overall the proposed method allows for 2.5-3.75x shorter scan times compared to GRAPPA-4 parallel imaging acquisition at 3 Tesla without compromising image quality.
\end{abstract}
%that exploits all the degrees of freedom offered by MR gradient coils


%\IEEEpeerreviewmaketitle

\section{Introduction}

The quest for efficient sampling strategies has been a major challenge in MRI since its invention. The theory of compressed sensing~(CS)~\cite{Lustig2007} boosted this quest by providing significant theoretical insights. \revcomment{It was proved and observed empirically that for under-sampled acquisitions and approximately sparse signals in an orthogonal basis, an efficient implementation relies on trajectories with a variable density in k-space: the lower frequencies~(center of k-space) have to be sampled more densely than the higher at the borders of k-space~\cite{puy2011variable,Chauffert_SIAM2014,adcock2017breaking,boyer2017compressed}.}{r:optimal_criteria} Non-Cartesian k-space trajectories~(e.g. spiral, radial, rosette, etc.) ~\cite{ahn1986high,meyer1992fast,jackson1992twisting,noll1997multishot,law2009interleaved,lustig2005faster} have been proposed for accelerated and robust-to-motion 2D imaging, prior to the existence of theoretical foundations. \revcomment{While being compliant with scanner hardware constraints on the gradients, these trajectories do not sample the k-space according to a well controlled target sampling density. For instance, in spiral imaging, fulfilling these constraints transforms an initially prescribed density into another one~\cite[p.~97]{Chauffert_PhDThesis2015}.}{r:nc_classical} 
Recently, the SPARKLING algorithm~\cite{Boyer2016, chauffert2017projection, Lazarus_MRM_19} has been shown to automatically generate optimized non-Cartesian sampling patterns compatible with MR hardware constraints on maximum gradient amplitude and slew rate. SPARKLING optimally samples the k-space (see \cite{puy2011variable,Chauffert_SIAM2014}) with a controlled distribution of samples (e.g., variable density) and a locally uniform k-space coverage. 

However, for the sake of signal-to-noise ratio~(SNR), 3~dimensional~(3D) imaging is preferred to reach isotropic high-resolution imaging~(e.g. 600\textmu m isotropic). In this regard, multiple approaches have been utilized to efficiently down-sample 3D k-space. Some of these involve a combination of a readout in the z-direction with a 2D under-sampled mask based on Poisson disk sampling~\cite{vasanawala2010improved}. Additional attempts on full 3D readouts were proposed like 3D radial trajectory~\cite{larson2008anisotropic}, 3D cones~\cite{irarrazabal1995fast}, twisted projections~\cite{boada1997fast} and hybrid radial-cones~\cite{johnson2017hybrid}. However, these trajectories were primarily based on parameterizing a k-space sampling curve, and the final sampling pattern was not optimized with respect to the reconstruction quality. Some recent studies explored how to optimize the sampling pattern~\cite{dale2004optimal,mir2004fast,kumar2008durga}, but did not include a clear sampling criterion in order to maximize the image reconstruction quality.

Other methodologies in the literature involved stacking a 2D under-sampled trajectory like stack of stars~\cite{song2004dynamic,lin2008respiratory}, stack of spirals~\cite{irarrazabal1995fast,thedens1999fast} and stack of 2D SPARKLING~\cite{Lazarus_NMRB_20}. Uniform~(i.e. cylindrical) stacking is usually implemented even though a spherical strategy, with a number of shots varying as a function of the latitude plan, was shown to be beneficial on image quality for SPARKLING trajectories~\cite{Lazarus_NMRB_20}. Further in \cite{Lazarus_NMRB_20}, 
%we made attempts at solving for a
a local 3D SPARKLING approach was proposed by designing a single trajectory within a cone obtained from a parcellation of the 3D spherical k-space. Then all the cones covering a given elevation plane were filled up using the replication of the resulting trajectory. However, this method did not ensure a locally uniform sampling pattern at the boundaries of cones as the problem was solved locally.

The recent rise of machine and deep learning has impacted the literature on MRI sampling~\cite{oedipus,seeger2010,baldassarre2016learning,gozcu2018learning,bahadir2020deep,sherry2020learning,weiss2019pilot,vedula20203d}. These approaches rely on supervised learning techniques, which means that they need a ground truth corresponding to fully sampled data (like fastMRI dataset \cite{zbontar2018fastmri}), to learn an optimal under-sampling pattern, whether it is Cartesian or not.
\multirevcomment{In~\cite{oedipus,seeger2010}, the authors explore the use of experimental design to choose the best subset of prescribed trajectories. Although there are substantial differences between these two methodologies~(deterministic vs Bayesian, offline vs online design, etc.) they share a similar theoretical background with ours in that sparsity is the key underlying hypothesis. In particular, \cite{seeger2010,oedipus} use the Cram\'er-Rao bound for sparse signals~\cite{haim2010} as a tailored optimality criteria. However, such methods are computationally demanding as they try to solve a nonconvex integer programming problem. Hence, in a given time budget this limits the exploration and the potential number of prescribed trajectories.}{\ref{r:oedipus} \ref{r:optimal_criteria}}

In ~\cite{baldassarre2016learning,gozcu2018learning}, the authors proposed to step away from the theoretical consideration in CS and adopted a purely data-driven approach. The authors proposed to find an optimal subset of Cartesian sampling lines by using a greedy algorithm aimed at maximizing the SNR. This algorithm can automatically adapt to different reconstruction algorithms and optimality criteria, but its use is limited to Cartesian imaging.
More recent approaches made some advance on learning gridded sampling patterns~\cite{bahadir2020deep,sherry2020learning}. \revcomment{Additionally, to the best of our knowledge, the only works that have learned a non-Cartesian trajectory under hardware constraints are PILOT~\cite{weiss2019pilot} BJORK~\cite{wang2021b} and 3D FLAT~\cite{vedula20203d} for 2D and 3D imaging, respectively. These works seem very promising despite significant theoretical and numerical challenges with a combinatorial number of local minimizers~\cite{degournay2021}. In contrast, our work is based on clear theoretical considerations with provable convergence~\cite{chizat2018} in short computing times~\cite{merigot2021}. Of interest, let us notice that the sampling patterns generated by these methods resemble the SPARKLING ones very much~\cite{weiss2019pilot,wang2021b}, suggesting that the main ideas behind are now reaching a mature and reliable state.}{r:nc_classical} Nonetheless, it is worth noting that none of these approaches has been prospectively validated on real 3D acquisitions. For all these reasons, these works won't be discussed any further in this paper.

In this paper, for the first time, we solve the SPARKLING optimization \revcomment{fully}{r:title} in 3 dimensions.
First, in Sec~\ref{sec:theory}, we remind the optimization problem to be solved for generating SPARKLING trajectories. Then we focus on major computational bottlenecks that prevented us from scaling the original solution to 3D and provide detail on our main contributions. One key ingredient in SPARKLING is the setting of the right target sampling density. The latter may vary as a function of the resolution, the acceleration factor and the object to be scanned. For that purpose, we parameterize radially decaying densities by two parameters~(cut-off, decay) and find the optimal density through grid search over pairs of parameters. This study can be conducted on retrospective analysis and then the sought optimal density can be used further in prospective acquisitions. In Sec.~\ref{sec:results}, we present the experimental data sets on which the numerical studies are performed later on for validation and comparison purposes. In this regard, we carry out retrospective \revcomment{and prospective}{r:resolution_loss} analysis on NIST phantom data collected at 3 Tesla~(3T). Then we perform prospective in vivo brain imaging acquisitions on a healthy adult volunteer still at 3T and compare the proposed full 3D SPARKLING with the existing spherical stack of 2D SPARKLING. \revcomment{We do not include any comparison with 3D radial sampling scheme or stack of spirals as this was already done in~\cite{Lazarus_NMRB_20}. However, we do compare our trajectories with improved 3D non-Cartesian trajectories, namely twisted projection imaging~(TPI)~\cite{boada1997fast}. TPI trajectories have better k-space coverage as compared to full 3D radial sampling scheme as these trajectories shift to pappus spirals after a fraction of readout.}{r:SOTA}

\section{Theory}
\label{sec:theory}

In this section we briefly introduce the SPARKLING algorithm as described in \cite{Lazarus_MRM_19}. We detail the particular steps involved in the optimization process. We point to some computational bottlenecks in each of these steps. Later, we describe the methods used to overcome these computational challenges, thereby allowing us to scale the problem to 3 dimensions. Most of the theoretical aspects are directly based on earlier works in~\cite{Chauffert_TMI_16,Boyer2016,chauffert2017projection}, which can be consulted for the problem description and derivations of~\eqref{eq:global_minimization}.

\subsection{3D K-space sampling}
\label{sec:3dkspace}
A 3D k-space sampling pattern $\mathbf{K}$ is usually composed of several shots or curves, say $N_c$, $\mathbf{K}=(\mathbf{k}_i)_{i=1}^{N_c}$, where each 3D shot $\mathbf{k}_i(t)=(k_{i,x}(t), k_{i,y}(t), k_{i,z}(t))$, is controlled by magnetic field gradients $\mathbf{G}_i(t)= (G_{i,x}(t), G_{i,y}(t), G_{i,z}(t))$ as follows: 
\begin{align}\label{eq:ktraj}
\mathbf{k}_i(t) &= \frac{\gamma}{2\pi} \int_0^t \mathbf{G}_i(\tau)d\tau \,,
\end{align}
with $\gamma$ the gyro-magnetic ratio~($\gamma=42.57$MHz/T for proton imaging). 
\revcomment{In practice, throughout the readout duration $T_{\rm obs}$, we sample each shot $\mathbf{k}_i(t)$ by a time period $\Delta t$, the gradient raster time as the scanner gradient hardware can play gradients at this pace. 
In the rest of the section, we refer to location of the k-space samples $\bK$ as the samples on gradient raster points. Then the number of gradient time steps is given by $N_s= \left\lfloor \dfrac{T_{\rm obs}}{\Delta t}\right\rfloor$ and the full 3D sampling pattern $\mathbf{K}$ finally consists of $p=N_c \times N_s$ points.}{r:samples_vs_loc}
%\ref{r:constraints}
\revcomment{Additionally, we limit ourselves to a long readout~($T_{\rm obs}\simeq 20$ms) for $T_2^*$-weighted imaging, as this allows the trajectory to be longer and maximally explore the k-space.}{r:T2star_TObs}

The k-space domain for a 3D MR volume of size $N_x\times N_y \times N_z$ over a field of view $\text{FOV}_x\times \text{FOV}_y \times \text{FOV}_z$, is defined within $[-K^x_{\rm max}, K^x_{\rm max}]\times[-K^y_{\rm max}, K^y_{\rm max}]\times [-K^z_{\rm max}, K^z_{\rm max}] $, with $K^\ell_{\rm max} = \frac{N_\ell}{2 FOV_\ell}$ and $\ell=x,y,z$. For the sake of simplicity, in what follows we assume the same spatial resolution along the three dimensions so $K^x_{\rm max}=K^y_{\rm max}=K^z_{\rm max}=K_{\rm max}$ even though we meet different matrix and FOV dimensions~($N^z \neq (N^x=N^y)$ and $\text{FOV}^z\neq (\text{FOV}^x=\text{FOV}^y)$).
Hereafter, the 3D k-space domain will be normalized to $\Omega=[-1,1]^3$.

Hardware constraints on the maximum gradient amplitude~($G_{\rm max}$) and slew rate~($S_{\rm max}$) induce limitations in trajectory speed and acceleration. These limits can be expressed as box constraints on the amplitude of the discrete derivatives of the k-space trajectory $(\mathbf{k}_i[n])_{n=0}^{N_s-1}$. \revcomment{These hardware constraints can be applied on a per dimension basis, giving rotation variant~(RV) constraints, whose resulting trajectories cannot be run on the scanner if the FOV is rotated. Due to this limitation, in this work, we focus on rotation invariant speed and acceleration constraints which can be expressed as follows: }{r:constraints}\footnote{In~\cite{Chauffert_TMI_16}, we have also dealt with the case of RV constraints where the $\ell_{\infty}$-norm replaces the mixed $\ell_{2,\infty}$-norm used here.}

\begin{equation}
\label{eq:constraints}
\mathcal{Q}_{\alpha, \beta}^{N_c} = 
\begin{Bmatrix}
\forall i=1,\ldots, N_c,\quad \bk_i \in \R^{3\times N_s},\\ \mathbf{A}\bk_i = \mathbf{v} , \\
\|\bk_i\|_\infty\leq 1, \: \|\dot{\bk}_i\|_{2,\infty} \leq \alpha, \: \|\ddot{\bk}_i\|_{2,\infty} \leq \beta,  
\end{Bmatrix}
\end{equation}

\noindent where 
\begin{align*}
\dot{\bk}_i[n]&=\frac{\bk_i[n] - \bk_i[n-1]}{\Delta t}\\
\ddot{\bk}_i[n]&=\frac{\bk_i[n+1] -2 \bk_i[n] + \bk_i[n-1]}{\Delta t^2}\\
\|\bc\|_{2,\infty}&= \sup_{0\leq n\leq N_s-1}\left(|c_x[n]|^2+|c_y[n]|^2 +|c_z[n]|^2\right)^{1/2},
\end{align*}
for all $\bc\in\Omega^{N_s}$ and $(\alpha,\beta)$ are obtained by normalizing hardware and Nyquist constraints to the sampling domain $\Omega$ (from \cite{Lazarus_MRM_19}): % of maximum gradient strength ($G_{\rm max}$) and maximum slew rate ($S_{\rm max}$) 
 \begin{subequations} 
%\left\{ 
\begin{align}
%\begin{gathered}
\alpha &= \frac{1}{ K_{\rm max}}\min\left( \frac{\gamma G_{\rm max}}{2\pi}, \frac{1}{FOV \cdot \delta t}\right) \label{eq:alpha_constraint}\\
\beta &= \frac{\gamma S_{\rm max}}{2\pi K_{\rm max}} 
\end{align}
\label{eq:constraints_alpha_beta}
%\right.
%\end{gathered}
\end{subequations}

The purpose of $\mathbf{A}$ and $\mathbf{v}$ are to model linear constraints on the trajectory, like the TE point constraint, which ensures that each trajectory passes through center of k-space at echo time~(TE). More sophisticated linear constraints~(e.g. gradient moment nulling) can be modeled too, see details in~\cite{Chauffert_TMI_16}.
The normalized constraint $\alpha \delta t\leq \frac{1}{FOV \times K_{\rm max}}$ ensures that the distance between k-space locations associated with two consecutive measurements, sampled by the analog-to-digital converter~(ADC) at the dwell time period $\delta t$~(see Subsection~\ref{sec:from_traj_to_data} for the relationship between $\Delta t$ and $\delta t$), is lower than the Nyquist rate, which is essential to discard some undesired filtering effects~\cite{lazarus2020correcting}.

%Nevertheless, for the purpose of this algorithm, we only care about the gradient raster points as these points actually control the trajectory and need to fulfill the hardware constraints. We also map the Nyquist criteria constraints on k-space locations to our k-space trajectory $\bK$ in \eqref{eq:alpha_constraint} as described below. In the rest of the paper, we refer to k-space samples $\bK$ as the gradient raster points sampled at the $\Delta t$ period.

\subsection{3D SPARKLING formulation}

Let $\rho:\Omega\to \mathbb{R}$ denote a target sampling density, with $\rho(x)\geq 0$ for all $x$ and $\int \rho(x) \,dx=1$. 
Following previous works~\cite{graf2012quadrature,halftoning,curve_based_approx,Lazarus_MRM_19}, we obtain $\mathbf{K} \in \Omega^p$ by solving:
\begin{equation}
\label{eq:global_minimization}
\widehat{\mathbf{K}}= \arg \min_{\mathbf{K} \in \mathcal{Q}_{\alpha, \beta}^{N_c}}  F_p(\mathbf{K}) = \bigcro{F_p^{\rm a}(\mathbf{K}) - F_p^{\rm r}(\mathbf{K})} 
\end{equation}
with $\mathcal{Q}_{\alpha, \beta}^{N_c}$ being the constraint set for the $N_c$ shots. Here we remind that $p$ refers to the total number of k-space samples~(or particles), so $p=N_c\times N_s$.

The term $F_p^{\rm a}(\mathbf{K})$ corresponds to an attraction term which ensures that the final distribution of the k-space sampling points follows the target density $\rho$ and $F_p^{\rm r}(\mathbf{K})$ is the repulsion term to ensure that the sampling is locally uniform and that we don't have any local clusters. These terms are defined as:
 \begin{subequations} 
\begin{align}
\label{eq:attraction_term}
F_p^{\rm a}(\mathbf{K})& =  \frac{1}{p}\sum\limits_{i=1}^p \int_{\Omega} H(x-\bK[i]) \rho(x) \,dx \\
F_p^{\rm r}(\mathbf{K}) &= \frac{1}{2 p^2}\sum\limits_{1\leq i,j \leq p} H(\bK[i] -\bK[j])\label{eq:repulsion_term}
\end{align}
\end{subequations}
where $\bK[i]\in \Omega$ describe the locations of k-space samples in a shot-based lexicographical order $\left[\mathbf{k}_1, \dots \mathbf{k}_{N_c}\right]$. The function $H$ is a well chosen kernel, typically $H(x)=\|x\|_2$. Note that alternative choices such as $H(x)=\log (x)$ have been also investigated in~\cite{teuber2011dithering}. 
The minimization problem \eqref{eq:global_minimization} can be attacked by various nonlinear programming procedures. In this work, we propose to use a projected gradient descent as described below:
\begin{align}
\mathbf{K}^{(t+1)}&= \Pi_{\mathcal{Q}_{\alpha, \beta}^{N_c}} \left(\mathbf{K}^{(t)} - \eta^{(t)} \nabla{F_p(\mathbf{K}^{(t)})}\right) 
\label{eq:proj_grad_descent}
\end{align}
    
The computational bottlenecks in \eqref{eq:global_minimization} involve the calculation of $\nabla F_p(\bK)=\nabla F_p^{\rm a}(\bK) - \nabla F_p^{\rm r}(\bK)$, and the projection of each shot onto the constraint set $\mathcal{Q}_{\alpha, \beta}^{N_c}$.

\subsection{Gradient Descent Step}
In what follows, we provide details about the calculation of $F_p$ and $\nabla{F}_p$.

\subsubsection{Evaluating $F_p^{\rm a}$ and its gradient}
To calculate the attraction term and its gradient, we can re-write \eqref{eq:attraction_term} as: 
\begin{equation}
\label{eq:Fa_rewritten}
F_p^{\rm a}(\mathbf{K})=\frac{1}{p} \sum_{i=1}^p (H\star \rho) (\bK[i])
\end{equation}
where $\star$ denotes the convolution-product in the continuous setting. The main difficulty is thus to quickly evaluate $(H\star \rho) (x)$ (optional, if we want to compute the cost function) and its derivatives. To this end, we discretize the target sampling distribution $\rho$ as follows:
\begin{align}
\label{eq:discritize}
\brho[i,j,k] =  \rho (i/N,j/N,k/N) 
\end{align}
where $i,j,k\in[-N,N]$, and $N\in\N$ describes the number of discretization points. We typically take $N$ twice as large as $\max(N_x,N_y,N_z)$ to define the density at a better resolution than the image size.
Similarly we compute a discrete version of the filter $H$ as:
\begin{equation}
 \bH[i,j,k]  = H(i/N,j/N,k/N)
\end{equation}
Letting $*$ denote the discrete convolution-product, we use the following approximation
\begin{equation}
(H\star \rho) (\bK[i])\simeq \mathcal{I}(\bH\ast \brho)(\bK[i]),
\end{equation}
where $\mathcal{I}:R^{(2N+1)^3}\to C^0(\Omega)$ denotes a tri-linear interpolant function. 
Hence, the computation of $F_p^{\rm a}(\mathbf{K})$ requires to precompute $\bH * \brho$ on a discrete grid with fast Fourier transforms once for all. 
The computation of the sum in~\eqref{eq:Fa_rewritten} then has a complexity $O(p)$, which is linear in the number of particles.

%As regards the computation of 
Similarly, the computation of $\nabla F_p^{\rm a}(\bK)$ involves the calculation of the partial derivatives
$\partial_{i,l} F_p^{\rm a}(\bK)$ where $1\leq i \leq p$ is the index of a particle and $1\leq \ell \leq 3$ the index of a dimension. 
According to~\eqref{eq:Fa_rewritten}, the partial derivative is:
\begin{equation}
\partial_{i,\ell} F_p^{\rm a}(\bK) = \frac{1}{p} (\partial_\ell H \star \rho)(\bK[i]) 
\label{eq:attraction_derivative_basic}
\end{equation}
Thus, letting $\mathbf{\nabla H}\in \R^{(2N+1)^3\times 3}$ denote a discretization of $\nabla H$,  we can  precompute the discrete vector field $\mathbf{\nabla H} \ast \brho$ on a Cartesian grid using fast Fourier transforms and then use a tri-linear interpolant to evaluate it off the grid.

\subsubsection{Evaluating $F^{\rm r}$ and its gradient}

The problem addressed here is to compute $F_p^{\rm r}$ and $\nabla F_p^{\rm r}(\bK)$.
For purposes of simplification, we introduce $r_{ij} = \|\bK[i] - \bK[j]\|_2$ and consider H to be a radial function depending on $r_{ij}$ only. Letting $K_\ell$ denote the spatial components of $\bK=[K_1,K_2,K_3]$, we get:
\begin{subequations}
\begin{gather}
F_p^{\rm r}(\bK) = \sum_{1\leq i,j\leq p} H(r_{ij}) \label{eq:Fr}\\
\partial_{i,\ell} F_p^{\rm r}(\bK)  = \frac{1}{p^2} \sum_{j\neq i} \left(\frac{K_\ell[i]-K_\ell[j]}{r_{ij}}\right) \partial_\ell H(r_{ij})
\label{eq:Fr_grad}
\end{gather}
\end{subequations}

The evaluation of all the components of the gradient require $O(p^2)$ computations, where $p$ can reach $10^8$ for high resolution imaging. An efficient implementation is therefore critical. In this work, we explored two possibilities. 

\paragraph{Brute force calculation using PyKeops}

The computation of \eqref{eq:Fr} and \eqref{eq:Fr_grad} can be highly parallelized, which is amenable to efficient GPU implementations. Carrying out such computations on array centric frameworks like PyTorch and Tensorflow would require the use of huge $p \times p$-dimensional matrices. This would result in a large memory footprint, much larger than what is typically available on current modern GPUs. For the sake of efficient memory usage, we used \texttt{PyKeops}, a library that permits low cost calculations of large kernel operations~\cite{charlier2020kernel}. \texttt{PyKeops} carries out the naive and direct computations using online map reduce schemes from CUDA routines for summations. Due to this, the whole matrices are not stored in the GPU memory, but rather just the final results.

\paragraph{Fast Multipole Methods}

Sums of the form \eqref{eq:Fr} and \eqref{eq:Fr_grad} appear in many n-body problems and can be computed efficiently using Fast Multipole Methods~(FMM)~\cite{fong2009black}. 
Given a set of positions $\bK[i]$, a kernel $\Psi:\R^d\times \R^d \to \R$ and a set of weights $\bw \in \R^p$, the FMM method allows for the efficient computation of vector $\bv$ of the form
\begin{align*}
\bv[i] &= \sum_{j=1}^p \Psi(\bK[i],\bK[j]) \bw[j] \, .
\end{align*}
The FMM utilizes a multipole expansion of the kernel $\Psi$, which allows for a hierarchical grouping of closely spaced k-space points and treat them as a single source. This results in a massive acceleration of the above computation with a complexity $O(p \log{p}/\epsilon)$, where $\epsilon$ is a user-prescribed precision.
For our implementations, we used the Parallel Black box FMM \cite{pbbfmm3d,darve2009} in 3D~(\texttt{PBBFMM3D}\footnote{see \url{https://github.com/ruoxi-wang/PBBFMM3D}.}), which can be run with any arbitrary kernel $\Psi$. 

To evaluate the cost function $F_p^{\rm r}$, we only need to set 
\begin{align*}
\Psi(\bK[i],\bK[j]) &= H(r_{ij}) \mbox{ and } \bw[j]=1, \forall j.
\end{align*}
To evaluate the gradient $(\partial_{i,\ell}F_p^{\rm r}(\bK) )_i$, we set $\bw[j]=1$ and
\begin{align*}
\Psi(\bK[i],\bK[j]) &= \left(\frac{K_\ell[i]-K_\ell[j]}{r_{ij}}\right) \partial_\ell H(r_{ij})\, .
\end{align*}

\paragraph{Comparisons}

\begin{figure}
	\centering
	\includegraphics[width=0.4\textwidth]{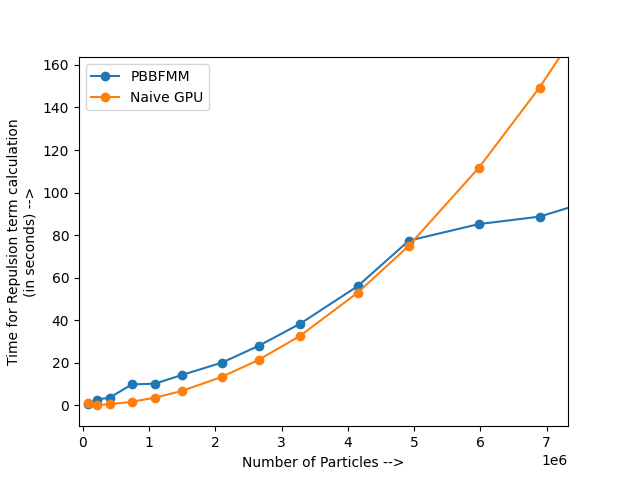}
	\caption{Computation times for the repulsion term $F^{\rm r}$ as a function of the number of particles $p$.}
	\label{fig:repulsion_benchmark}
\end{figure}

From Fig~\ref{fig:repulsion_benchmark}, we see that naive GPU implementations on \texttt{PyKeops} outperforms the \texttt{PBBFMM3D} implementation for $p<5\times 10^6$. Beyond this value, \texttt{PBBFMM3D} gets faster.
It is likely that faster computations with the FMM would be obtained with a GPU implementation. Unfortunately, we did not find any robust and efficient GPU implementation of FMM.

\subsubsection{Choice of step size}
In our implementation, we use a combination of two step sizes. 
In the first 20 iterations, we use a fixed step size: $\eta^{(t)}= \eta$. As analyzed in \cite{chauffert2017projection}, this strategy provides a convergence guarantee to a local minimizer of the cost function given that:
\begin{itemize} 
	\item The kernel $H$ has a $L$-Lipschitz continuous gradient. 
	\item The step size is inversely proportional to the Lipschitz constant.
\end{itemize}
These conditions are satisfied with a regularized norm of the form $H(r)=\sqrt{r^2+\epsilon^2}$. We can then set $\eta^{(t)}$ proportional to $\epsilon$~(i.e. $6.25^{-2}$). The value of $\epsilon$ can be chosen as a fraction of the minimal distance between two points at a stationary point. 

A constant step size is too conservative and a faster convergence can be obtained using a second-order dynamics close to the minimizer. This justifies switching to a Barzilai$-$Borwein \cite{barzilai_borwein_step} after first few iterations. Few theoretical guarantees are available for this technique, but it significantly accelerates the convergence empirically.

\subsection{Projection step}
\label{sec:proj_step}
\begin{figure}
\rem{	\begin{subfigure}{0.24\textwidth}
		\hspace*{.2cm}\includegraphics[width=.95\textwidth]{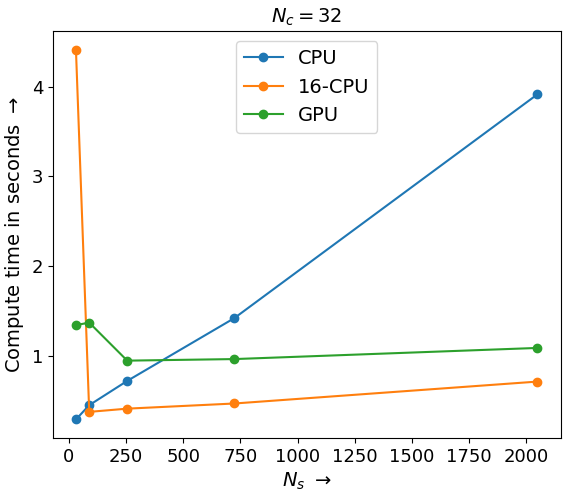}
	\end{subfigure}
	\begin{subfigure}{0.24\textwidth}
		\includegraphics[width=\textwidth]{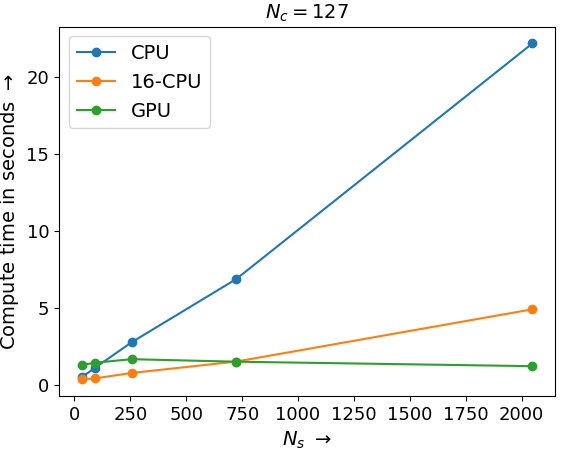}
	\end{subfigure}
	\begin{subfigure}{0.24\textwidth}
		\includegraphics[width=\textwidth]{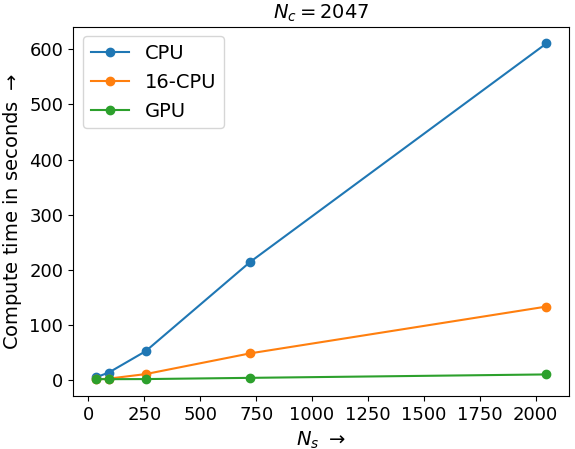}
	\end{subfigure}
	\begin{subfigure}{0.24\textwidth}
		\includegraphics[width=\textwidth]{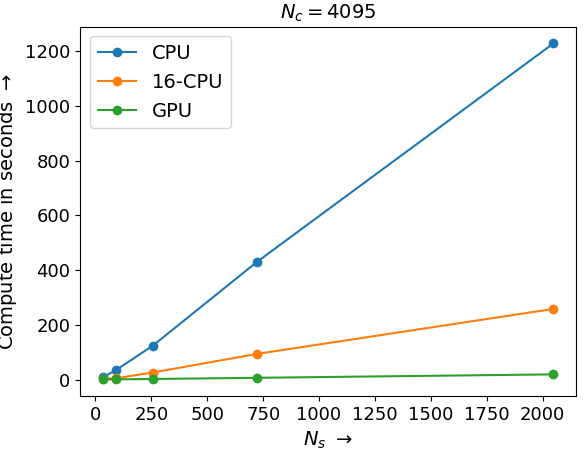}
	\end{subfigure}
}
\begin{center}
\begin{tabular}{@{}c@{}c}
\hspace*{.15cm}	\includegraphics[width=.225\textwidth]{./images/theory/projection_time/N_c32_cropped.png} & 
			\includegraphics[width=.24\textwidth]{./images/theory/projection_time/N_c127_cropped.png} \\
	\includegraphics[width=.24\textwidth]{./images/theory/projection_time/N_c2047_cropped.png} & 
\includegraphics[width=.24\textwidth]{./images/theory/projection_time/N_c4095_cropped.png}	
	\end{tabular}
\end{center}\vspace*{-.5cm}
	\caption{Computation times for varying $N_s$ and $N_c$ for the projection step $\Pi_{\mathcal{Q}_{\alpha, \beta}^{N_c}}(\cdot, n_{pit})$ that was run over $n_{pit}=200$ iterations with $G_{\rm max}=40$mT/m and $S_{\rm max}=180$T/m/s.} 
	\label{fig:projection_compute_time}
\end{figure}

The projection step in~\eqref{eq:proj_grad_descent} for a general single k-space shot onto a given constraint set parameterized by $(\alpha, \beta)$ has been explored in \cite{Chauffert_TMI_16, chauffert2017projection}. For our implementations, we needed to extend the single shot iterative procedure called Algorithm 1 in~\cite{Chauffert_TMI_16} to projecting $N_c$ shots~(see Sec.~\ref{supp-app:projection} for details). Note that the actual projection of a k-space shot is independent of other shots, and thus the computation can be done in parallel. Hence, we have implemented this step both on multi-CPU and GPU. To efficiently utilize a GPU, we used the CuPy module~\cite{cupy}. The computation times with different implementations for varying $N_c$ and $N_s$ are shown in Fig.~\ref{fig:projection_compute_time}. We found that the computation times vary linearly with $N_s$ and are drastically reduced for the GPU implementation compared to the CPU versions~(single and multicore).
\revcomment{At lower $N_s$, we found that m-CPU and GPU implementations are latency bound, giving anomalously higher computation times. However, the speedup obtained for larger $N_s$ offsets these anomalous cases, giving an overall efficient implementation.}{r:cpu_cost}

\subsection{Multi-resolution strategy for faster convergence}\label{multi-resolution}

In order to allow for the algorithm to reach faster convergence and lead to a better approximation of the target density, a multi-resolution approach as described in~\cite{lebrat_optimal_2019} was implemented. Under this methodology, the optimization of the sampling pattern was carried out on down-sampled curves. The interpolated solution was later used as a warm restart for the up-sampled problem. Our implementations involved dyadic scaling and up-scaling through simple linear interpolation of k-space shots. Let the linear interpolator be of the form $\mathcal{L}_{2d} : \Omega^d \rightarrow \Omega^{2d}$. We define the parameter $N_d$ as the number of decimation steps in the algorithm. Note that the constraint space needs to be equally scaled with the problem, which results in scaling the $\alpha$ and $\beta$ constraints mentioned in~\eqref{eq:constraints_alpha_beta} to:
\begin{align}
\label{eq:nd_constraints}
\alpha &= \frac{\gamma G_{\rm max} 2^{N_d}}{2\pi K_{\rm max}}, \quad 
\beta = \frac{\gamma S_{\rm max} 2^{N_d}}{2\pi K_{\rm max}} 
\end{align}
 
As we move through the dyadic decimation steps and up-sample the curve, these constraints are halved.

\subsection{Overall algorithm}
Algorithm~\vref{alg:overall_algorithm} summarizes how to concretely compute the SPARKLING solution along with multi-resolution steps described in Sec.~\ref{multi-resolution}. Further, for the sake of completeness, we have briefly summarized the multi-shot projection step $\Pi_{\mathcal{Q}_{\alpha, \beta}^{N_c}}$ in Sec.~\ref{supp-app:projection} of Supplementary Material.  However, for more details on the iterative procedure involved in the projection step, the reader can refer to~\cite{Chauffert_TMI_16}.

\begin{algorithm}
\caption{Multi-resolution implementation of SPARKLING}
\label{alg:overall_algorithm}
\SetAlgoLined
\SetKwInOut{Inputs}{Inputs}
\Inputs{$\rho$, $G_{max}$, $S_{max}$, $N_c$, $N_s$, $N_d$, $n_{git}$, $n_{pit}$} 
\KwOut{$\bK$, the k-space sampling pattern}
\textbf{Initializations:} $\mathbf{K}^{(0)} \in \Omega^{\frac{N_c \times N_s}{2^{N_d}}} = \Omega^p$\\
$\alpha \gets \frac{\gamma G_{\rm max} 2^{N_d}}{2\pi K_{\rm max}} $, 
$\beta \gets \frac{\gamma S_{ \rm max} 2^{N_d}}{2\pi K_{\rm max}} $\\

\While{$N_d > 0$}
{
	$p \gets \frac{N_c \times N_s}{2^{N_d}}$\\
    \For{$t = 1 \dots n_{git}$}
    {
        $\mathbf{K}^{(t-1/2)} = \mathbf{K}^{(t-1)} - \eta^{(t)} \nabla{F_p(\mathbf{K}^{(t-1)})}$\\
        $\mathbf{K}^{(t)}= \Pi_{\mathcal{Q}_{\alpha, \beta}^{N_c}} \left(\mathbf{K}^{(t-1/2)}, n_{pit}\right)$ 
    }
	\tcp{Warm restart next decimation step with linear interpolation}
	\tcp{The dimension of k is doubled at each decimation step}
	\For{$s = 1 \dots N_c$}
	{
		$\mathbf{k}_s^{(0)} \gets \mathcal{L}_{\frac{N_s}{2^{N_d-1}}}\left(\mathbf{k}_s^{(n_{git})}\right)$
	}
	$\mathbf{K}^{(0)} \gets \left[\mathbf{k}_1^{(0)},\ldots, \mathbf{k}_{N_c}^{(0)} \right]$\\
	\tcp{Scale constraints}
    $\alpha \gets \frac{\alpha}{2}, \beta \gets \frac{\beta}{2}$\\
    $N_d \gets N_d - 1$
}
\end{algorithm}

\section{Numerical experiments and data acquisition}
\label{system_specs}
The sampling patterns were obtained by carrying out projected gradient descent as described above. With the above described improvements, the SPARKLING Generation time was just 10 minutes for 2D and nearly 6-9 hours for 3D on NVIDIA V100 with 5120 CUDA cores and 16GB DDR5X memory. 

\subsection{SPARKLING: a Python package}
In the ethos of reproducible research and to move forward into better optimized patterns for MRI acquisition, all the implementations as described above is present in a Python package at the private repository\footnote{\url{https://gitlab.com/cea-cosmic/CSMRI_sparkling}}. All codes in the package scale to 2 and 3 dimensions directly, and most codes are agnostic and can be run on CPU or GPU with some change in run parameters. All the scanner constants and trajectory specification can be provided through a configuration file, and most of the codes are modular in nature. Interested researchers are requested to contact the authors for obtaining access to this package\footnote{It cannot be made open source given patent application.}.

\subsection{Acquisition parameters}

With a goal of 600\textmu m isotropic resolution in 3D MRI acquisitions, we planned to obtain a volume of $(N_x  \times N_y  \times N_z) = (384 \times 384 \times 208)$ size in order to cover the whole brain. For the sake of consistency, we used the same matrix size and resolution for our acquisitions on the NIST phantom\footnote{\url{https://www.nist.gov/programs-projects/quantitative-mri}}. The trajectories were generated for a clinical 3T MR system (Magnetom $\rm Prisma^{\rm FIT}$, Siemens Healthcare, Erlangen, Germany) with maximum gradient strength $G_{\rm max}=40$mT/m and peak slew rate $S_{\rm max}=180$T/m/s.
\revcomment{As the readout time was set to
$T_{\rm obs}=20.48$ms and the gradient raster time is $\Delta t=10\mu$s, the number of samples per shot $\bk_i$, was $N_s=2048$.}{r:constraints}
%while the associated k-space data measurements $\by_i$, were eventually acquired at the dwell time of $\delta t = 2\mu$s. We actually used linear interpolation to map the $\Delta t$-sampled optimized trajectory to the faster sampling period $\delta t$.}{r:constraints}
%The number of samples per shot was set to $N_s=2048$
The number of shots $N_c$ was varied based on the study described hereafter. For our in vivo studies, the k-space data was acquired on a Siemens 64 channel Head/Neck coil, while using 44 channels around head during acquisition. The echo time (TE) was $20$ms and repetition time (TR) was $37$ms. The flip angle was set to $15$\textdegree, and the slice excitation was slab selective. We also obtained a reference volume collected using a 4-fold accelerated Cartesian acquisition~(acquisition time or TA=15min~13sec) based on GRAPPA parallel imaging technique~\cite{Griswold02} \revcomment{with the same TE, TR and $T_{\text{obs}}$}{r:off_res}. The projected gradient descent was carried out with multi-resolution decimation steps $N_d=6$ for faster convergence. The algorithm was run for $n_{git}=100$ outer gradient descent iterations with $n_{pit}=100$ steps in the inner projection loop.

\subsubsection{Choice of target sampling density}
\label{target_distribution}
\begin{figure}
	\centering
	\includegraphics[width=0.4\textwidth]{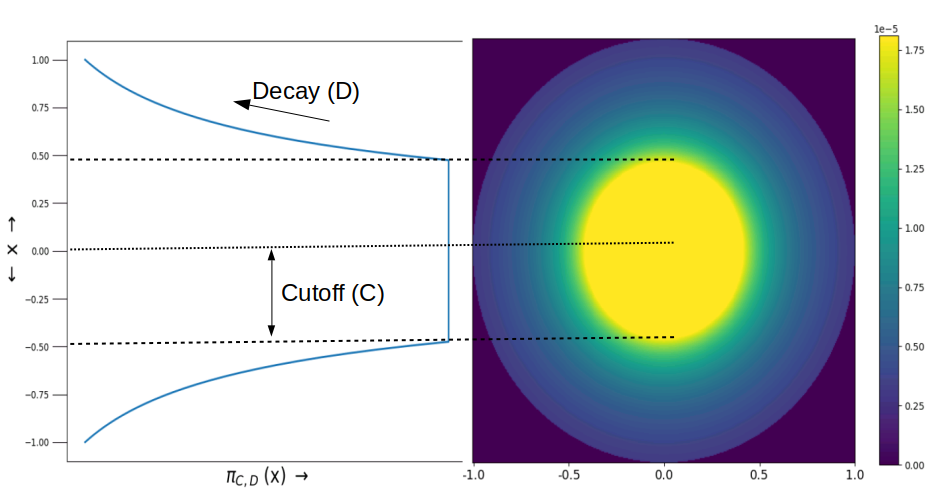}
	\caption{Parameterization of variable density with cutoff $C$ and decay $D$.}
	\label{fig:Density_Parametrization}
\end{figure}

The target sampling density was chosen to be radially isotropic, which decays as an inverse polynomial with a constant plateau in the center of k-space. The density was defined with $C$, the cutoff frequency in k-space center having a constant density and $D$, the rate of decay for higher frequencies. Mathematically, we define the target density $\pi_{C,D}(x): \Omega^d \rightarrow \mathbb{R}_+$ as follows:
\begin{equation}
\label{eq:target_distribution}
    \pi_{C,D}(x) = 
    \begin{cases}
        \kappa &\quad|x| < C \\
        \kappa \left(\frac{C}{|x|}\right)^D &\quad|x| > C
    \end{cases}
\end{equation}\\

\noindent where $\kappa$ is a constant obtained through normalization as $\kappa = \frac{1-D}{2 C (C^{D-1} - D) }$. The resulting density is radially symmetric and is of the form described in Fig.~\ref{fig:Density_Parametrization}. The choice of a radial density was motivated by the wish to provide rotation invariant reconstruction results. Notice that the recent learning based approaches~\cite{sherry2020learning,bahadir2020deep} result in non-symmetric densities. This is probably due to the fact that brains or knees databases such as fastMRI~\cite{zbontar2018fastmri} used for training have boundaries which are dominantly vertical or horizontal. However, the fine details may be in arbitrary orientations.

The correct choice of density was carried out by grid searching for the optimal parameters $\hat{C}$ and $\hat{D}$ on the target sampling distribution as defined in~\eqref{eq:target_distribution}. We performed retrospective reconstruction on \multirevcomment{complex Cartesian reference (with phase to account for off-resonance artifacts by phase accrue) in vivo brain data.}{\ref{r:complex_images}, \ref{r:blurr_r3}, \ref{r:blurry_r4}} For this study, the number of shots was set to $N_c=4096$. \revcomment{We found that $\hat{C}=25\%$ and $\hat{D} = 2$ provided the highest and most robust SSIM score of 0.97.}{r:blurr_r3} The SSIMs for varying values of $C$ and $D$ are presented in Supplementary Material (see Tab.~\ref{supp-fig:VD_Study_Retro}).

\begin{figure}[h]
	\centering
	\resizebox{\linewidth}{!}{
		\begin{tabular}{@{}c@{\hspace*{1mm}}c@{\hspace*{1mm}}c@{\hspace*{1mm}}c}
			&{\bf Initialization} & {\bf Generated Trajectory} &
			\\
			\rotatebox[origin=c]{90}{(a) P=0.25}&
			\parbox[m]{.35\linewidth}{
				\includegraphics[trim={7cm 3.2cm 5.5cm 4cm},clip, width=\linewidth]{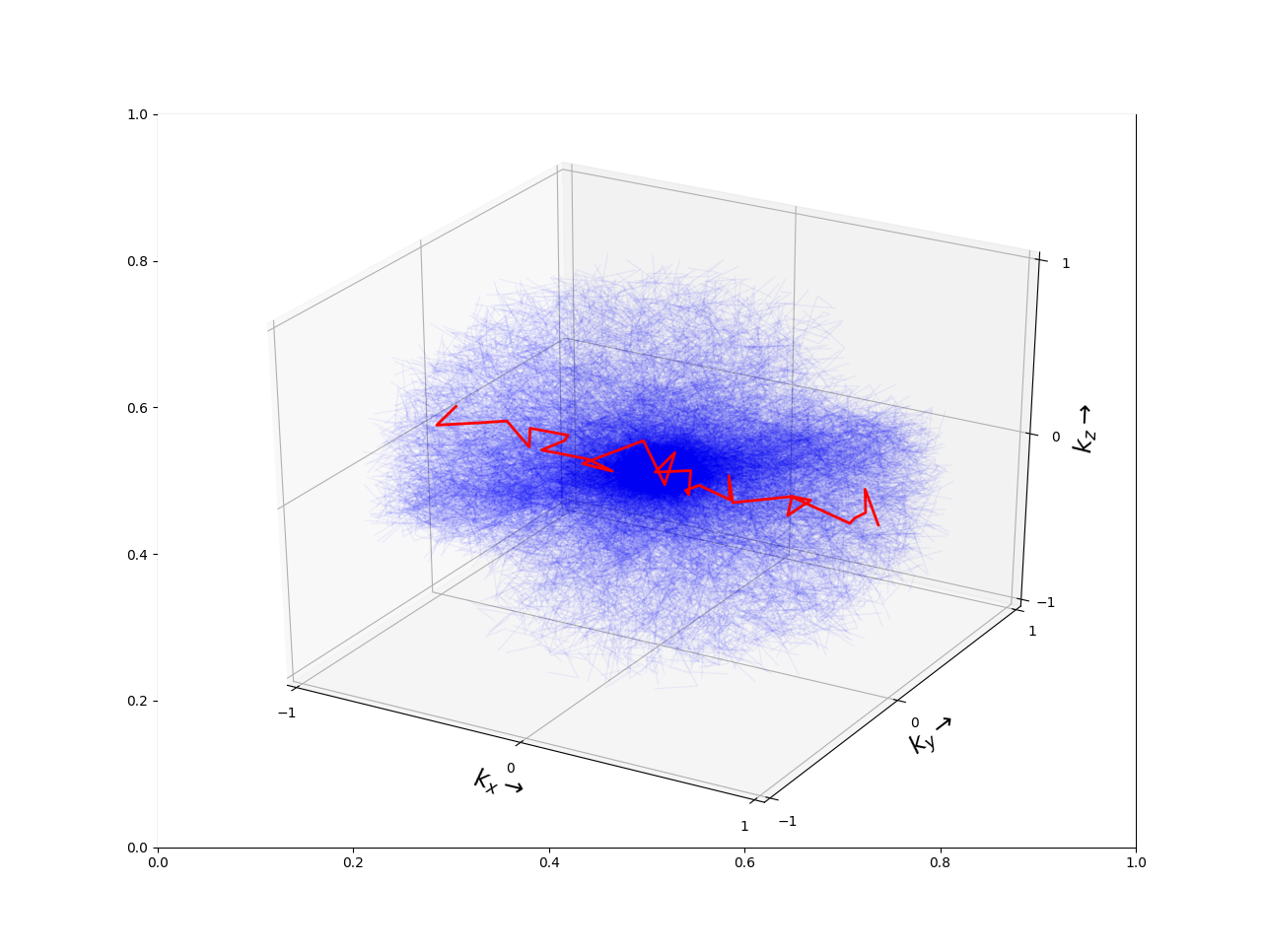}}&
			\parbox[m]{.55\linewidth}{
				\includegraphics[trim={7cm 3.2cm 5.5cm 4cm},clip, width=\linewidth]{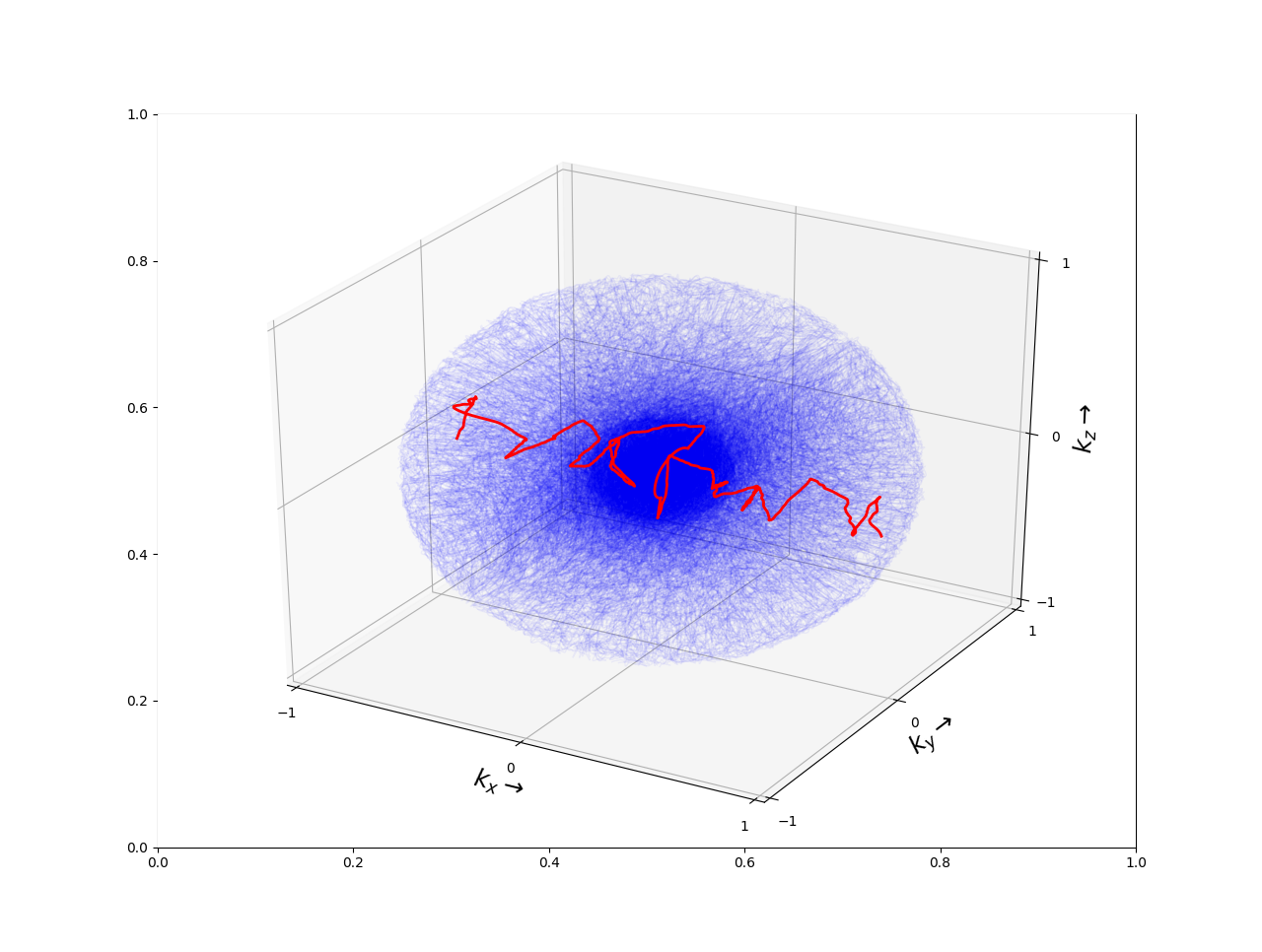}} &
			\rotatebox[origin=c]{270}{\footnotesize \bf Cost = 0.000757}
			\\
			\rotatebox[origin=c]{90}{(b) P=0.75}&
			\parbox[m]{.35\linewidth}{
				\includegraphics[trim={7cm 3.2cm 5.5cm 4cm},clip, width=\linewidth]{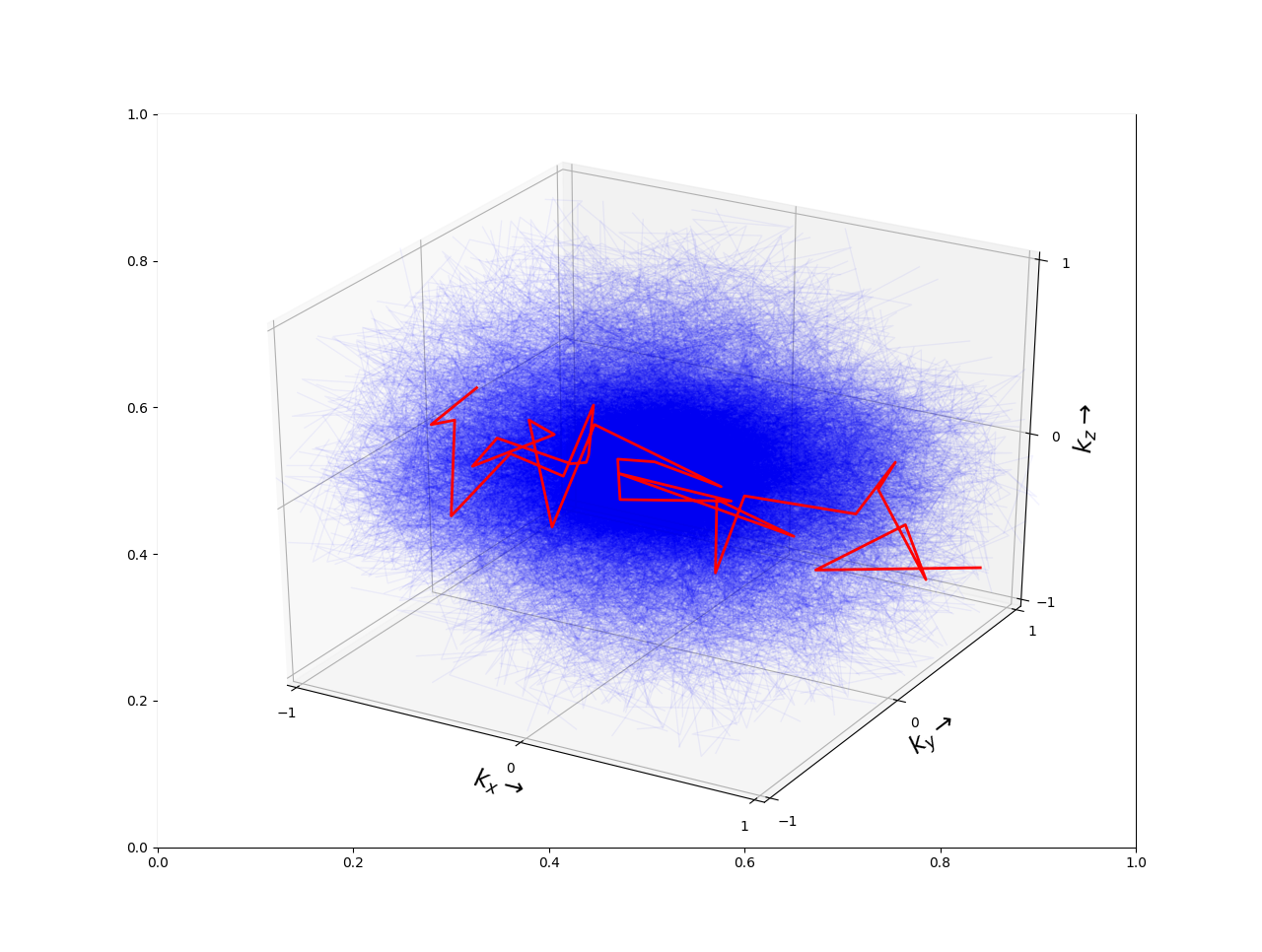}}&
			\parbox[m]{.55\linewidth}{
				\includegraphics[trim={7cm 3.2cm 5.5cm 4cm},clip, width=\linewidth]{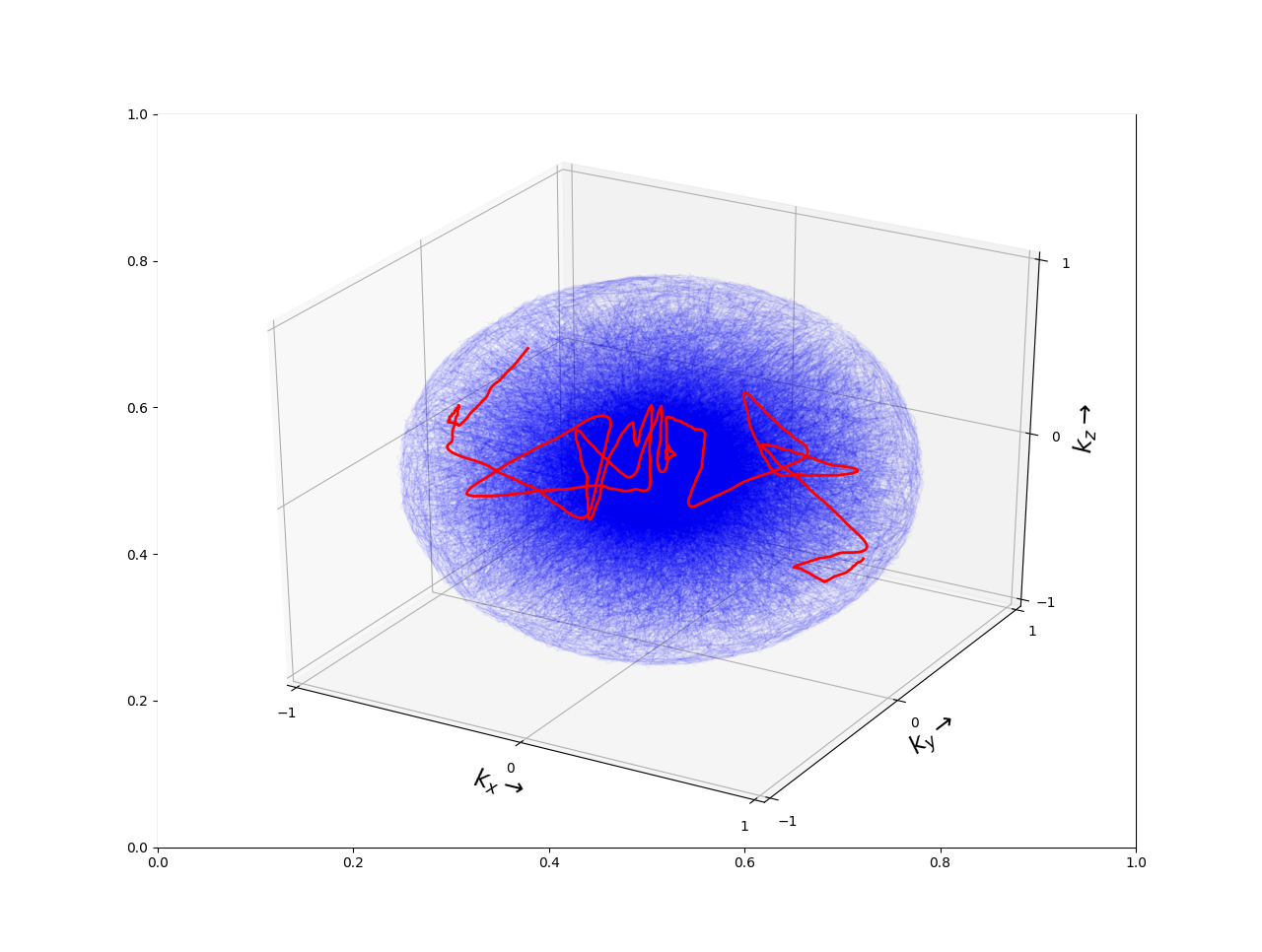}} &
			\rotatebox[origin=c]{270}{\footnotesize \bf Cost = 0.000748}
			\\
		\end{tabular}}
	\caption{Effect of adding a perturbation (P) to the initial k-space trajectory in  $\Omega = [-1, 1]^3$ as zero mean uniform random noise at each trajectory sample. Trajectories are generated with maximum displacement of k-space point to (a) $0.25$ and (b) $0.75$ in the initialization. The left side of the figure is the initialization to SPARKLING algorithm and the right is the output of the algorithm. We also present the values of the cost obtained with \eqref{eq:global_minimization}.
		\label{fig:add_perturbationb}}
\end{figure}

\subsubsection{Initialization and Perturbation}
\label{sec:init_n_perturb}
As the problem being solved is non-convex, different choices of initialization would lead to different solutions. In~\cite{Lazarus_MRM_19} for 2D imaging we observed that radial initialization performed the best for exploring the k-space. Hence, here for 3D imaging we stick to 3D radial initialization too. \multirevcomment{For the sake of simplicity, and also to ensure radially symmetric initialization, we set up the trajectories with $\sqrt{N_c}$ shots in x-y plane and then rotate each shot $\sqrt{N_c}$ times along an in-plane axis orthogonal to the shot. More generic solution can be obtained by solving for the minimum electrostatic potential energy configuration of $N_c$ electrons over the surface of a unit sphere, however this approach was not pursued in this work.}{\ref{r:other_init}\\\ref{r:radial_init}}

As the optimization problem in \eqref{eq:global_minimization} is non-convex, the resulting solution is heavily dependent on the initialization. For best reconstructed image results, we would want each k-space shot to maximally explore the k-space. The 3D radial initialization is too structured with each k-space shot traveling only from end to end of k-space. To enable a broader k-space exploration and obtain a better minimizer of the original problem, we added a perturbation to each initial shot. To achieve this, we perturbed each trajectory sample point in k-space by adding zero mean uniform random noise along each dimension. Particularly, we compared the resulting optimized trajectory obtained after a perturbation as a random motion of each k-space point with maximum amplitude set at $0.1$ and $0.75$ (we remind that the sampling domain is normalized to $\Omega = [-1, 1]^3$). The optimized trajectory patterns are presented in Fig.~\ref{fig:add_perturbationb}. We clearly show that with more perturbation, the k-space trajectory tends to explore a broader part of k-space giving a better coverage overall. \revcomment{We also notice quantitatively that with more perturbation the value of the cost function converges to a lower local minimum.}{r:perturb_quantify} Further, we would like to emphasize that these trajectories are particularly useful in cases of high receiver sampling rates, as they would then sample more of the k-space per shot and would overall prevent the presence of any hole in the sampling pattern.

\section{Results}
\label{sec:results}

\subsection{From trajectories to actual k-space data}
\label{sec:from_traj_to_data}
 
\multirevcomment{The k-space data $\bY=(\by_i)_{i=1}^{N_c}$ is sampled by the ADC at the dwell time period $\delta t$. In practice, the dwell-time $\delta t$ is a fraction of the raster time $\Delta t$ and was set to $\delta t = 2\mu$s. This means that $\by_i \in \mathbb{C}^{m}$ with $m= N_s \left\lfloor\frac{\Delta t}{\delta t}\right\rfloor$ the number of measurements per shot. Overall, we collect $M = N_c\, m $ k-space data points in $\bY$.
Consequently, during the image reconstruction process, we obtain the k-space locations of $\bY$ by linearly interpolating the optimized trajectory $\widehat{\mathbf{K}}$ originally sampled at $\Delta t$, to the $\delta t$ period.
}{\ref{r:constraints}\\\ref{r:samples_vs_loc}}

\subsection{Assessment of point spread function}
\label{sec:traj_n_psf}
\begin{figure}[h!]
	\includegraphics[width=0.5\textwidth]{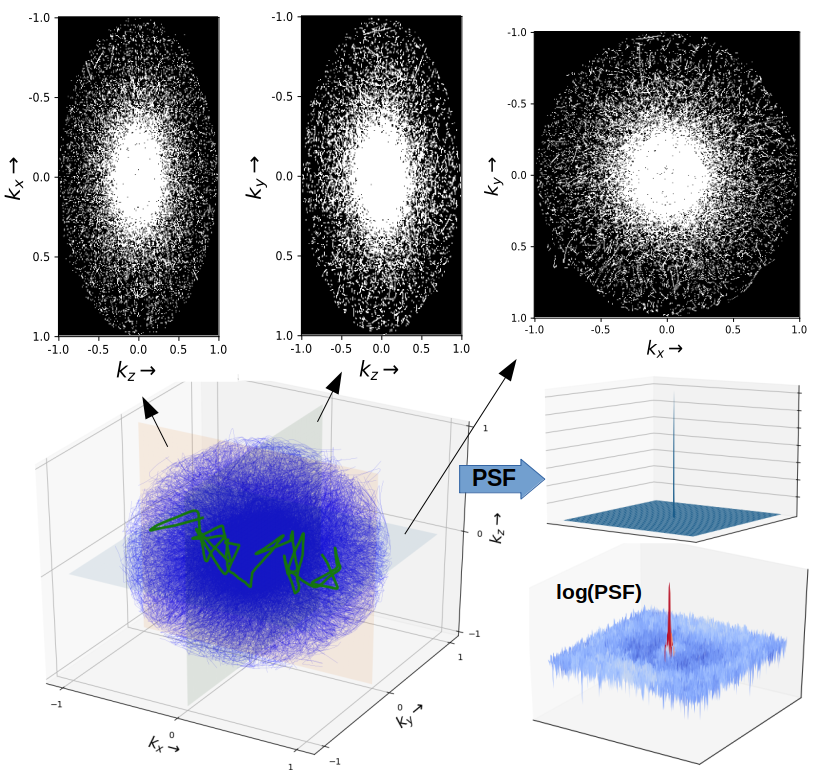}
	\caption{Full 3D SPARKLING Trajectory for $N_c = 4096, N_s=2048$ and the point spread function~(PSF) along the mid z-plane computed from the sampling mask~(measurements sampled at the dwell-time period $\delta t$).}
	\label{fig:full3D_trajectory}
\end{figure}

\begin{figure*}[h]
	\resizebox{\linewidth}{!}{\centering
		\begin{tabular}{c@{\hspace*{1mm}}c@{\hspace*{1mm}}c@{\hspace*{1mm}}c@{\hspace*{1mm}}c}
			\stackunder[1pt]{\parbox[m]{.07\linewidth}{
				\includegraphics[clip,width=\linewidth]{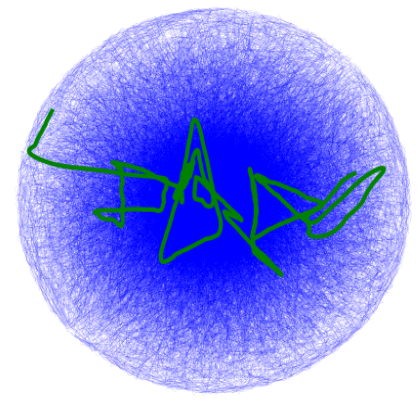}}}{\thead{\bf Full 3D \\ \bf SPARKLING}}&
			\parbox[m]{.15\linewidth}{
				\includegraphics[trim={2.5cm 0.5cm 3.2cm 1cm},clip,width=\linewidth]{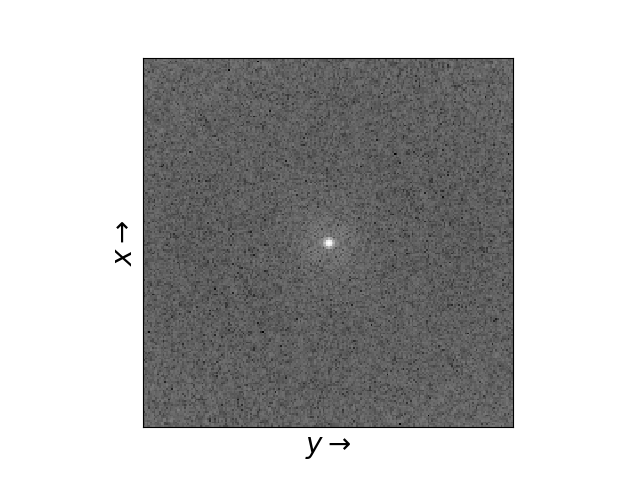}}&
			\parbox[m]{.15\linewidth}{
				\includegraphics[trim={2.5cm 0.5cm 3.2cm 1cm},clip,width=\linewidth]{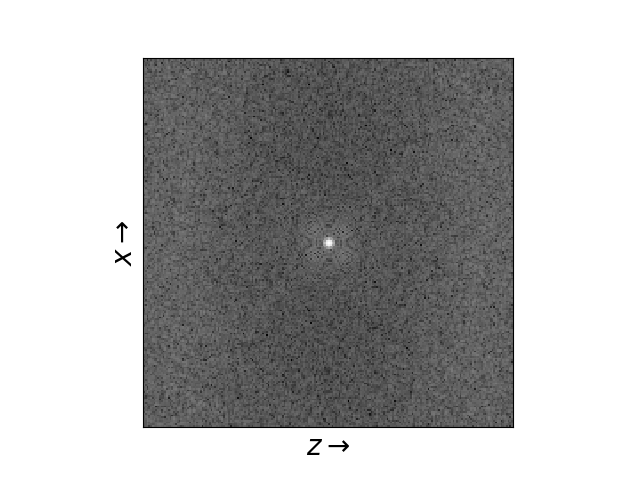}}&
			\parbox[m]{.15\linewidth}{
				\includegraphics[trim={2.5cm 0.5cm 3.2cm 1cm},clip,width=\linewidth]{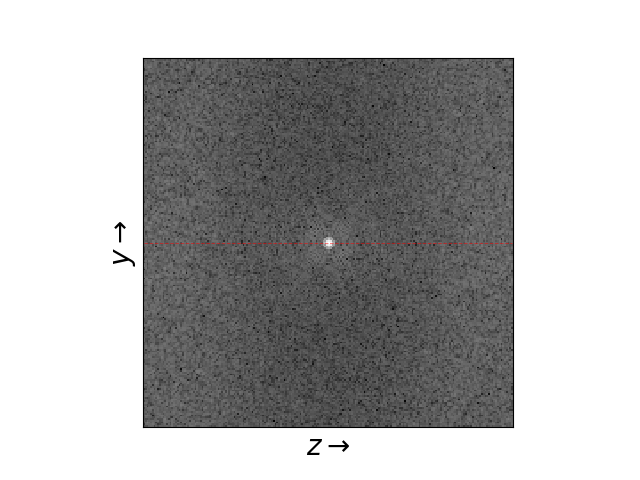}}&
		\multirow{2}{*}[0.4in]{\parbox[]{.45\linewidth}{
				\includegraphics[trim={1.5cm 0.6cm 2.4cm 1cm},clip,width=\linewidth]{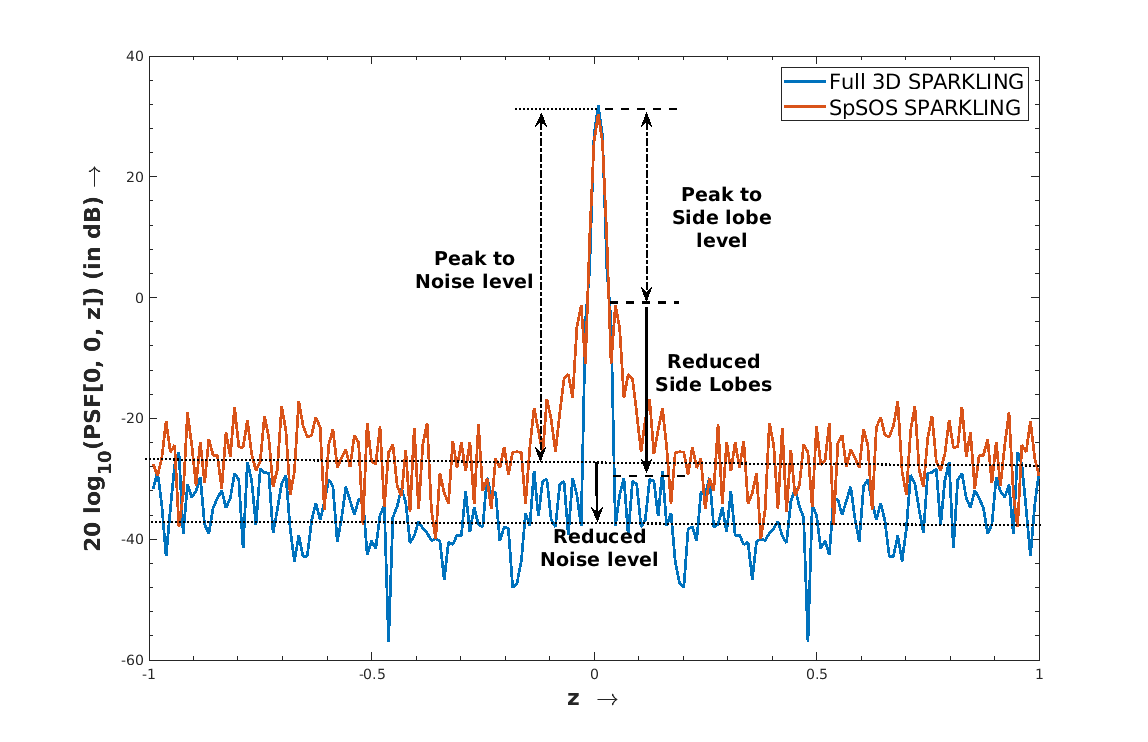}}}
			\\
			\stackunder[1pt]{\parbox[m]{.07\linewidth}{
				\includegraphics[clip,width=\linewidth]{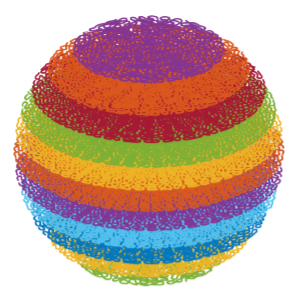}}}{\thead{\bf SpSOS \\ \bf SPARKLING}}&
			\parbox[m]{.15\linewidth}{
				\includegraphics[trim={2.5cm 0.5cm 3.2cm 1cm},clip,width=\linewidth]{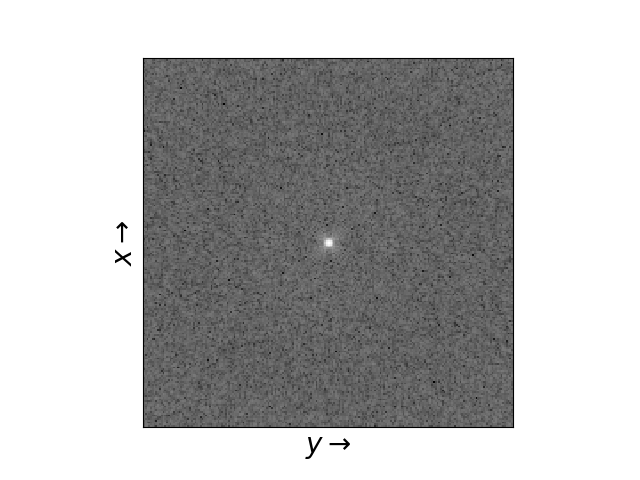}}&
			\parbox[m]{.15\linewidth}{
				\includegraphics[trim={2.5cm 0.5cm 3.2cm 1cm},clip,width=\linewidth]{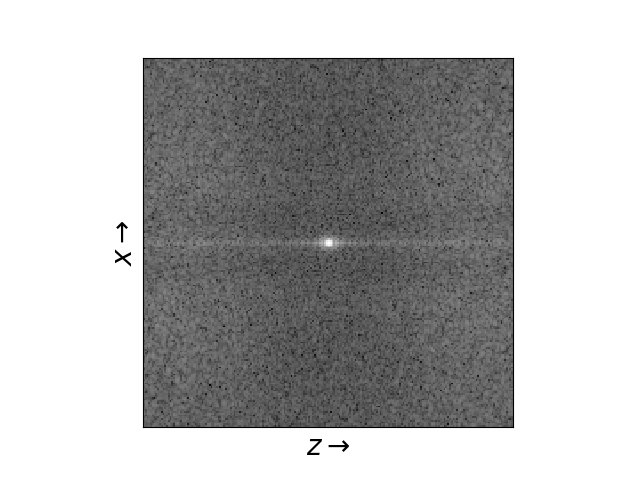}}&
			\parbox[m]{.15\linewidth}{
				\includegraphics[trim={2.5cm 0.5cm 3.2cm 1cm},clip,width=\linewidth]{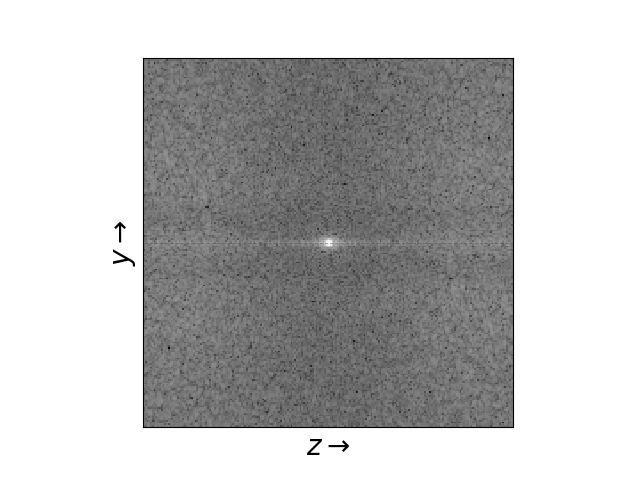}}&
			\\\\
			& {\bf (a)}& {\bf (b)}& {\bf (c)}& {\bf (d)}
	\end{tabular}}
	\caption{Comparison of PSFs between full 3D SPARKLING and SpSOS sampling masks~(measurements collected at the dwell-time period over the corresponding trajectories). The logarithm of 3D PSFs (in voxel units) are viewed along the mid-slices in {\bf (a)} axial plane (x, y, 0), {\bf (b)} sagittal plane (x, 0, z) and {\bf (c)} coronal plane (0, y, z). {\bf (d)} The PSFs are compared in logarithmic scale along the $z$ direction.}
	\label{fig:compare_psf}
	
\end{figure*}

We present the full 3D SPARKLING, obtained with $N_c=4096$ in Fig.~\ref{fig:full3D_trajectory}. We visualize the trajectory along the mid-planes of 3 orthogonal orientations and provide an approximate sampling mask in these planes. Further, to understand why these trajectories are expected to yield good image reconstructions, we measure the 3D point spread function~(PSF). \revcomment{Each point spread function was computed by taking a density compensated NUFFT adjoint of k-space measurements set to 1~($y_{i}[n]=1, \forall i=1,\cdots, N_c, \forall n=0, \cdots, m -1$) as described in \cite{pauly_gridding}.}{r:calc_psf} 
%\textbackslash\{\}footnote\{sampled along each shot at the dwell-time \$\textbackslash\{\}delta t\$ period.\}
In Fig. \ref{fig:compare_psf} we compare the PSFs with respect to earlier generated spherical stack of 2D SPARKLING~(SpSOS) trajectories. Particularly, we emphasize the reduction in sidelobes along the $z$ axis. Further, for the purpose of numerical comparison, we computed the full width at half maximum~(FWHM), peak-to-sidelobe level~(PSL) and peak-to-noise level~(PNL) in Tab.~\ref{table:compare_psf_metrics}. \revcomment{The FWHM is calculated as the width of the peak of PSF at half of the maximum value and the PSL and PNL are calculated in dB as presented in Fig.~\ref{fig:compare_psf}(d).}{r:psf_metrics}

\begin{table}[h!]
	\centering
	\caption{Comparing metrics of PSF with FWHM~(lower is better), PSL and PNL~(higher is better).}
	\begin{tabular}{*6c}
		\toprule
		Trajectory 	& \multicolumn{3}{c}{\parbox{2.3cm}{\centering FWHM \\ \footnotesize(in voxel units)}} 
					& \parbox{1cm}{\centering PSL \\ \footnotesize(in dB)} 
					& \parbox{1cm}{\centering PNL \\ \footnotesize(in dB)}\\
		\midrule
		{}   					&	x   &  y 	&  z	& 					&				\\
		Full 3D   			&  2.3 	& 2.4  	& 2.5  	& \textbf{35.80}	& \textbf{67.25}\\
		SpSOS  		&  2.4 	& 2.4  	& 2.5 	&   	31.65		& 		65.44	\\
		\bottomrule
	\end{tabular}
	\label{table:compare_psf_metrics}
\end{table}
As shown in Fig.~\ref{fig:compare_psf}, we see that the full 3D pattern provides us with much higher PSL~(4.15 dB more) and PNL~(1.81 dB more), two quantitative indices that demonstrate the full 3D SPARKLING methodology outperforms the spherically stacked version. \revcomment{In contrast, we observe that the FWHM is nearly the same for both methods, even though the FWHM along the $x$ axis is slightly lower for the full 3D pattern. However, this minor difference in FWHMs and the slight anisotropy in FWHM can be explained by the fact that the full 3D initialization was severely perturbed~(0.75) as described in Sec.~\ref{sec:init_n_perturb}.}{r:fwhm_isotropic}

\subsection{Non-Cartesian MR image reconstruction}

All MR images that rely on non-Cartesian k-space data in this paper were reconstructed using a self-calibrated synthesis-based CS reconstruction algorithm~\cite{pipe_dc, Knoll2014, gueddari:hal-02399267, ElGueddari_SAM18} whose details are provided in Supplementary Material~(cf. Sec.~\ref{supp-app:reconstruction}). For the sake of reproducibility, the code for MR image reconstruction is made open source in \texttt{pysap-mri}\footnote{\url{https://github.com/CEA-COSMIC/pysap-mri}}, a plugin of the PySAP software~\cite{farrens2019pysap}. Of course, future work will combine deep-learning based image reconstruction with full 3D SPARKLING.

In this work, we did not carry out off-resonance artifact corrections using \cite{sutton_b0}, as it is beyond the scope of this manuscript. However, note that this does not require any supplementary scan for obtaining $\Delta B0$ map as the latter can be directly estimated from phase information using~\cite{DavalFrerot_ISMRM21}. For the sake of completeness, we show in Supplementary Material the performance of our trajectory with off-resonance corrections for AF=10 in Fig.~\ref{supp-fig:off_res_correction}.

\subsection{Phantom}

\subsubsection{Retrospective studies}
We proceed by carrying out a retrospective study to assess the quality of reconstructed images. \revcomment{We varied the acceleration factor~($\text{AF} = \frac{N_y \times N_z}{N_c}$ for 3D MR imaging}{r:AF_def}, i.e. computed with respect to fully sampled data) from 10~(TA=4min~58sec) to 40~(TA=1min~16sec) compared to a fully sampled scenario or equivalently from 2.5 to 10 compared to the reference Cartesian p4~(i.e. AF=4) under-sampled acquisition, reconstructed using the GRAPPA algorithm~\cite{Griswold02}. Our motivation was to understand the degradation in image quality while decreasing the number of collected spokes. \revcomment{Further, a study was also carried out with the TPI~\cite{boada1997fast}, as a comparison with a non-Cartesian reference from the literature.}{r:SOTA} The results are presented in Tab.~\ref{table:compare_phantom_retrospective}. They clearly show that the optimized \revcomment{full}{r:title} 3D SPARKLING strategy is robust to high acceleration factors in terms of image quality as reflected by the higher SSIM scores. In contrast, the performances of the SpSOS approach start to get worse already for AF=20. \revcomment{Finally, the SSIM score for TPI for AF=10 is already significantly lower than that of SpSOS.}{r:SOTA} The reconstructed images are presented in the supplementary material (see Fig.~\ref{supp-fig:retrospective_phantom}). 

\begin{table}[h]
	\centering
	\caption{Comparing SSIM metrics of retrospective phantom image reconstruction.}
	\begin{tabular}{*6c}
		\toprule
		Trajectory &  AF10 & AF15 & AF20 & AF40\\
		\midrule
		Full 3D   &  \textbf{0.964} 	& \textbf{0.935}  	& \textbf{0.923}  & \textbf{0.816}	\\
		SpSOS  &  0.927 	& 0.864  	& 0.737 	& 0.575	\\
		TPI & 0.63 &  0.592 & 0.573 & - \\
		\bottomrule
	\end{tabular}
	\label{table:compare_phantom_retrospective}
\end{table}

\subsubsection{Prospective acquisition}

\label{sec:compare_res}
\multirevcomment{In order to understand how the effective spatial resolution compares to the target resolution~(here 0.6mm isotropic), we performed prospective acquisitions on the NIST phantom for full 3D SPARKLING and SpSOS trajectories with varying AF~(AF=15 and 20 for both, AF=40 for full 3D). The results are presented in Fig.~\ref{fig:compare_res}. Particularly, we show a slice that includes the resolution insets present on the NIST phantom~(coffin of plate 4). The latter can be used to estimate the effective resolution. This slice consists of 5 resolution insets, each having 2x16 circles. The dia\-meters of these circles vary linearly from $0.8$mm down to $0.4$mm. The inter-circle space~(measured between the centers of the circles) also reduces linearly from $1.6$mm down to $0.8$mm in steps of $0.2$mm.
%We remind that the target resolution was 0.6 mm isotropic resolution. 

\begin{figure*}[h]
	\centering
	\resizebox{0.8\linewidth}{!}{
		\begin{tabular}{c@{\hspace*{0mm}}c@{\hspace*{1mm}}c@{\hspace*{1mm}}c@{\hspace*{1mm}}c}
			& \multicolumn{1}{c!{\vrule width 2pt \hspace*{2mm}}}{\bf (a) Cartesian p4} && {\bf (b) Full 3D  SPARKLING} & {\bf (c) SpSOS} \\
			& 
			\multicolumn{1}{c!{\vrule width 2pt \hspace*{2mm}}}{\parbox[m]{.33\linewidth}{	
				\begin{tikzpicture}
				\node[anchor=south west,inner sep=0] at (0,0){\includegraphics[trim={8cm 5.1cm 6cm 5.5cm},clip, width=\linewidth]{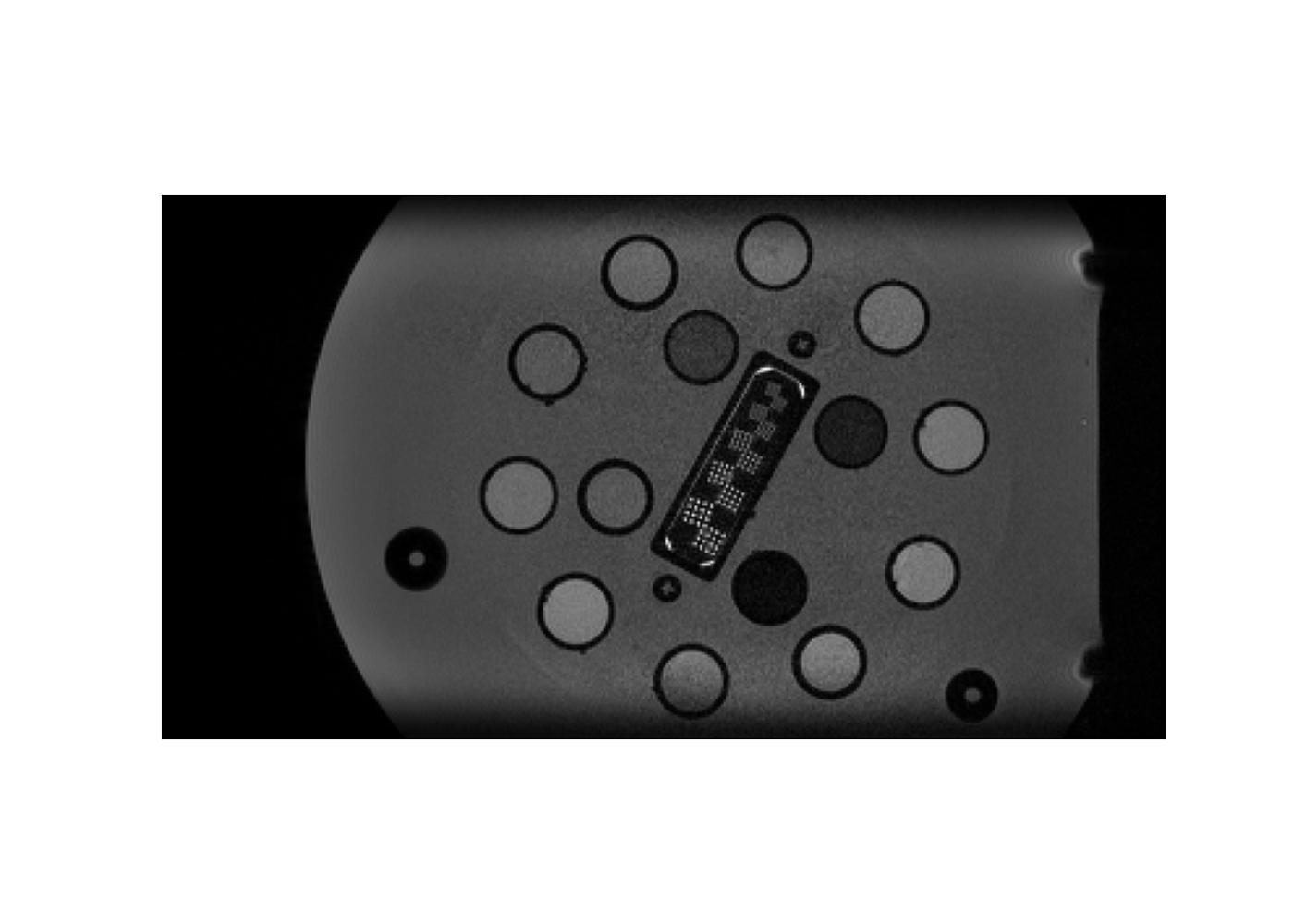}};
				\draw[red, thick, rounded corners](1.9cm, 1cm) rectangle (4.85cm, 3.22cm);
				\end{tikzpicture}}}&&
			\parbox[m]{.33\linewidth}{
				\begin{tikzpicture}
				\node[anchor=south west,inner sep=0] at (0,0){\includegraphics[trim={8cm 5.1cm 6cm 5.5cm},clip, width=\linewidth]{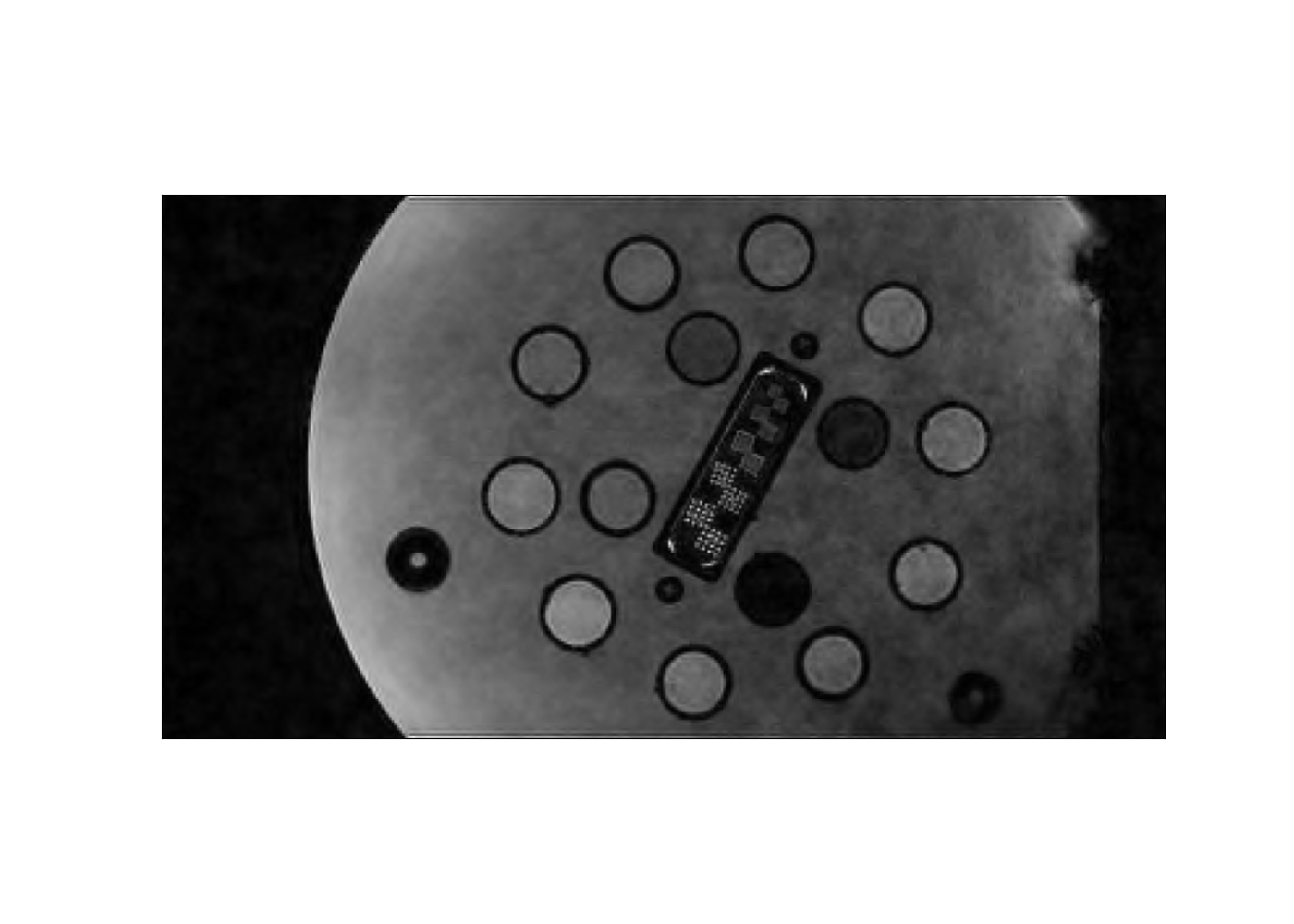}};
				\draw[red, thick, rounded corners](1.9cm, 1cm) rectangle (4.85cm, 3.22cm);
				\end{tikzpicture}}&
			\parbox[m]{.33\linewidth}{
				\begin{tikzpicture}{4.6cm 3cm 2.7cm 3cm}
				\node[anchor=south west,inner sep=0] at (0,0){\includegraphics[trim={8cm 5.1cm 6cm 5.5cm},clip, width=\linewidth]{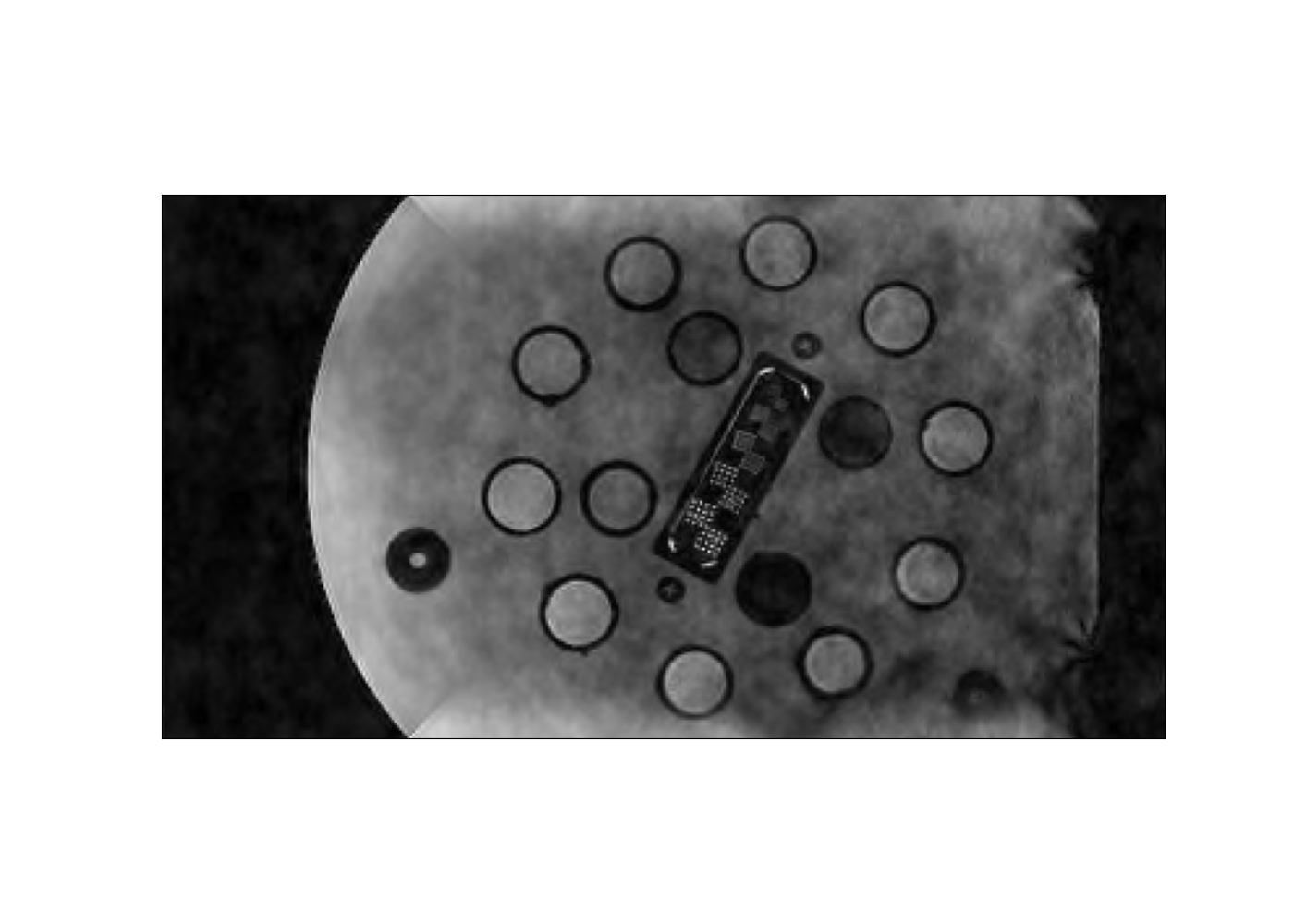}};
				\draw[red, thick, rounded corners](1.9cm, 1cm) rectangle (4.85cm, 3.22cm);
				\end{tikzpicture}}
			\\[2cm]
			& 
			\multicolumn{1}{c!{\vrule width 2pt \hspace*{2mm}}}{\parbox[m]{.33\linewidth}{
				\includegraphics[trim={15cm 9cm 10cm 9cm},clip,width=\linewidth]
				{images/results/prospective/resolution/00_cartesian.png}}}&
			\rotatebox[origin=c]{90}{\bf  (i) AF=15} &
			\parbox[m]{.33\linewidth}{
				\includegraphics[trim={15cm 9cm 10cm 9cm},clip,width=\linewidth]
				{images/results/prospective/resolution/N384_FOV230_Nz208_SqNc73_Ns2048_t02.png}}&
			\parbox[m]{.33\linewidth}{
				\includegraphics[trim={15cm 9cm 10cm 9cm},clip,width=\linewidth]
				{images/results/prospective/resolution/SOS_N384_FOV230_Nz208_Nc26_Ns2048.png}} 
			\\[2cm] 
			\cmidrule[2pt]{1-2}
			\rotatebox[origin=c]{90}{\bf (iii) Full 3D, AF=40}&
			\parbox[m]{.33\linewidth}{
				\includegraphics[trim={15cm 9cm 10cm 9cm},clip,width=\linewidth]
				{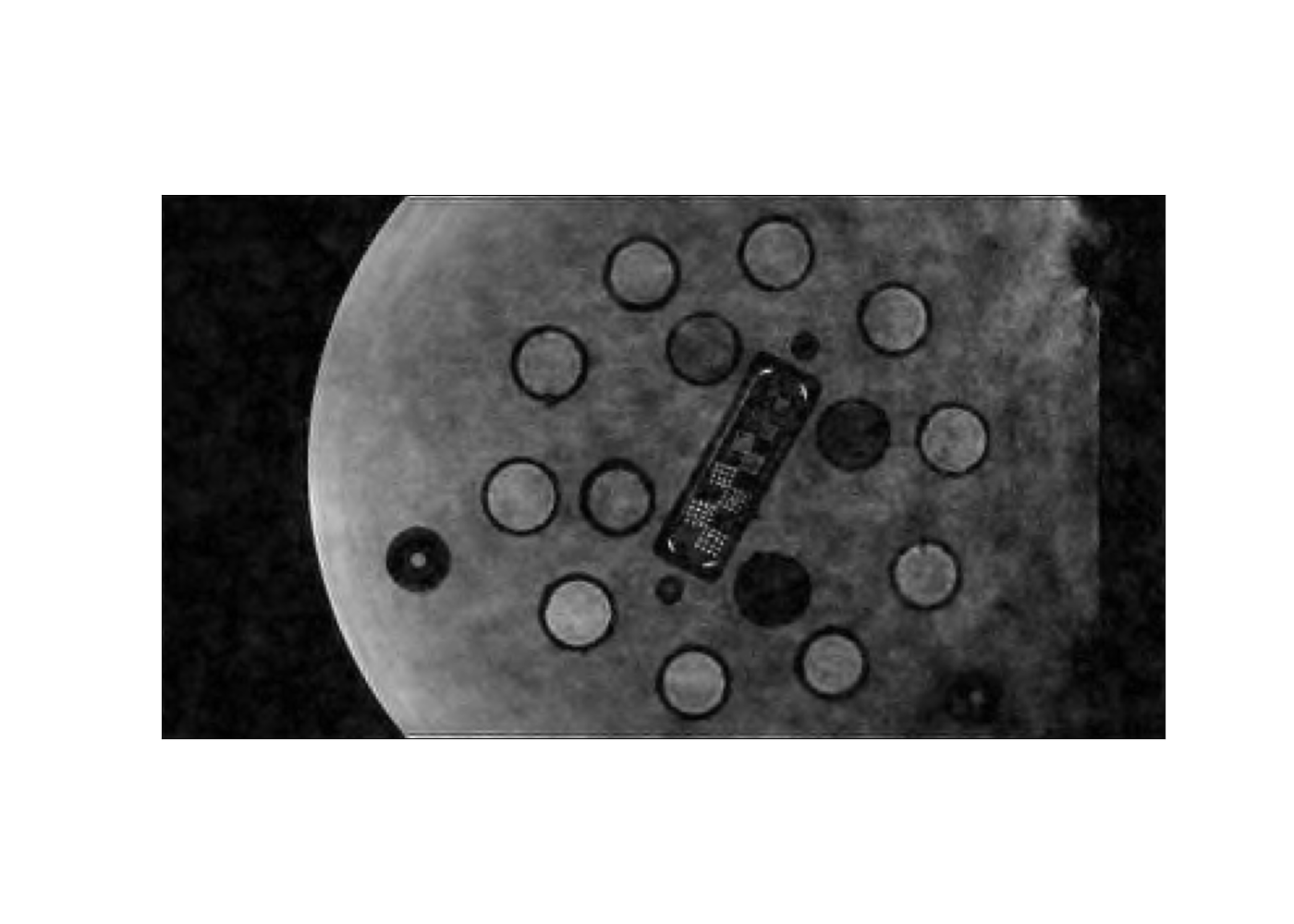}}&
			\rotatebox[origin=c]{90}{\bf (ii) AF=20} &
			\parbox[m]{.33\linewidth}{
				\includegraphics[trim={15cm 9cm 10cm 9cm},clip,width=\linewidth]
				{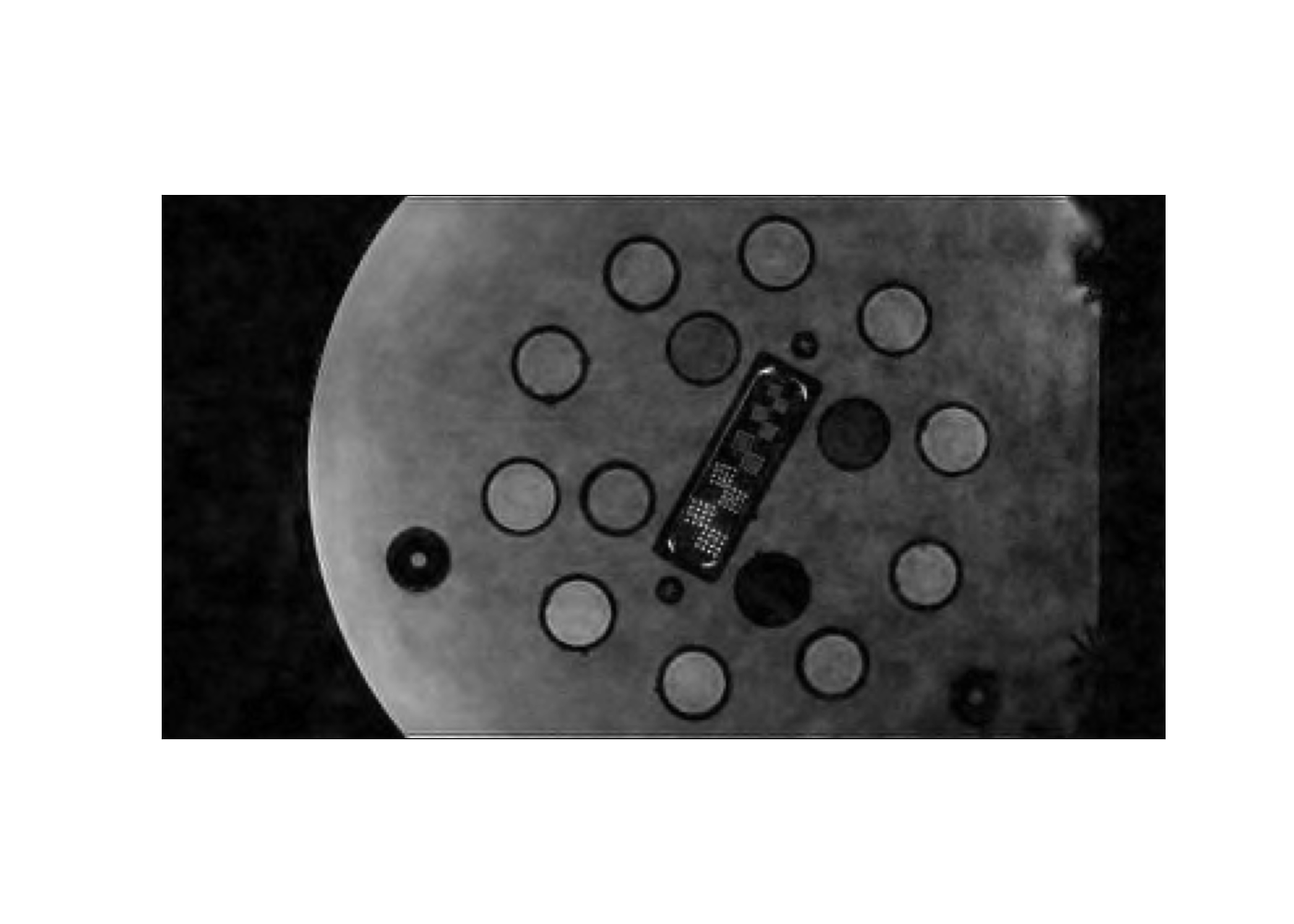}}&
			\parbox[m]{.33\linewidth}{
				\includegraphics[trim={15cm 9cm 10cm 9cm},clip,width=\linewidth]
				{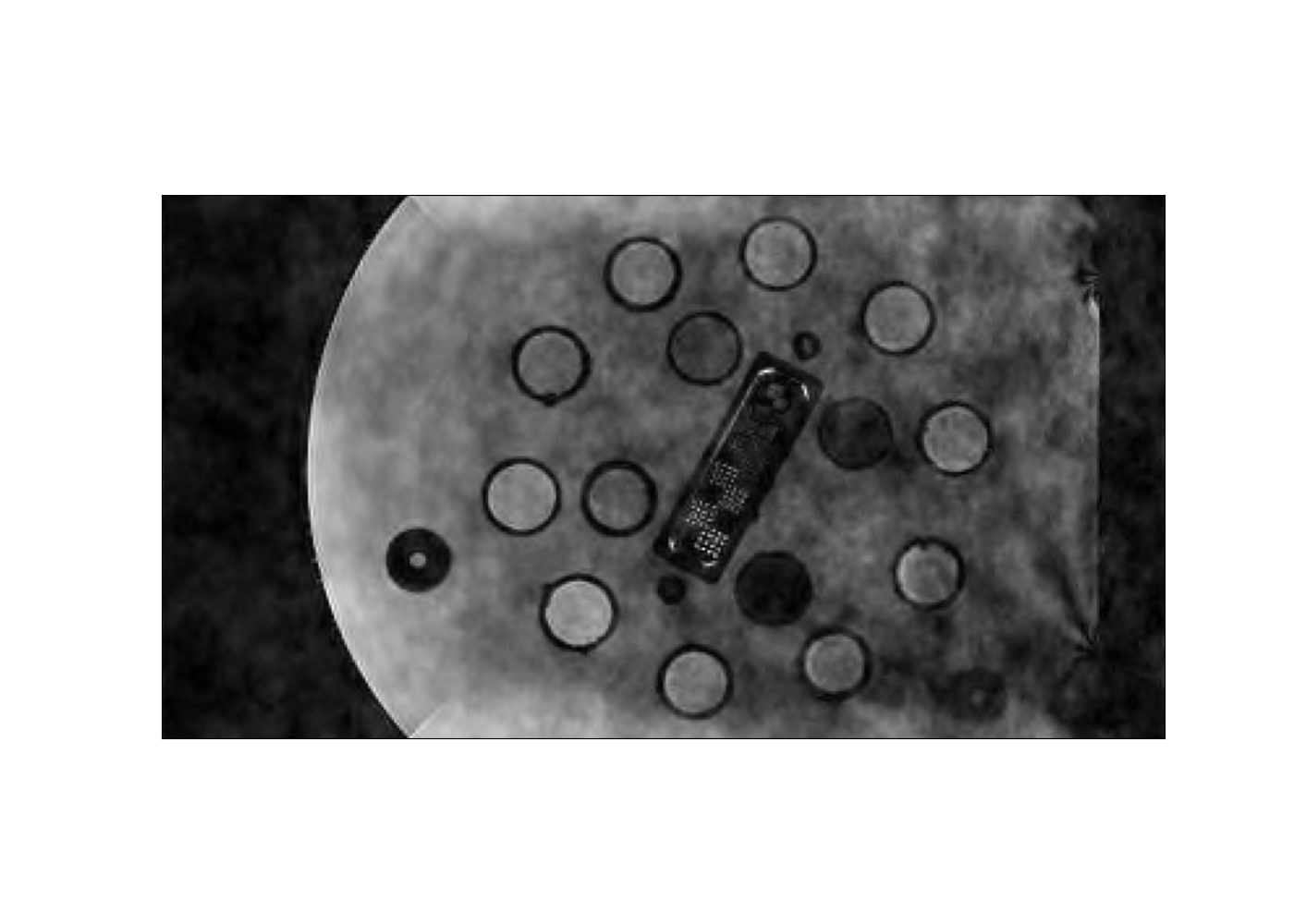}}
	\end{tabular}}
	\caption{\label{fig:compare_res} Comparison of resolution insets for Full 3D SPARKLING (b) and SpSOS (c) with prospective phantom scans at (i) AF=15, (ii) AF=20 as compared to Cartesian p4 (a). Additionally, we present the results for full 3D SPARKLING trajectory at (iii)~AF=40 at the bottom-left.}
\end{figure*}

We see that the intensity profile of our reconstructed MR images does not follow that of Cartesian reference as we did not carry out coil sensitivity normalization in our reconstructions. This can be performed using the rapid pre-scan coil sensitivity measurements done in a few seconds. This point will be addressed in future works. 
Overall, we observe that full 3D SPARKLING trajectories provide less noisy images compared to SpSOS ones. Further, it is worth noting that at AF=10 and 20, we can distinguish in between the resolution insets down to 0.7mm, with an increasing noise level over the image for a higher acceleration factor. However, we observe some resolution loss over the images based on SPARKLING trajectories~(more pronounced for SpSOS), where we see some blurring for circles of diameter 0.6mm separated by 1.2mm~(taken from the center of circle). This helps us understand the expected degradation in image resolution. Therefore, the effective image resolution is estimated to be 0.6-0.7mm isotropic at AF=10 and 20 and is evolving toward 0.7-0.8mm at AF=40 for full 3D SPARKLING.}{\ref{r:resolution_nist}, \ref{r:blurr_r3}, \ref{r:resolution_loss}, \ref{r:blurry_r4}}

\newcommand{\retroInvivo}{images/results/retrospective/invivo}
\begin{figure*}[h]
	\centering
	\begin{mdframed}[innertopmargin=2pt, innerbottommargin=2pt, innerleftmargin=0pt, innerrightmargin=0pt, backgroundcolor=black, leftmargin=0cm,rightmargin=0cm,usetwoside=false]
		\resizebox{\linewidth}{!}{
			\begin{tabular}{c@{\hspace*{1mm}}c@{\hspace*{1mm}}c@{\hspace*{1mm}}c@{\hspace*{1mm}}c@{\hspace*{1mm}}c@{\hspace*{1mm}}c@{\hspace*{1mm}}c@{\hspace*{1mm}}c@{\hspace*{1mm}}c}
				&\multicolumn{2}{c}{\Bw{\Large (a) AF = 10}}&
				\multicolumn{2}{c}{\Bw{\Large (b) AF = 15}}&
				\multicolumn{2}{c}{\Bw{\Large (c) AF = 20}}&
				\multicolumn{2}{c}{\Bw{\Large (d) AF = 40}}\\
				&
				\multicolumn{2}{c}{\Bw{\small	SSIM = 0.964}}&
				\multicolumn{2}{c}{\Bw{\small	SSIM = 0.937}}&
				\multicolumn{2}{c}{\Bw{\small	SSIM = 0.918}}&
				\multicolumn{2}{c}{\Bw{\small   SSIM = 0.792}}
				\\
				\multirow{2}{*}[0.4in]{\rotatebox[origin=c]{90}{\Bw{\Large (i) Full 3D SPARKLING}}}&
				\multirow{2}{*}[0.4in]{\parbox[m]{.23\linewidth}{
						\includegraphics[trim={6cm 4cm 5.5cm 5cm},clip,width=\linewidth]
						{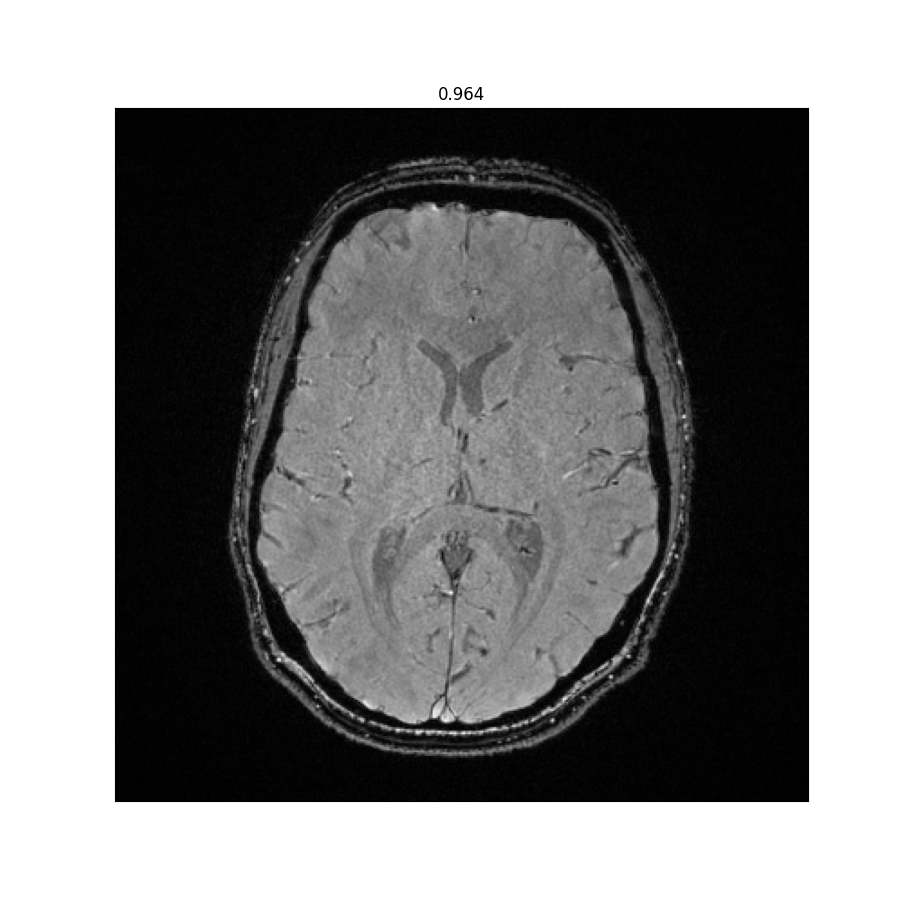}}}&
				\parbox[m]{.2\linewidth}{\includegraphics[trim={5.2cm 6.5cm 4cm 7cm},clip,width=\linewidth]{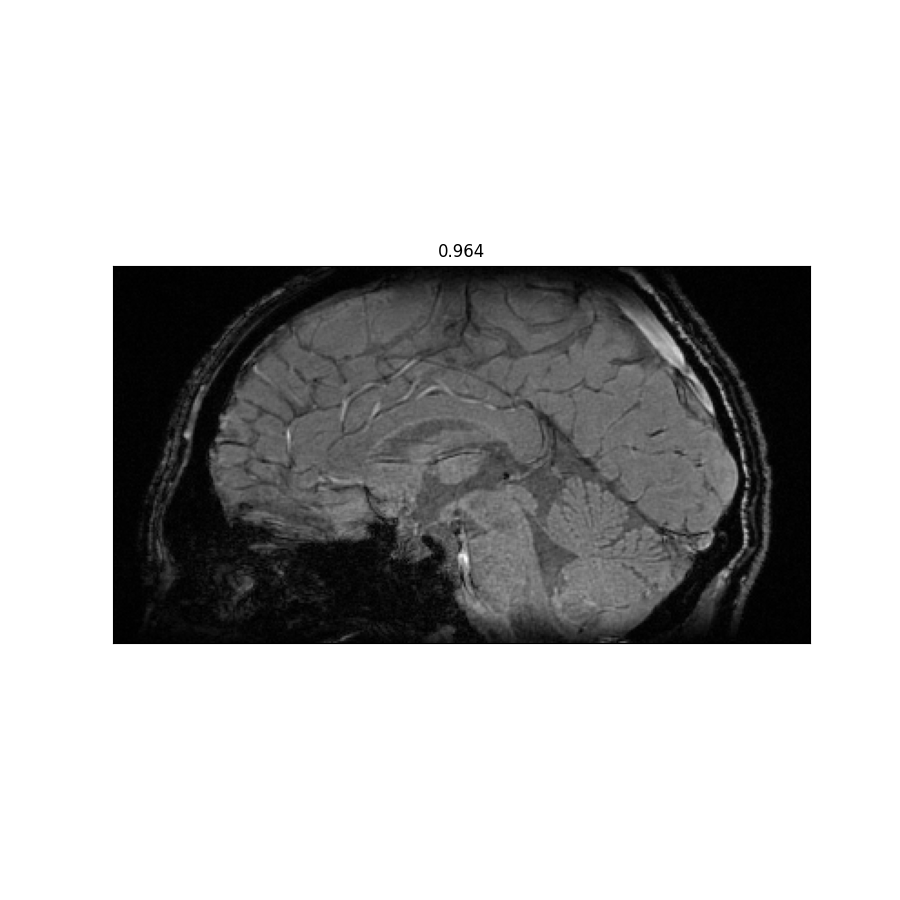}}&
				\multirow{2}{*}[0.4in]{\parbox[m]{.23\linewidth}{\includegraphics[trim={6cm 4cm 5.5cm 5cm},clip,width=\linewidth]
						{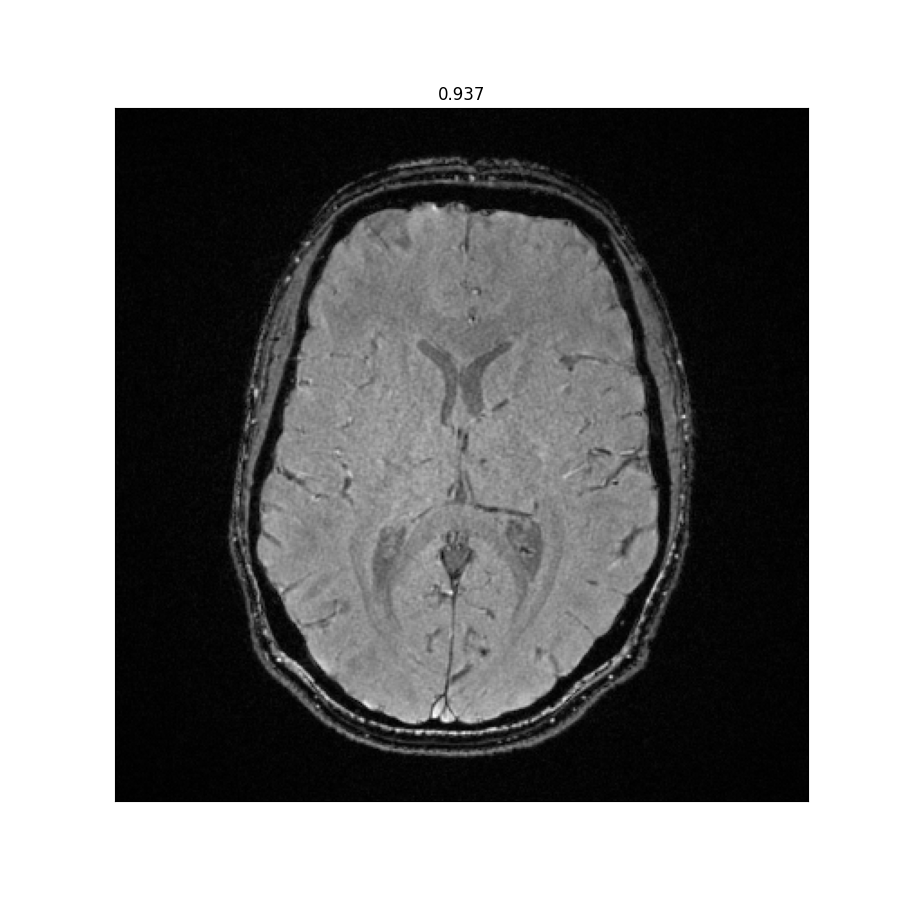}}}&
				\parbox[m]{.2\linewidth}{\includegraphics[trim={5.2cm 6.5cm 4cm 7cm},clip,width=\linewidth]	{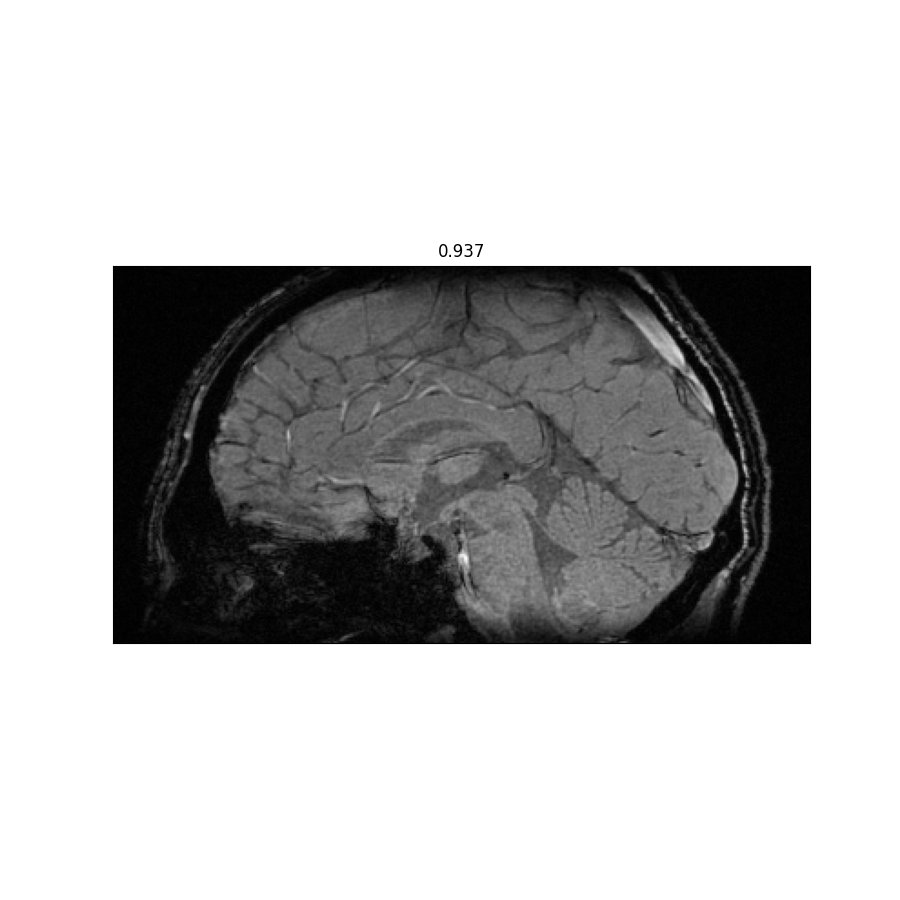}}&
				\multirow{2}{*}[0.4in]{\parbox[m]{.23\linewidth}{
						\includegraphics[trim={6cm 4cm 5.5cm 5cm},clip,width=\linewidth]
						{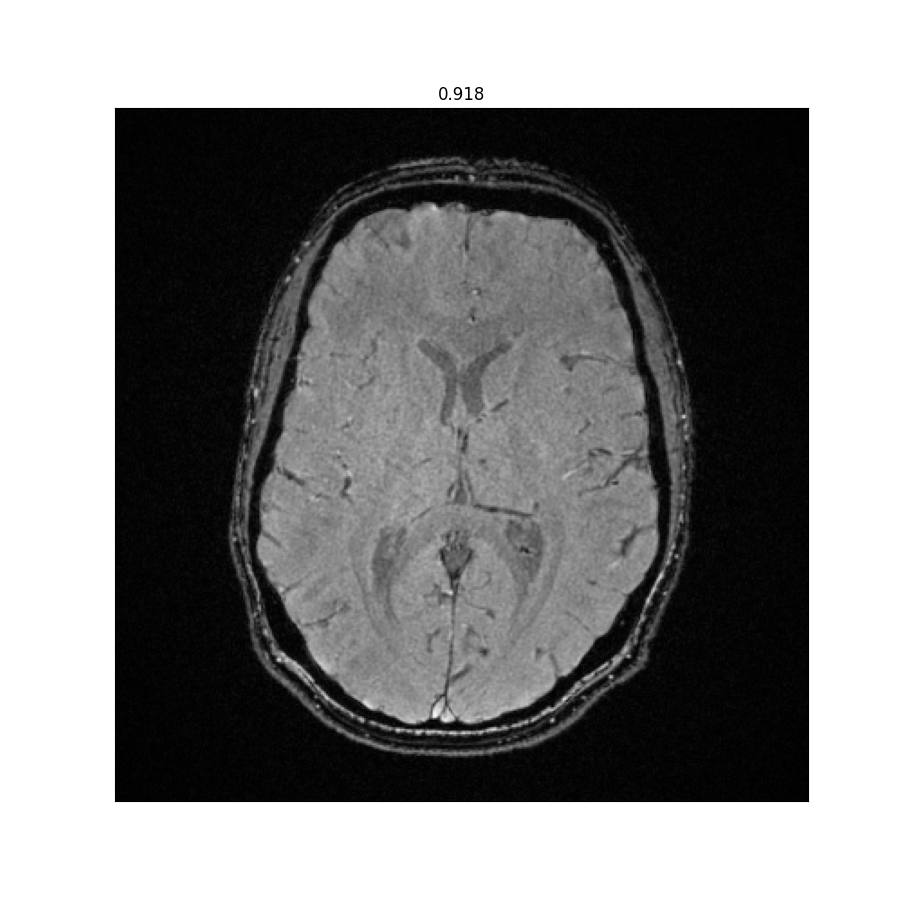}}}&
				\parbox[m]{.2\linewidth}{\includegraphics[trim={5.2cm 6.5cm 4cm 7cm},clip,width=\linewidth]{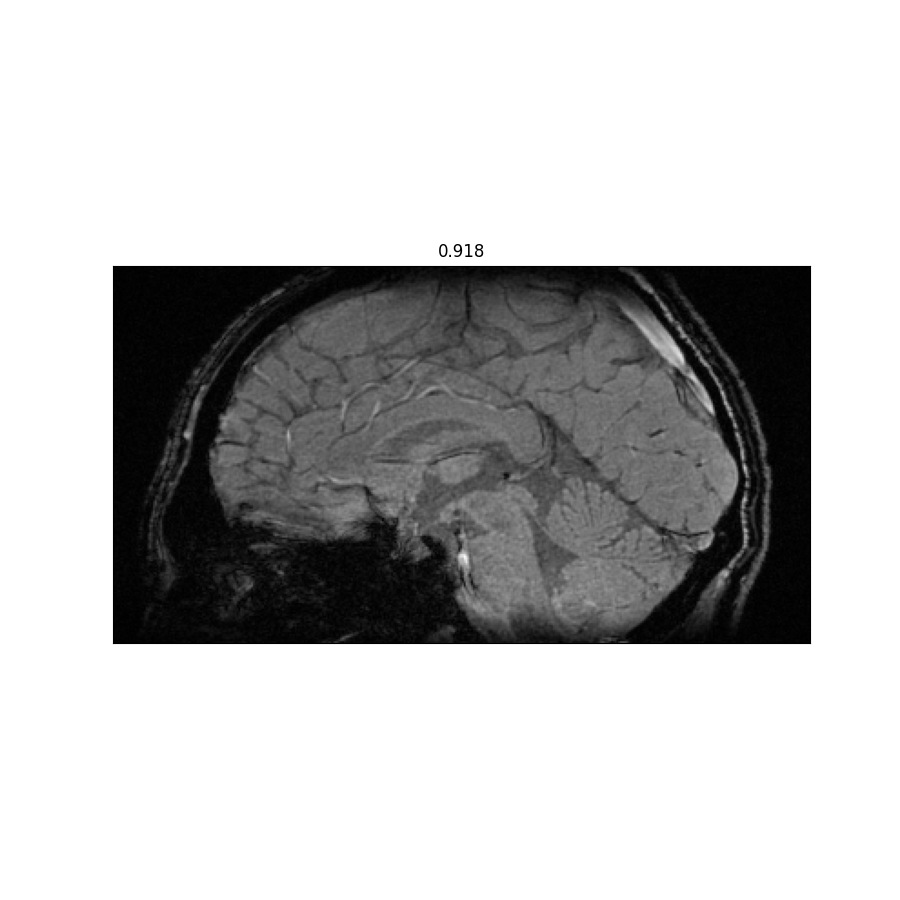}}&
				\multirow{2}{*}[0.4in]{\parbox[m]{.23\linewidth}{
						\includegraphics[trim={6cm 4cm 5.5cm 5cm},clip,width=\linewidth]
						{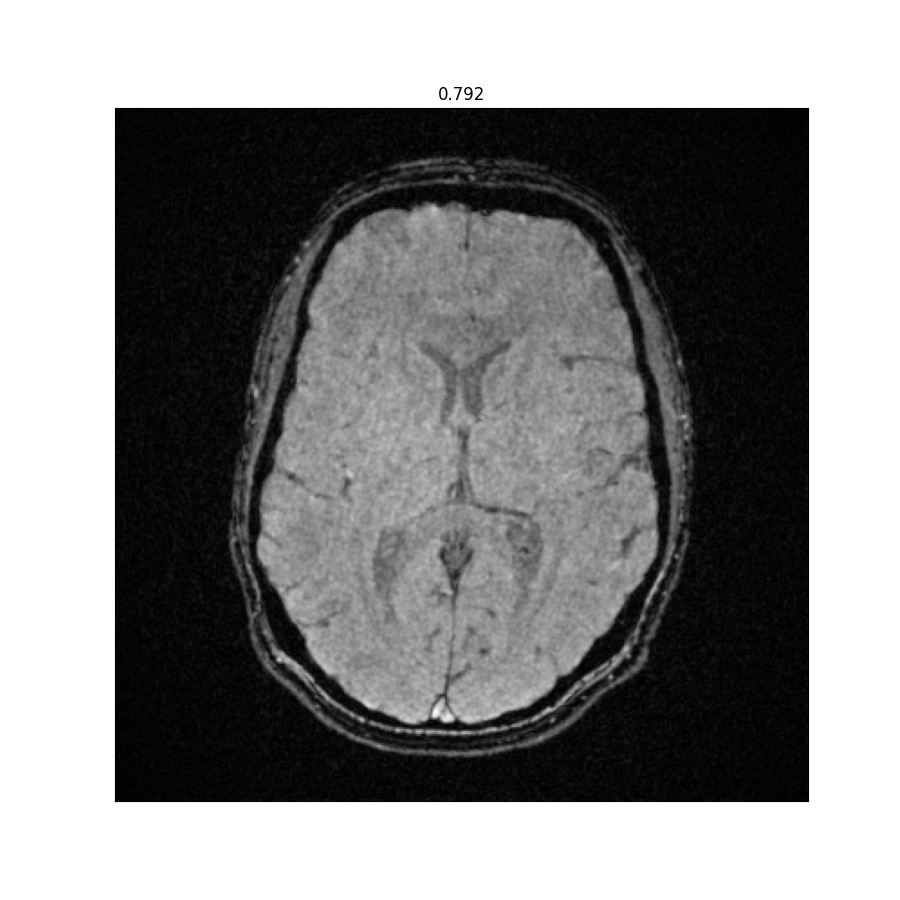}}}&
				\parbox[m]{.2\linewidth}{
					\includegraphics[trim={5.2cm 6.5cm 4cm 7cm},clip,width=\linewidth]
					{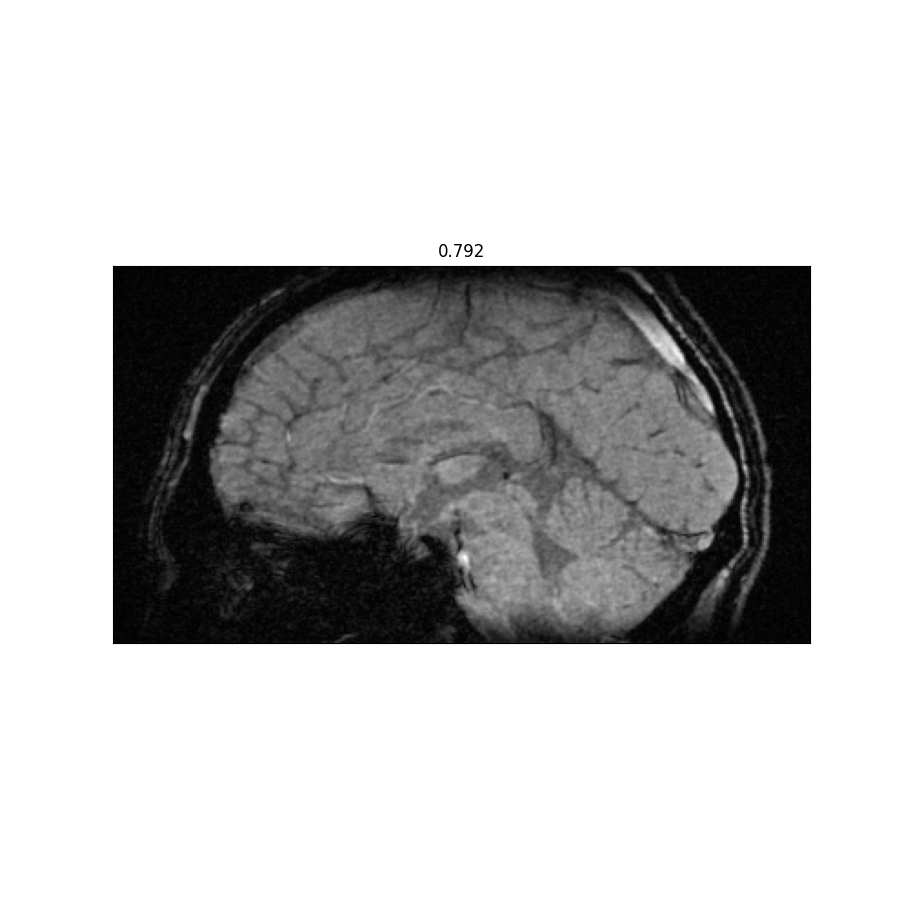}}
				
				\\
				&&\parbox[m]{.2\linewidth}{\includegraphics[trim={6.2cm 7cm 5.5cm 7cm},clip,width=\linewidth]{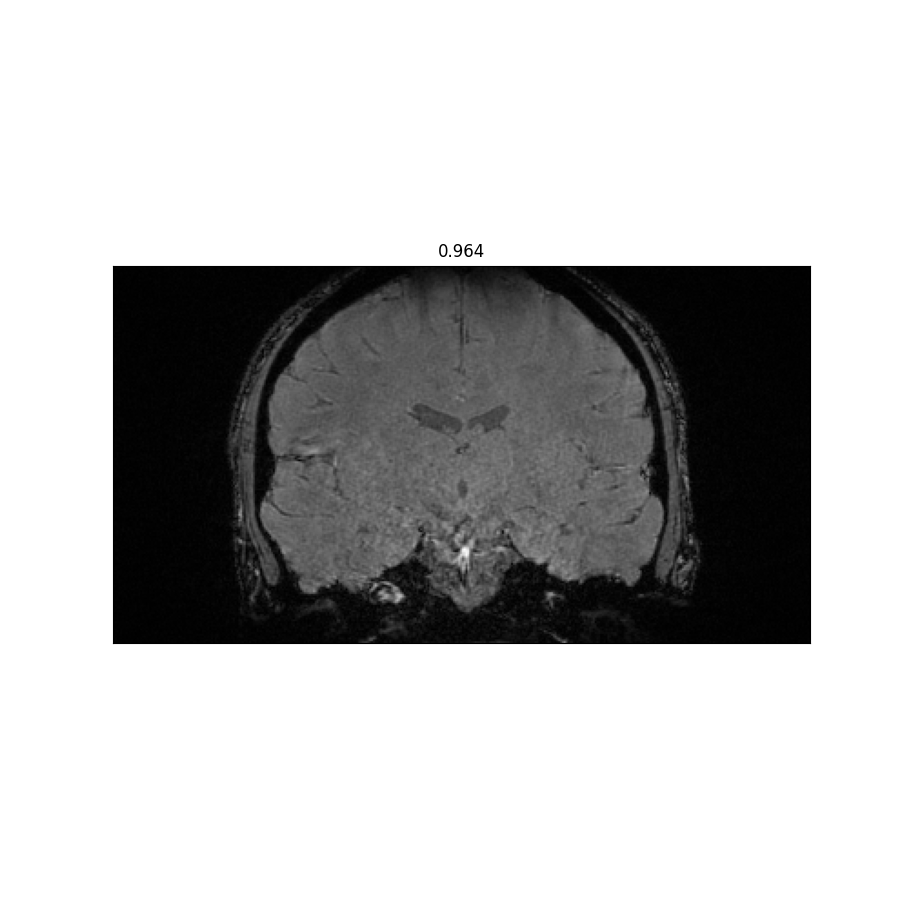}}&
				&\parbox[m]{.2\linewidth}{\includegraphics[trim={6.2cm 7cm 5.5cm 7cm},clip,width=\linewidth]{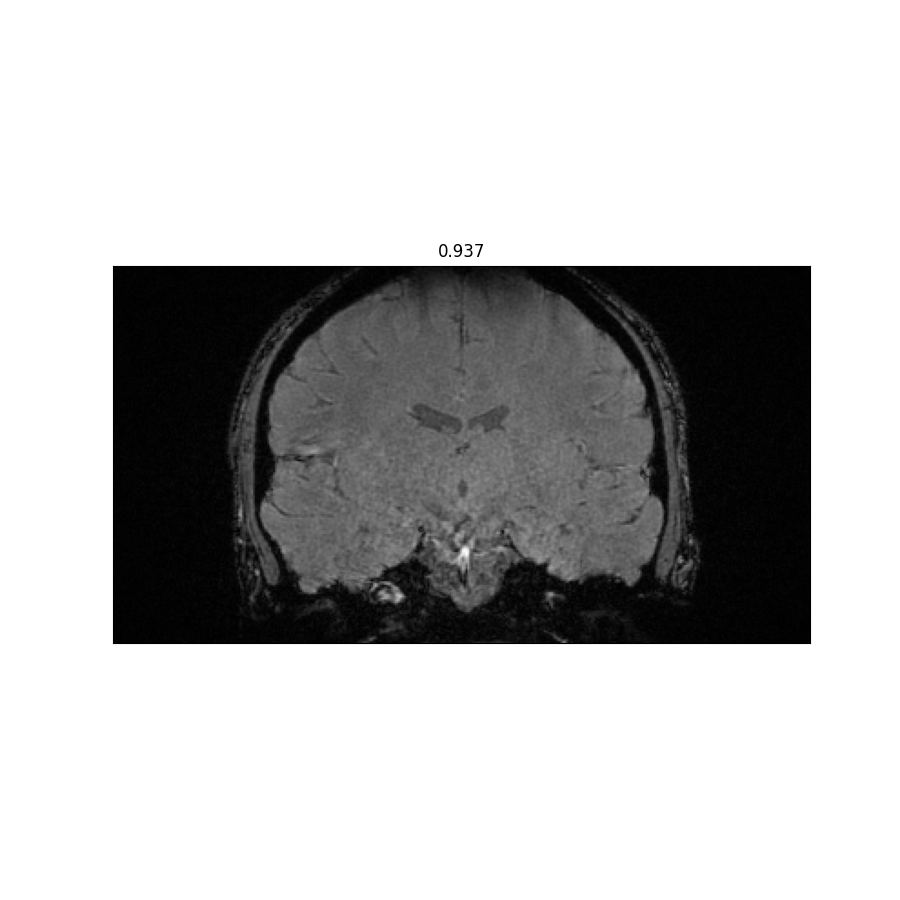}}&
				&\parbox[m]{.2\linewidth}{\includegraphics[trim={6.2cm 7cm 5.5cm 7cm},clip,width=\linewidth]{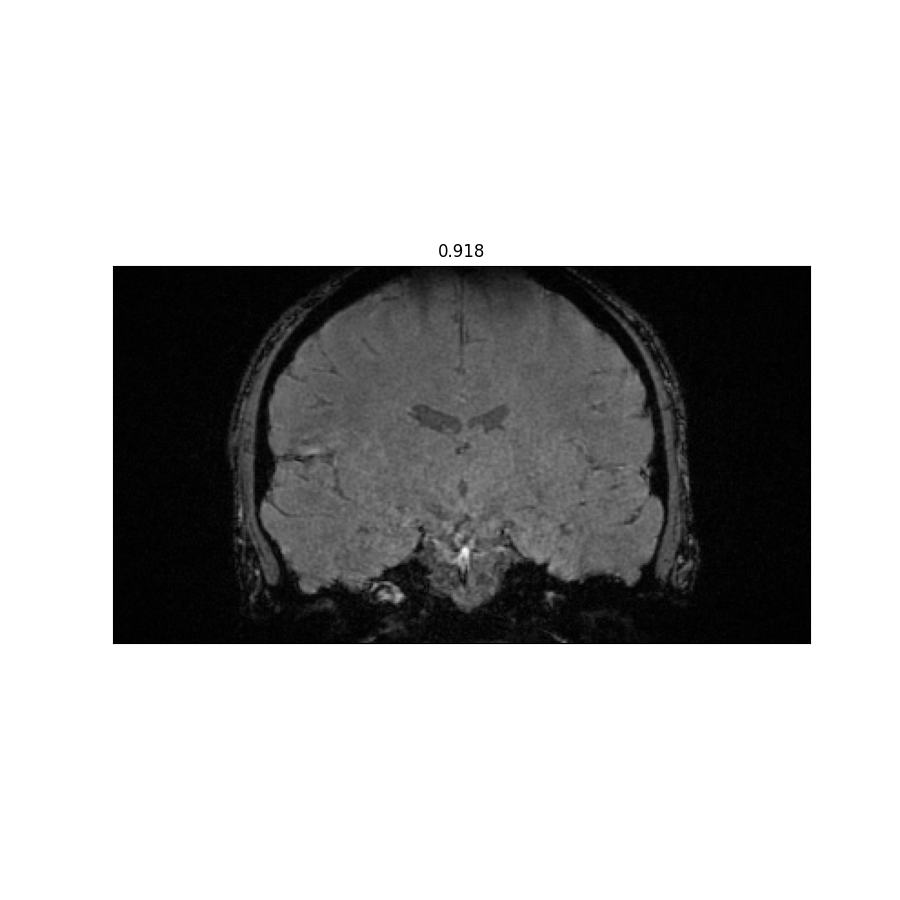}}&
				&\parbox[m]{.2\linewidth}{\includegraphics[trim={6.2cm 7cm 5.5cm 7cm},clip,width=\linewidth]{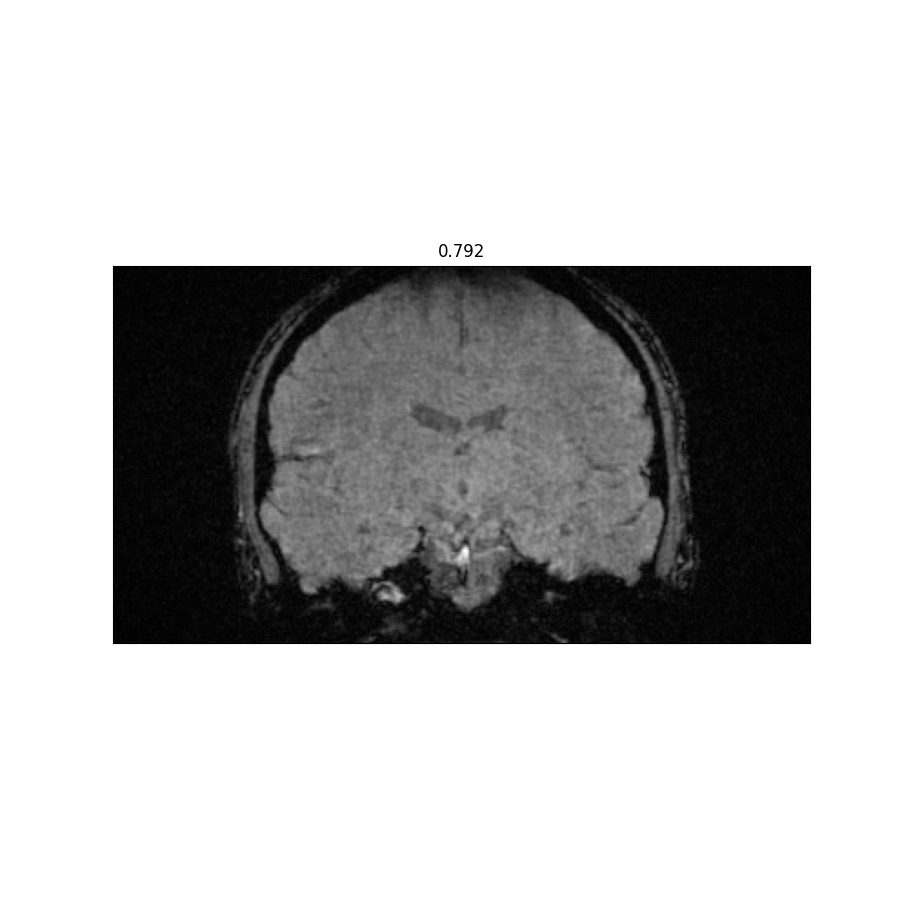}}
				\\
				\arrayrulecolor{white}\cmidrule[2pt]{8-9}
				&\multicolumn{2}{c}{\Bw{\small   SSIM = 0.93}}&
				\multicolumn{2}{c}{\Bw{\small	SSIM = 0.868}}&
				\multicolumn{2}{c!{\color{white}\vrule width 2pt \hspace*{2mm}}}{\Bw{\small	SSIM = 0.759}}&
				\multicolumn{2}{c}{\Bw{\Large  (e) Cartesian Reference}}
				\\
				\multirow{2}{*}[0.4in]{\rotatebox[origin=c]{90}{\Bw{\Large (ii) \texttt{SpSOS} SPARKLING}}}&
				\multirow{2}{*}[0.4in]{\parbox[m]{.23\linewidth}{\includegraphics
						[trim={6cm 4cm 5.5cm 5cm},clip,width=\linewidth]
						{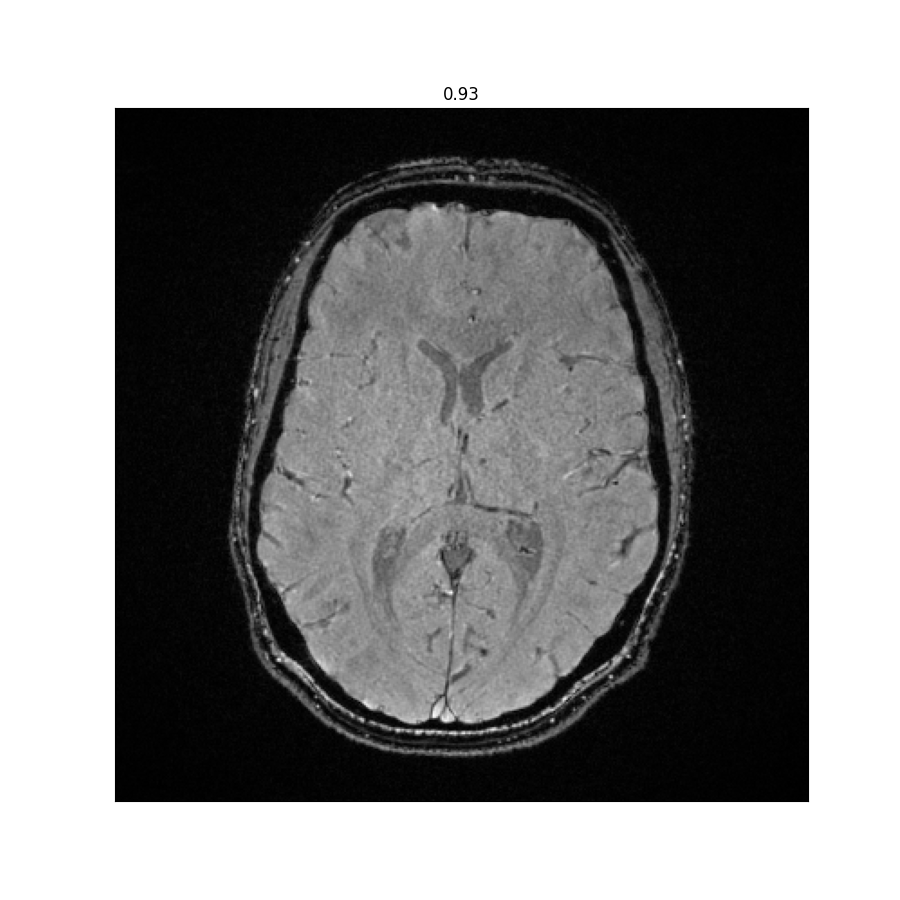}}}&
				\parbox[m]{.2\linewidth}{\includegraphics[trim={5.2cm 6.5cm 4cm 7cm},clip,width=\linewidth]{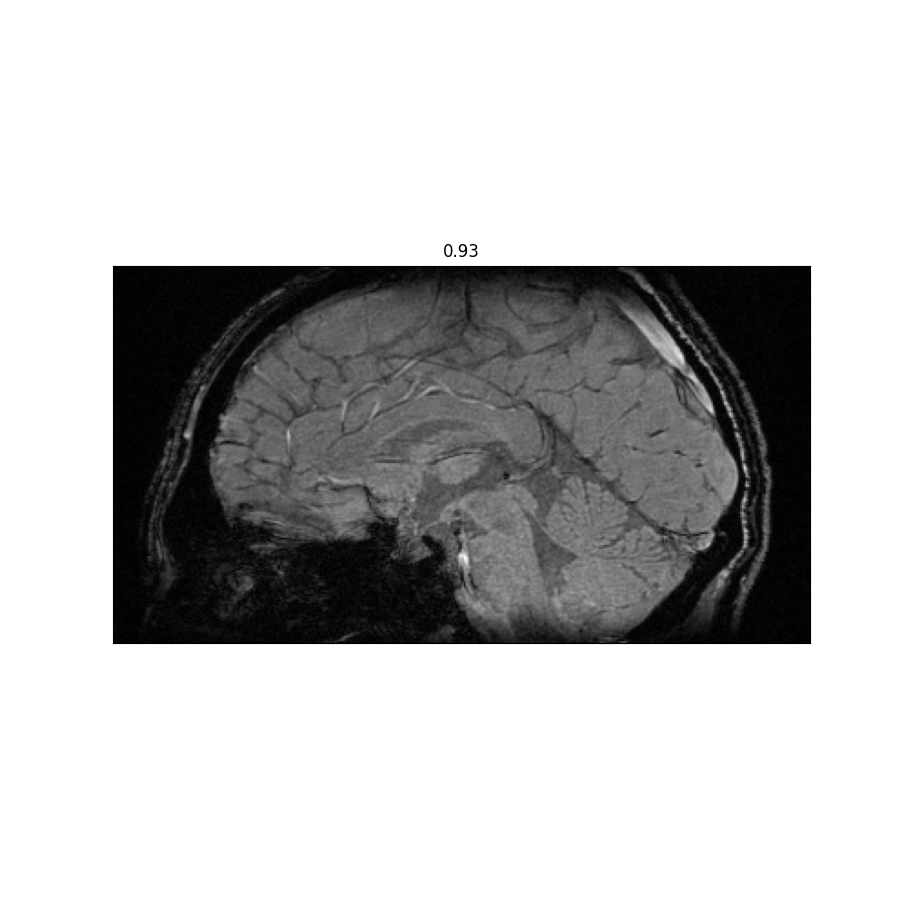}}&
				\multirow{2}{*}[0.4in]{\parbox[m]{.23\linewidth}{\includegraphics[trim={6cm 4cm 5.5cm 5cm},clip,width=\linewidth]
						{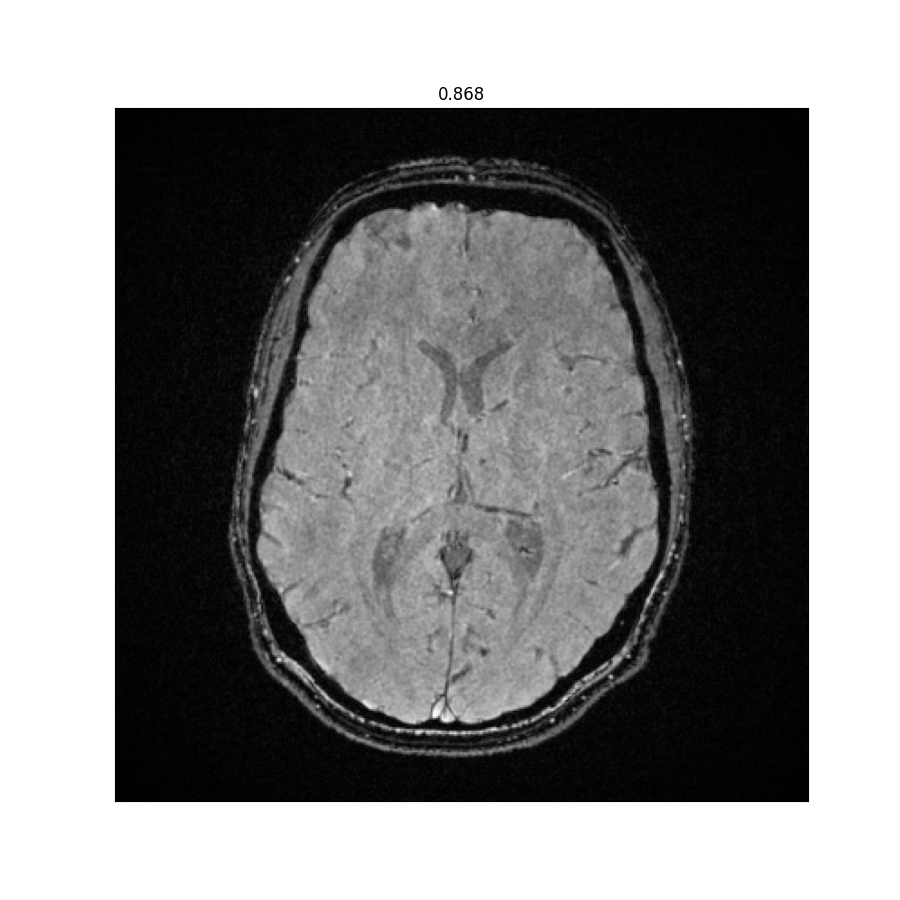}}}&
				\parbox[m]{.2\linewidth}{\includegraphics[trim={5.2cm 6.5cm 4cm 7cm},clip,width=\linewidth]{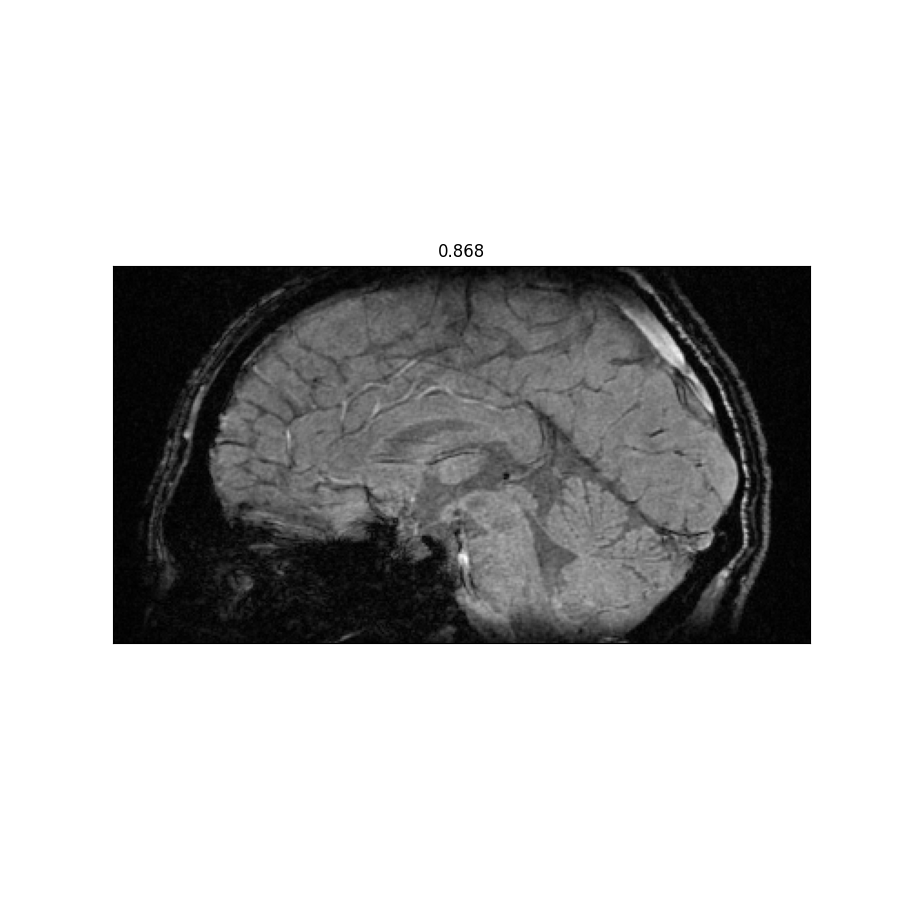}}&				\multirow{2}{*}[0.4in]{\parbox[m]{.23\linewidth}{\includegraphics[trim={6cm 4cm 5.5cm 5cm},clip,width=\linewidth]
						{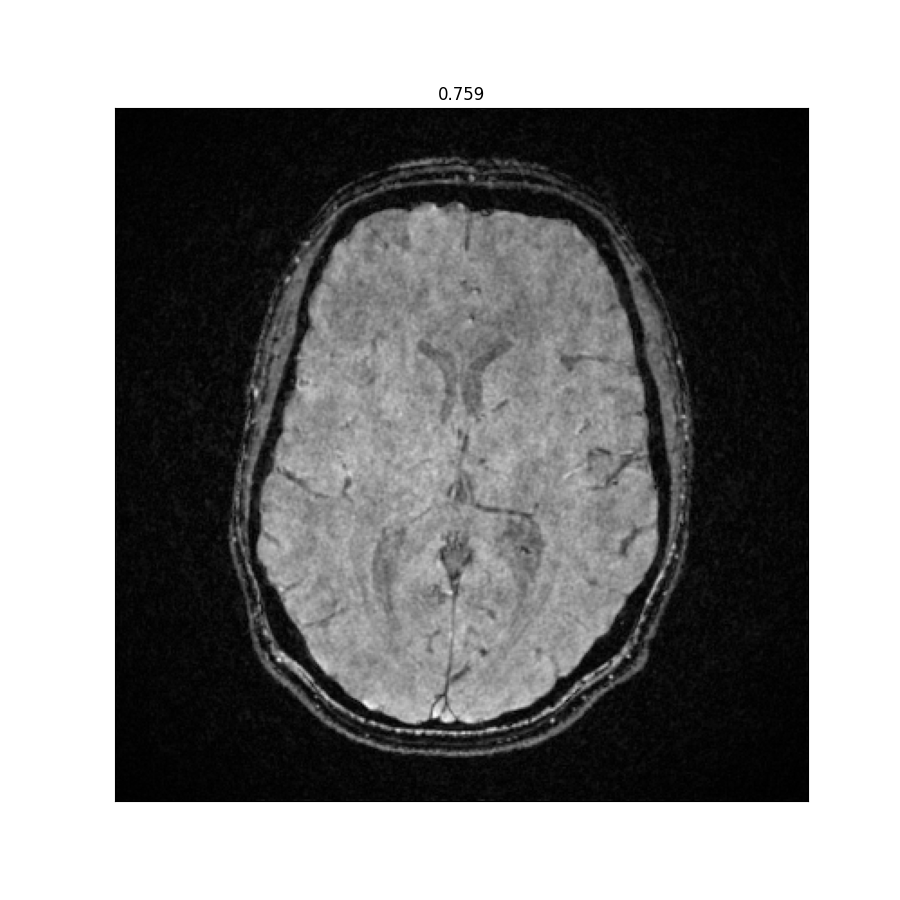}}}&
				\multicolumn{1}{c!{\color{white}\vrule width 2pt \hspace*{2mm}}}
				{\parbox[m]{.2\linewidth}{\includegraphics[trim={5.2cm 6.5cm 4cm 7cm},clip,width=\linewidth]{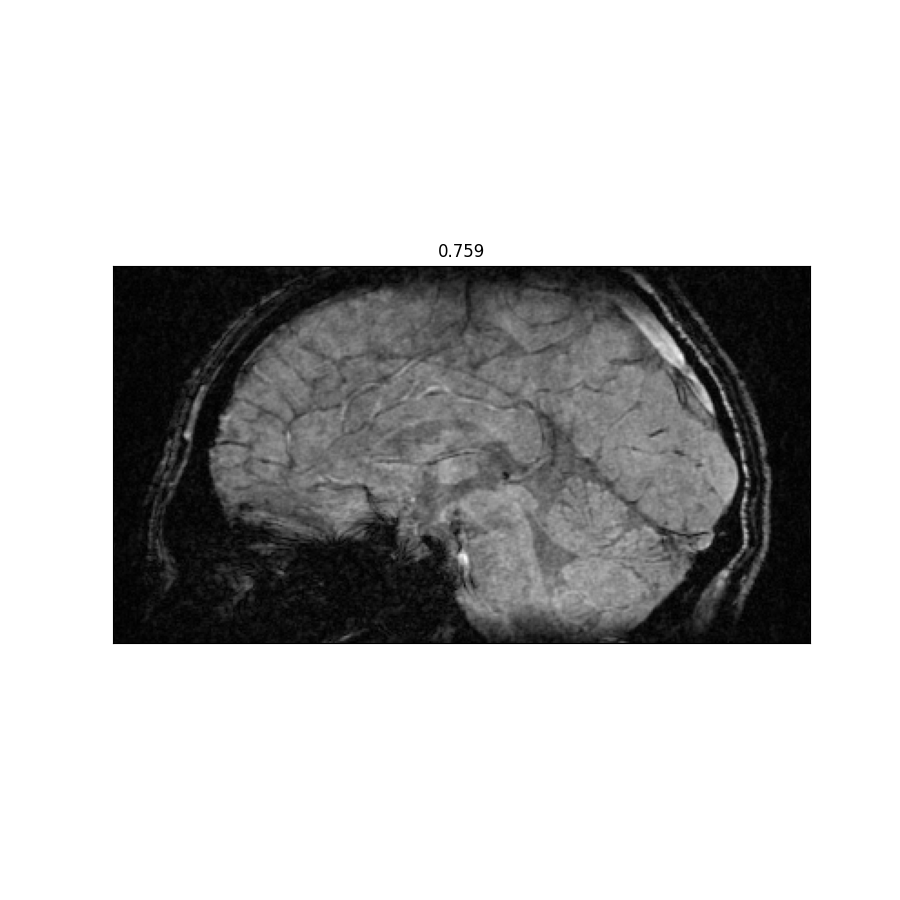}}}
				&
				\multirow{2}{*}[0.4in]{\parbox[m]{.23\linewidth}{\includegraphics[trim={6cm 4cm 5.5cm 5cm},clip,width=\linewidth]{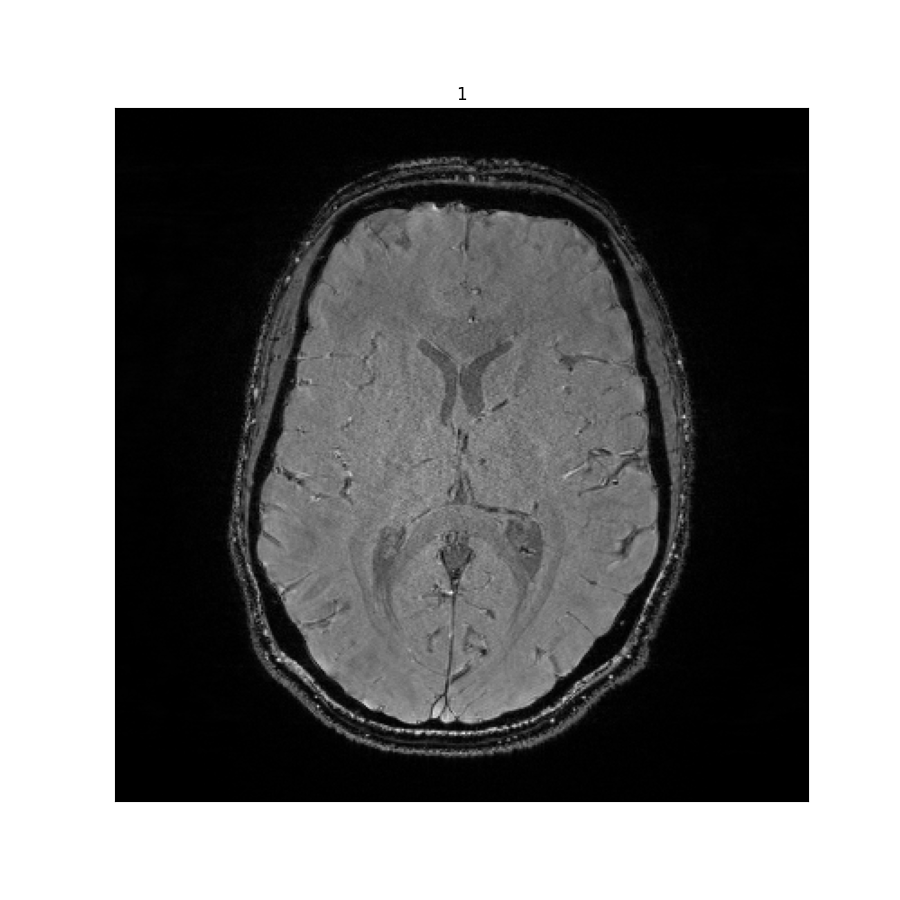}}}&
				\parbox[m]{.2\linewidth}{\includegraphics[trim={5.2cm 6.5cm 4cm 7cm},clip,width=\linewidth]{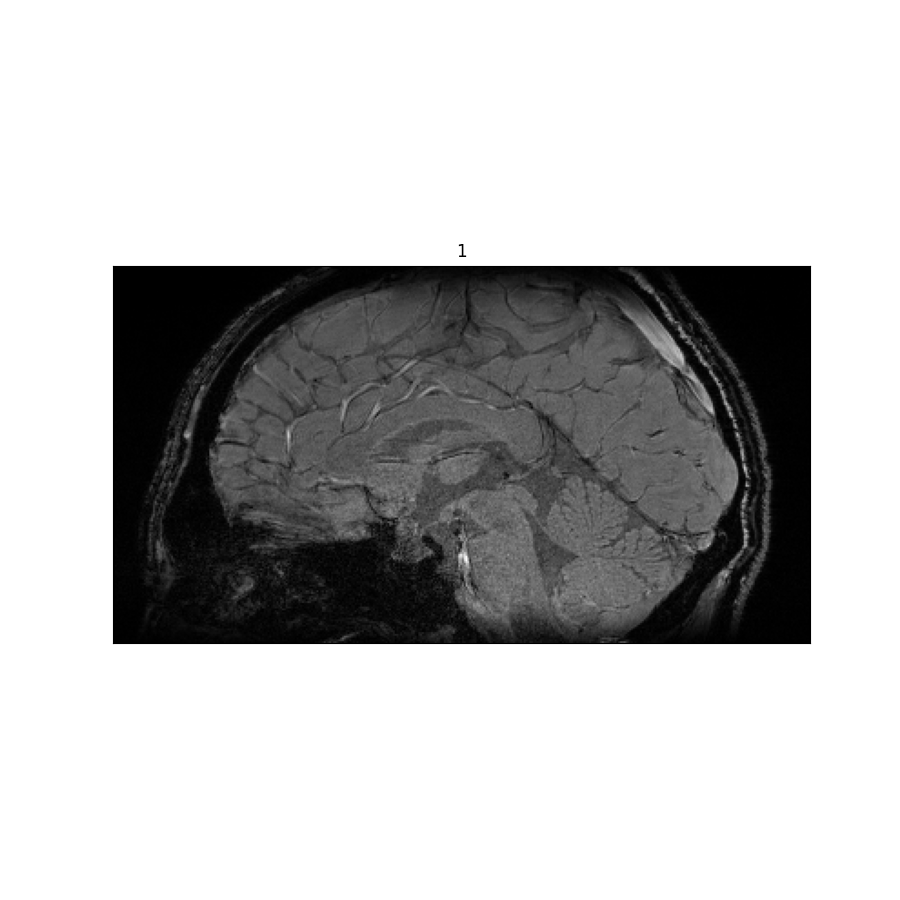}}&
				\\
				&&\parbox[m]{.2\linewidth}{\includegraphics[trim={6.2cm 7cm 5.5cm 7cm},clip,width=\linewidth]{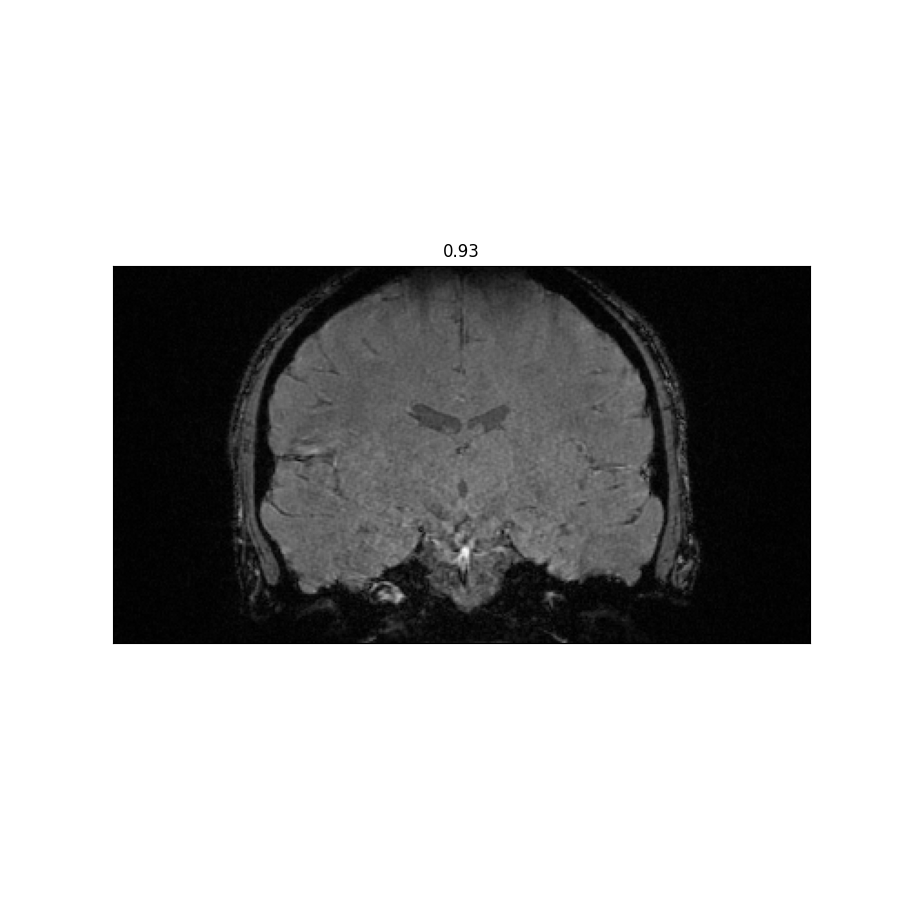}}&
				&\parbox[m]{.2\linewidth}{\includegraphics[trim={6.2cm 7cm 5.5cm 7cm},clip,width=\linewidth]{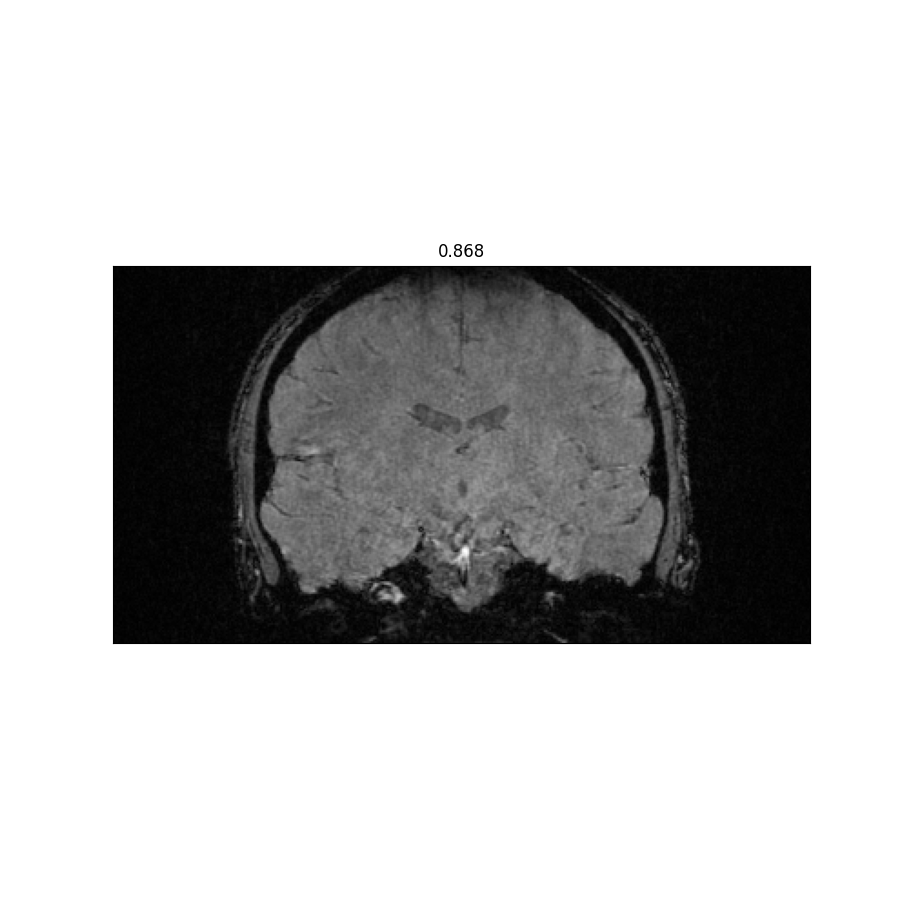}}&
				&\multicolumn{1}{c!{\color{white}\vrule width 2pt \hspace*{2mm}}}{\parbox[m]{.2\linewidth}{\includegraphics[trim={6.2cm 7cm 5.5cm 7cm},clip,width=\linewidth]{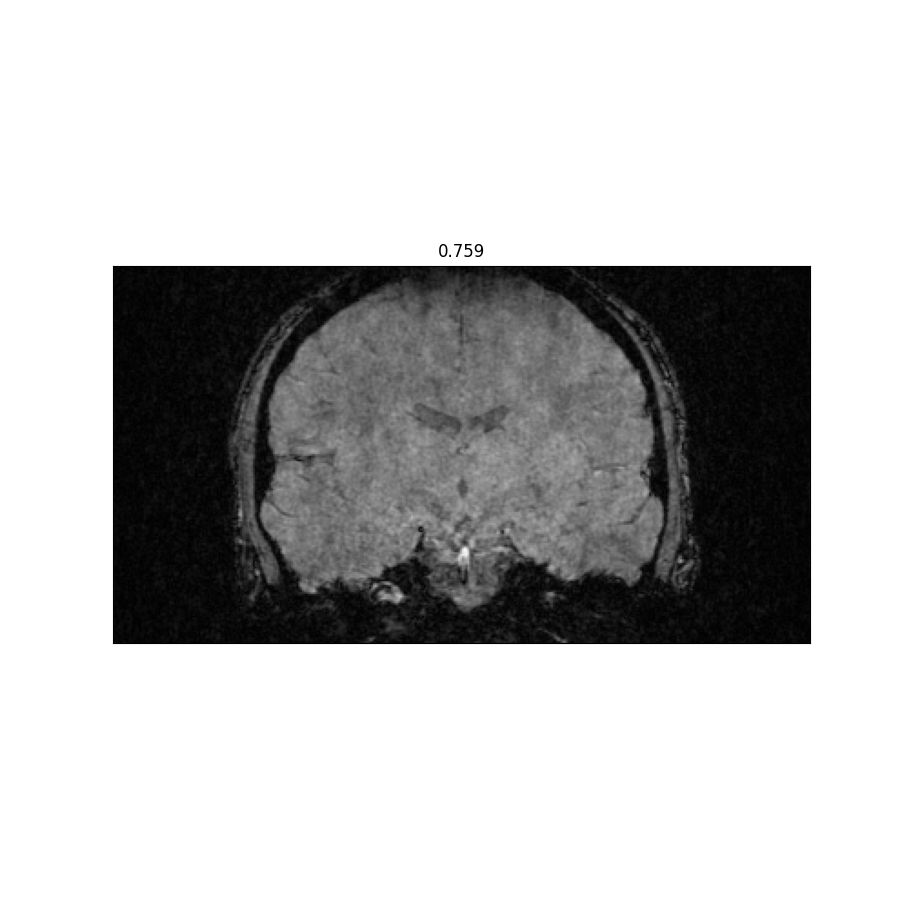}}}&
				&\parbox[m]{.2\linewidth}{\includegraphics[trim={6.2cm 7cm 5.5cm 7cm},clip,width=\linewidth]{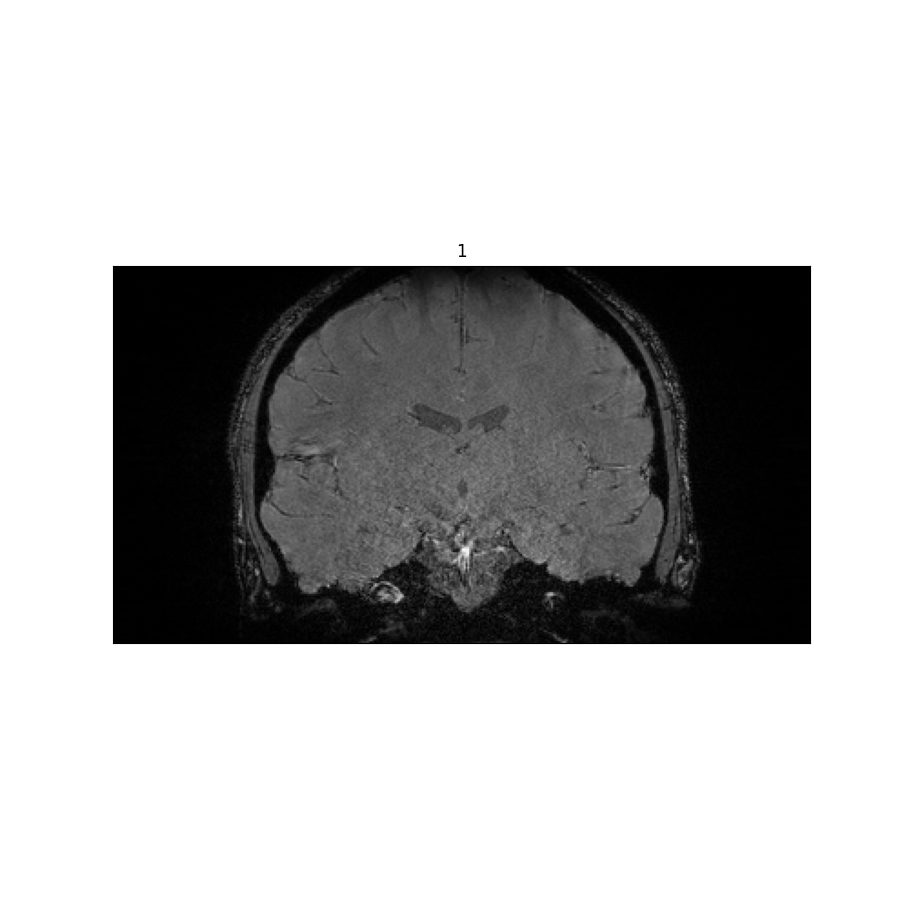}}
				\\
			\end{tabular}}
		\end{mdframed}
		\caption{Comparison of retrospective results for (i) fully optimized 3D SPARKLING~(top row) and (ii) SpSOS~(bottom row) for varying acceleration factors~(from left to right, AF=10~(a), 15~(b) and 20~(c)) on in vivo human brain scans. Cartesian Reference (e) is provided for comparison and results for full 3D trajectory at AF 40~(d) is also presented. SSIM scores are reported for each setup.}
		\label{fig:retropective_invivo}
\end{figure*}

\subsection{In vivo}
We collected in vivo data with full 3D SPARKLING and SpSOS trajectories for brain imaging on a healthy volunteer~(male, 25 y.o.). This study was approved by a national ethics committee (CPP 100048). The volunteer signed a written informed consent form. %The MR image  associated with this k-space data was reconstructed by the method \cite{pipe_dc, Knoll2014, gueddari:hal-02399267, ElGueddari_SAM18} as described in supplementary material~(see Sec.~\ref{supp-app:reconstruction}).

\subsubsection{Retrospective studies}
To understand how the trajectories perform for in vivo brain data, we repeat the earlier retrospective study on Cartesian p4 scans acquired on the volunteer. The results of the scans are presented in Fig.~\ref{fig:retropective_invivo}. We see that \revcomment{full}{r:title} 3D SPARKLING trajectories outperforms the SpSOS trajectories both visually and quantitatively in SSIM metrics with maximum SSIM of 0.964~(AF=10). \revcomment{Moreover we show that both SPARKLING trajectories outperform TPI in  Fig.~\ref{supp-fig:compare_tpi_invivo} in Supplementary Material.}{r:SOTA} Additionally, the SSIM metrics follow the similar trend as seen for phantom data, with the SSIMs for SpSOS dropping off more rapidly from 0.93~(AF=10) to 0.759~(AF=20). In contrast, the \revcomment{full}{r:title} 3D SPARKLING trajectories tend to preserve the structures~(SSIM scores above 0.9 at AF=15 and 20) and show some blurring artifacts only at AF=40 where SSIM drops to 0.792. Particularly, it is interesting to note that full 3D SPARKLING at AF=40 outperforms SpSOS at AF=20.

\subsubsection{Prospective acquisition}
\label{prospective_results}
\begin{figure*}[h]
	\centering
	\begin{mdframed}[innertopmargin=2pt, innerbottommargin=2pt, innerleftmargin=0pt, innerrightmargin=0pt, backgroundcolor=black, leftmargin=0cm,rightmargin=0cm,usetwoside=false]
		\resizebox{\linewidth}{!}{
			\begin{tabular}{c@{\hspace*{1mm}}c@{\hspace*{1mm}}c@{\hspace*{1mm}}c@{\hspace*{1mm}}c@{\hspace*{1mm}}c@{\hspace*{1mm}}c@{\hspace*{1mm}}c@{\hspace*{1mm}}c@{\hspace*{1mm}}c}
				&\multicolumn{2}{c}{\Bw{\Large (a) AF = 10}}&
				\multicolumn{2}{c}{\Bw{\Large (b) AF = 15}}&
				\multicolumn{2}{c}{\Bw{\Large (c) AF = 20}}&
				\multicolumn{2}{c}{\Bw{\Large (d) AF = 40}}
				\\
				&\multicolumn{2}{c}{\Bw{\small  4min 58sec}}&
				\multicolumn{2}{c}{\Bw{\small	3min 22sec}}&
				\multicolumn{2}{c}{\Bw{\small	2min 36sec}}&
				\multicolumn{2}{c}{\Bw{\small   1min 16sec}}&
				\\
				\multirow{2}{*}[0.4in]{\rotatebox[origin=c]{90}{\Bw{\Large (i) Full 3D SPARKLING}}}&
				\multirow{2}{*}[0.4in]{\parbox[m]{.23\linewidth}{
						\includegraphics[trim={6cm 4cm 5.5cm 5cm},clip,width=\linewidth]
						{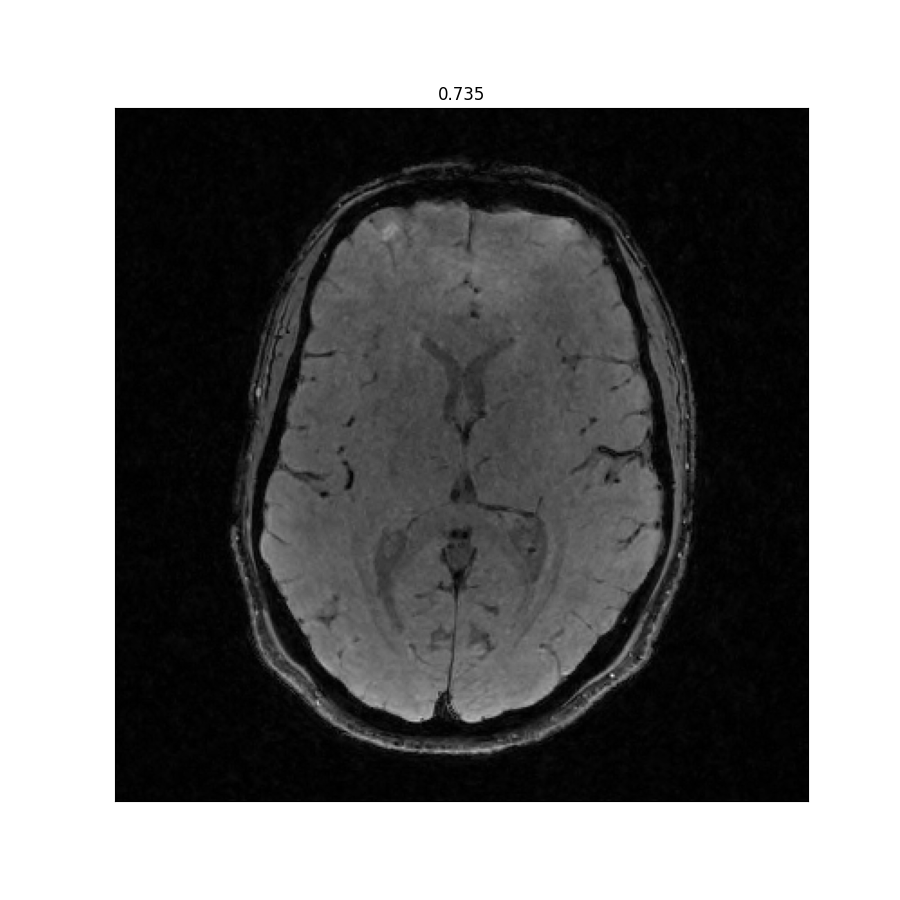}}}&
				\parbox[m]{.2\linewidth}{\includegraphics[trim={5.2cm 6.5cm 4cm 7cm},clip,width=\linewidth]{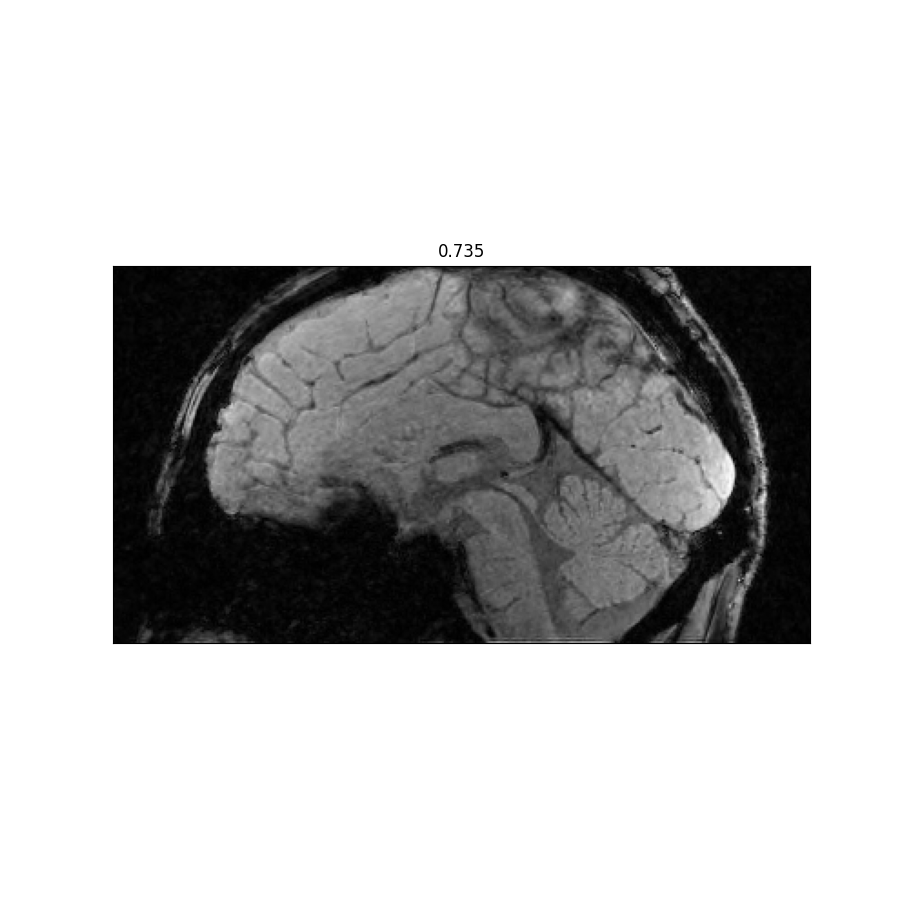}}&
				\multirow{2}{*}[0.4in]{\parbox[m]{.23\linewidth}{\includegraphics[trim={6cm 4cm 5.5cm 5cm},clip,width=\linewidth]
						{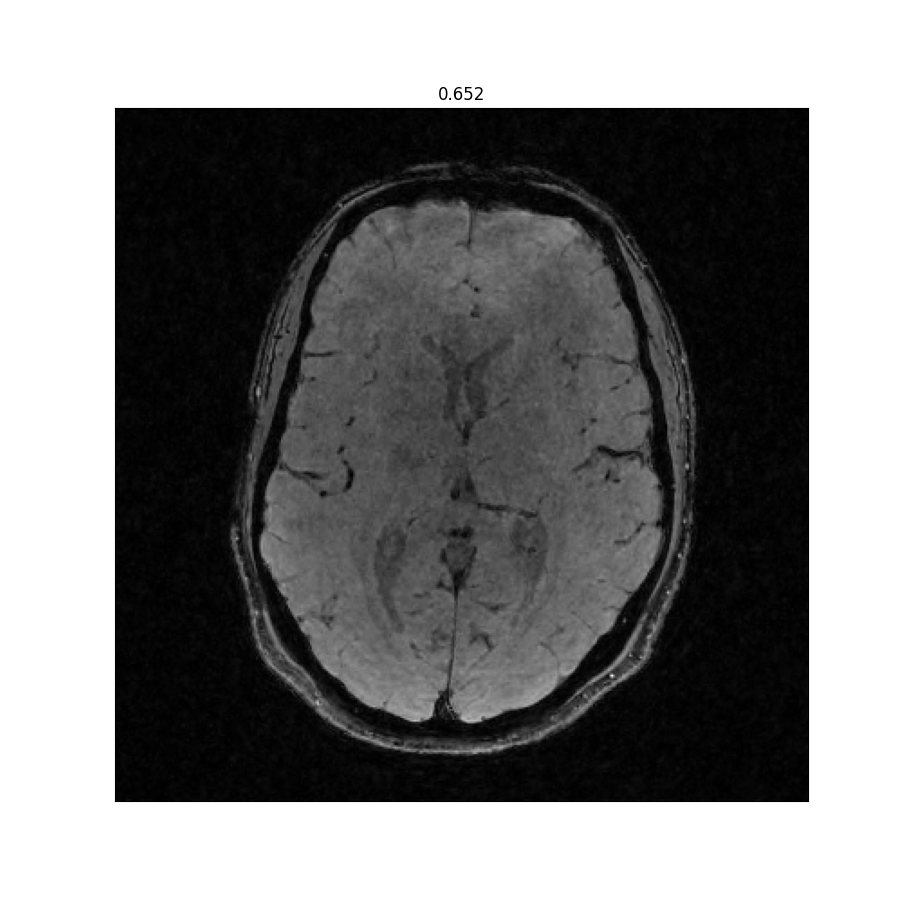}}}&
				\parbox[m]{.2\linewidth}{\includegraphics[trim={5.2cm 6.5cm 4cm 7cm},clip,width=\linewidth]	{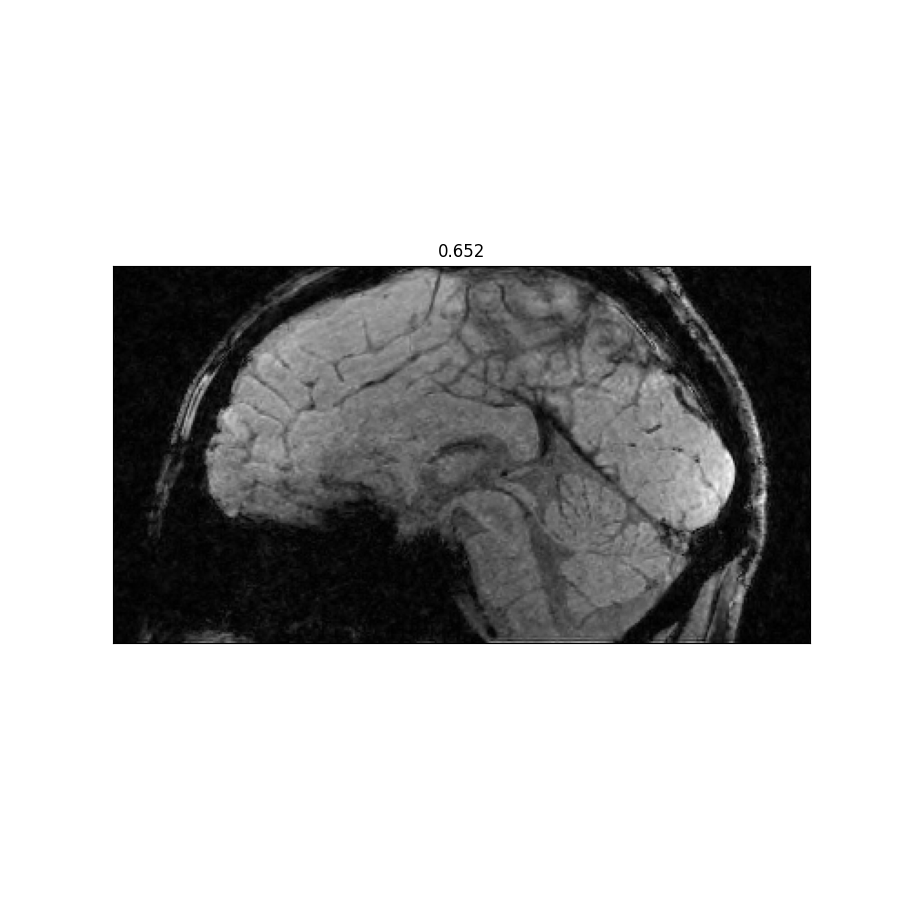}}&
				\multirow{2}{*}[0.4in]{\parbox[m]{.23\linewidth}{
						\includegraphics[trim={6cm 4cm 5.5cm 5cm},clip,width=\linewidth]
						{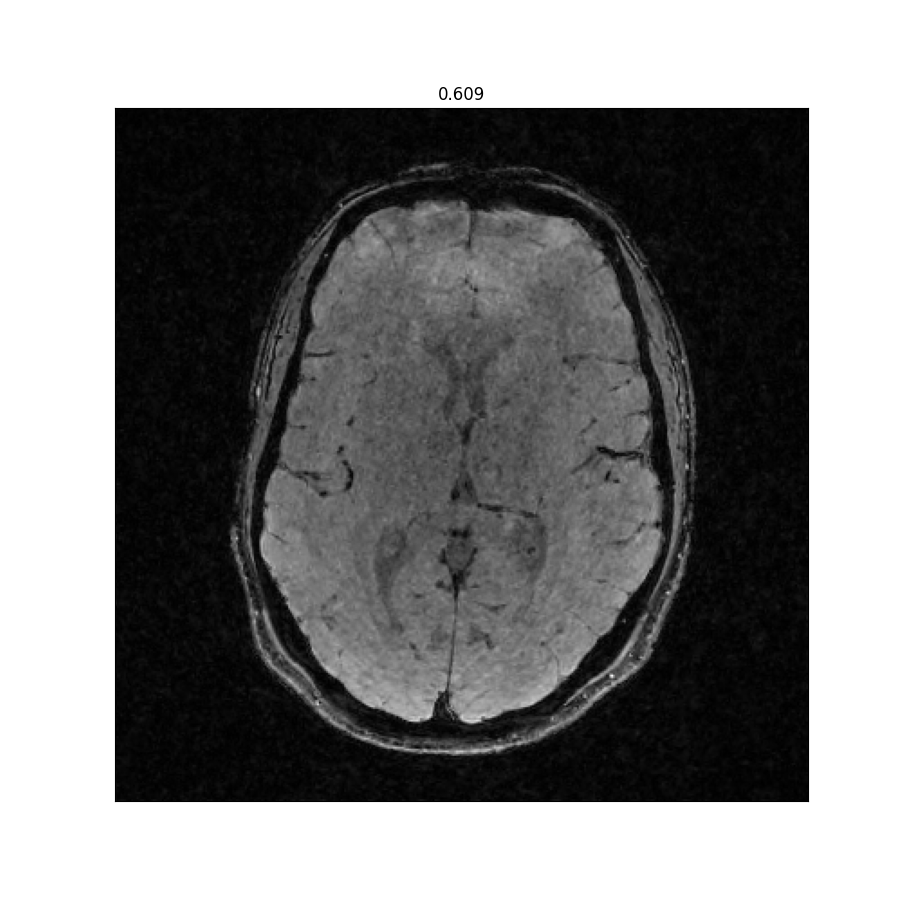}}}&
				\parbox[m]{.2\linewidth}{\includegraphics[trim={5.2cm 6.5cm 4cm 7cm},clip,width=\linewidth]{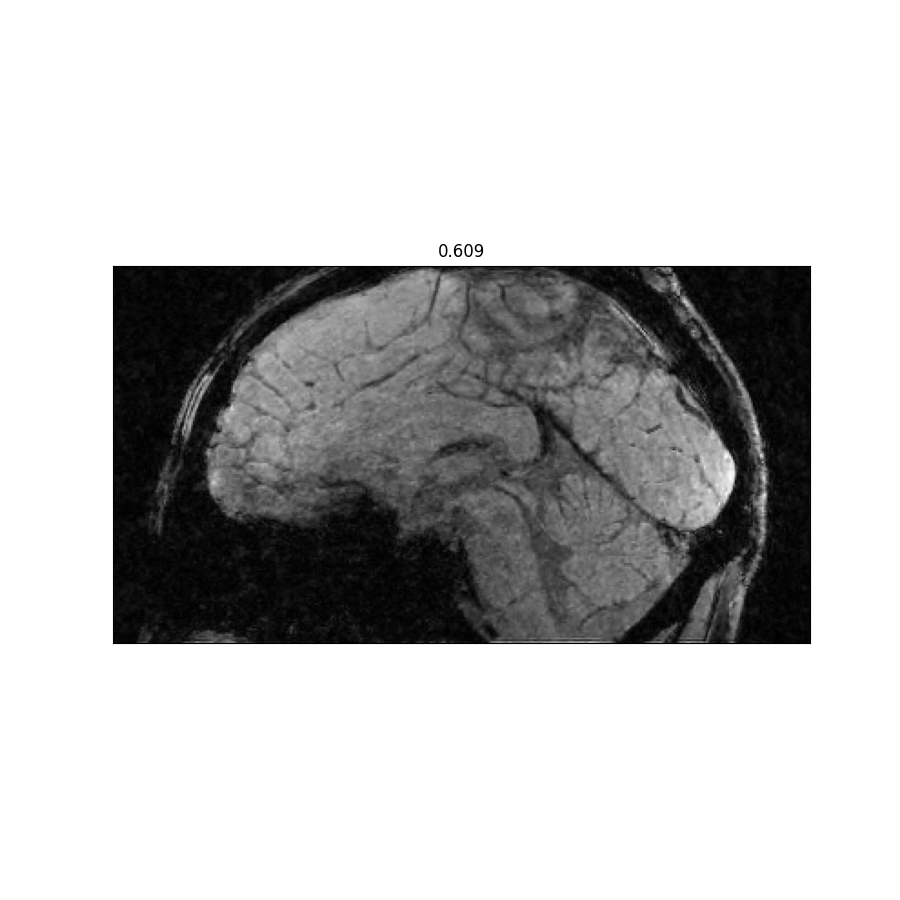}}&
				\multirow{2}{*}[0.4in]{\parbox[m]{.23\linewidth}{
						\includegraphics[trim={6cm 4cm 5.5cm 5cm},clip,width=\linewidth]
						{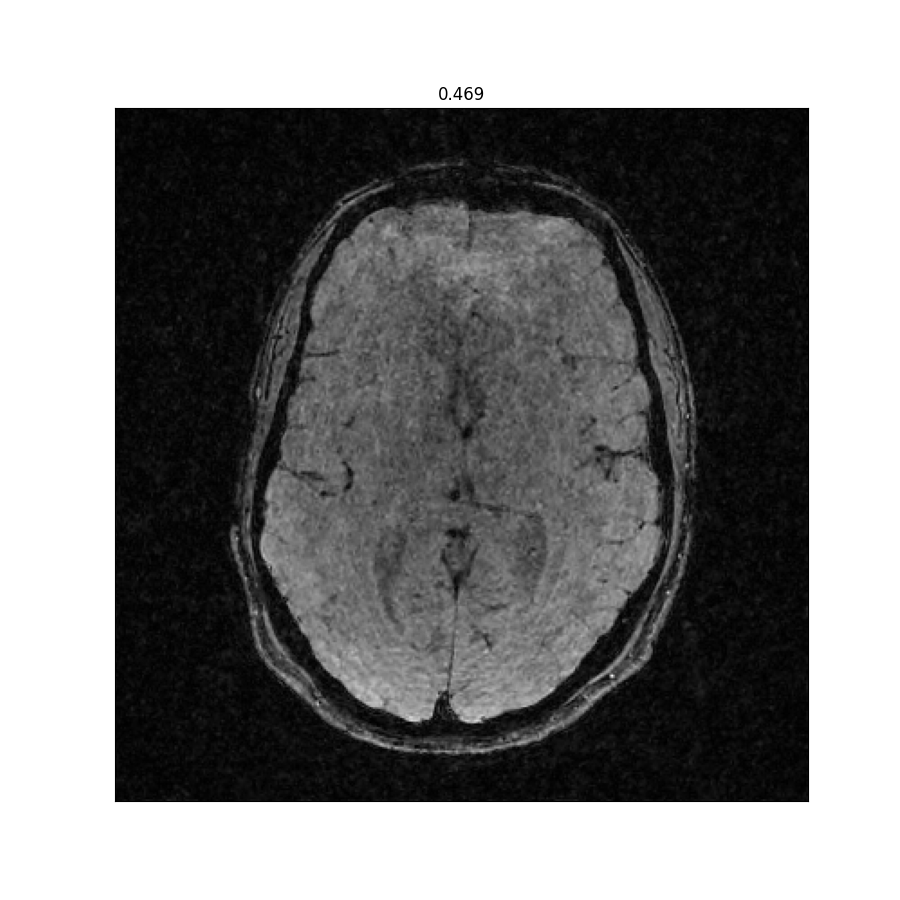}}}&
				\parbox[m]{.2\linewidth}{
					\includegraphics[trim={5.2cm 6.5cm 4cm 7cm},clip,width=\linewidth]
					{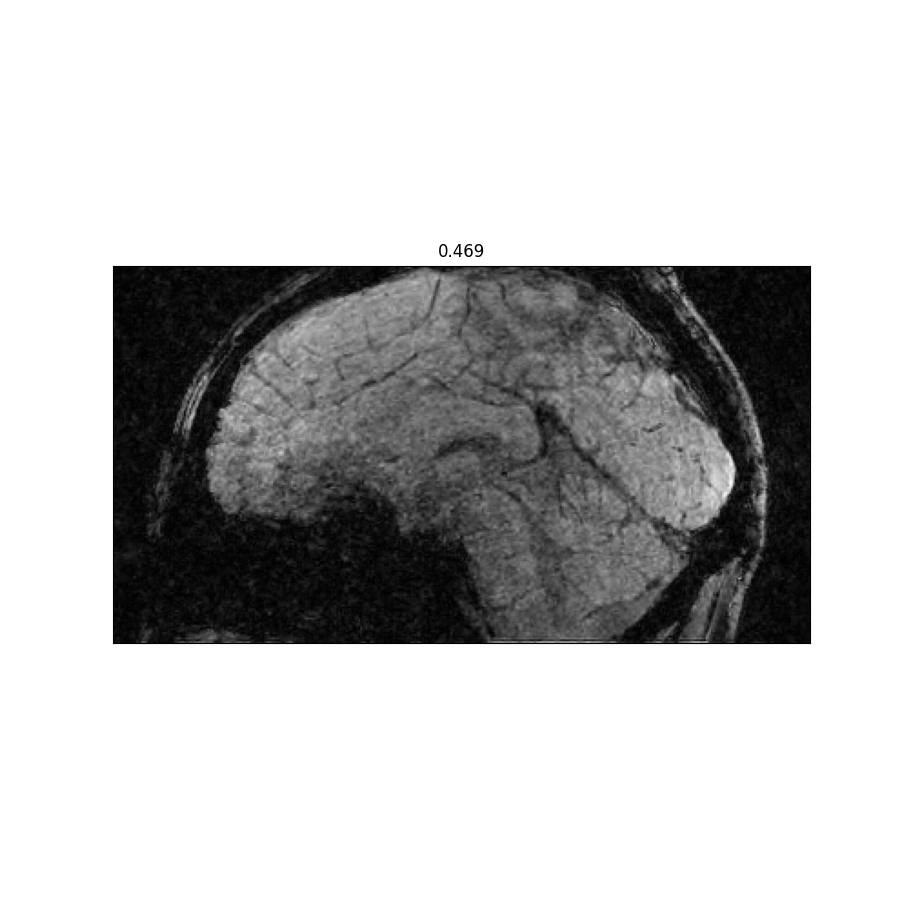}}
				\\
				&&\parbox[m]{.2\linewidth}{\includegraphics[trim={6.2cm 7cm 5.5cm 6.8cm},clip,width=\linewidth]{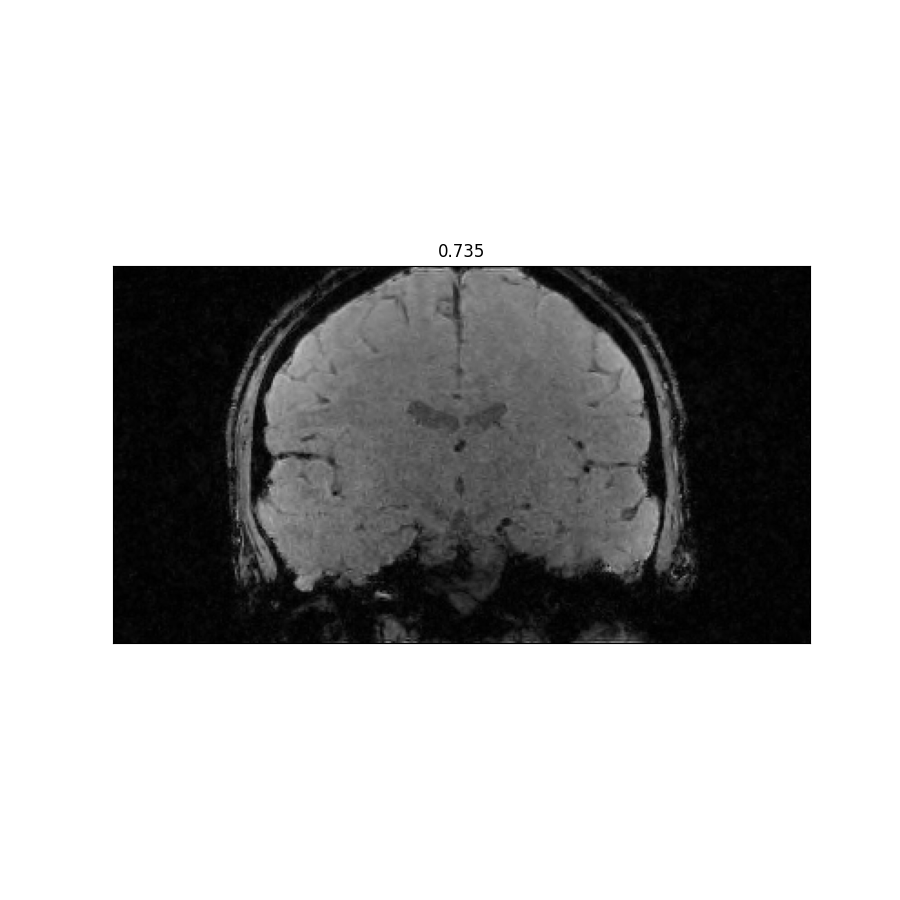}}&
				&\parbox[m]{.2\linewidth}{\includegraphics[trim={6.2cm 7cm 5.5cm 6.8cm},clip,width=\linewidth]{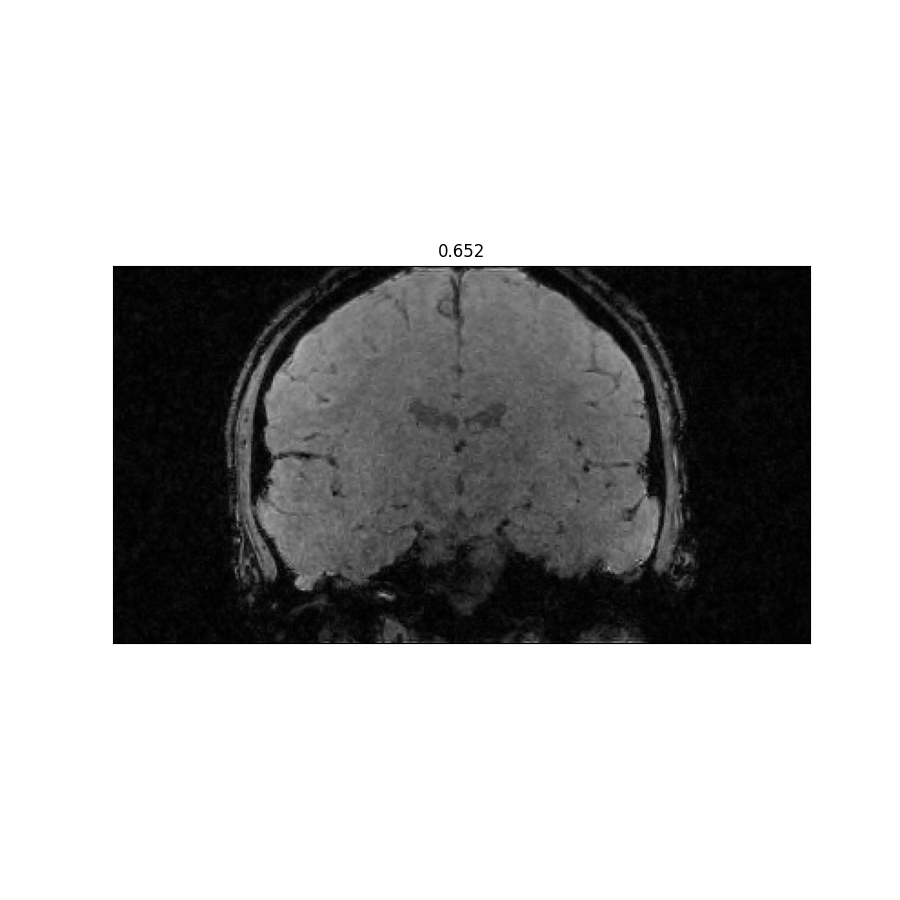}}&
				&\parbox[m]{.2\linewidth}{\includegraphics[trim={6.2cm 7cm 5.5cm 6.8cm},clip,width=\linewidth]{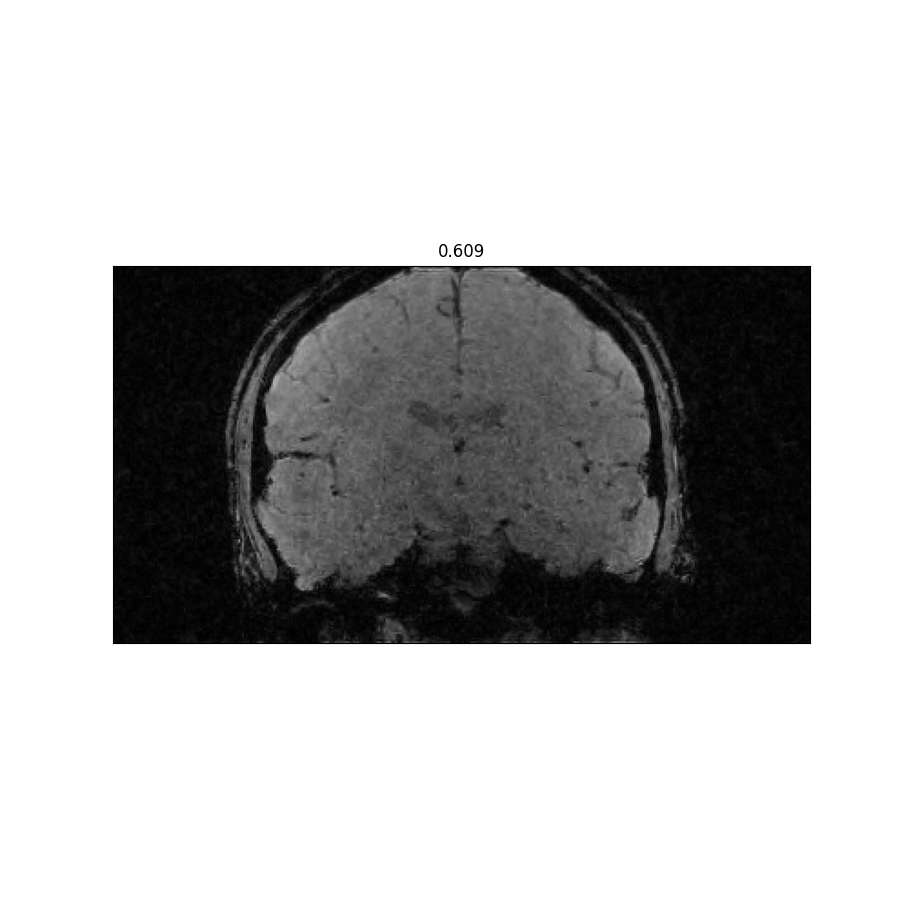}}&
				&\parbox[m]{.2\linewidth}{\includegraphics[trim={6.2cm 7cm 5.5cm 6.8cm},clip,width=\linewidth]{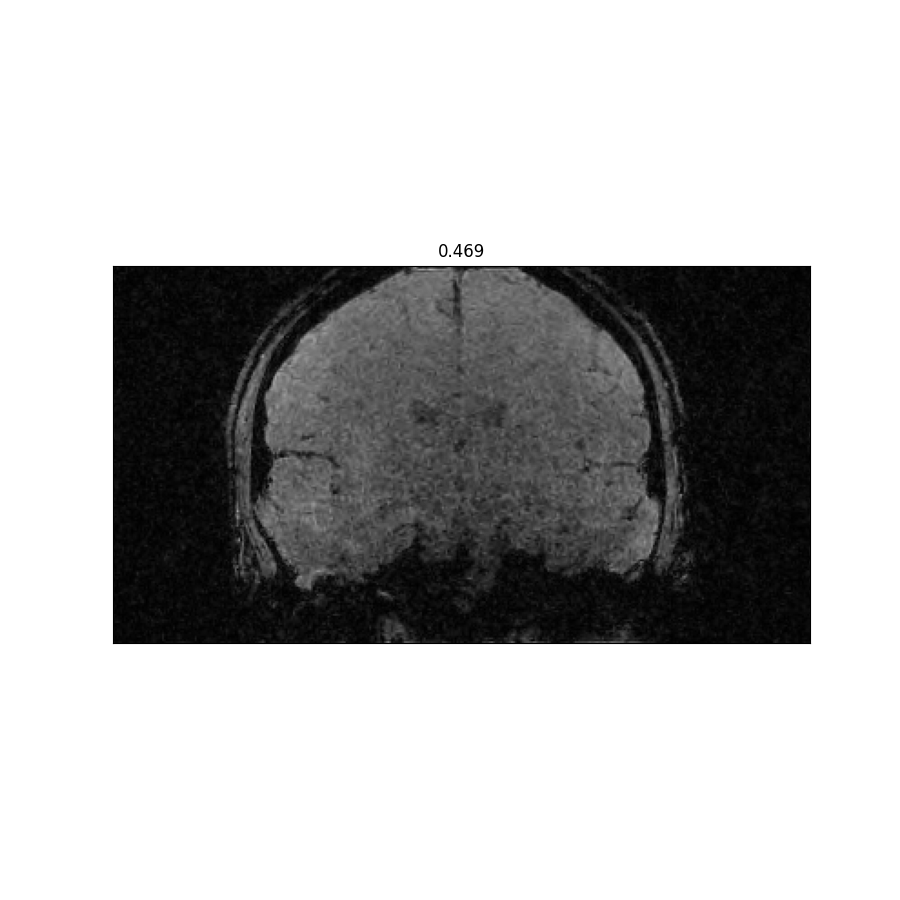}}
				\\
				\arrayrulecolor{white}\cmidrule[2pt]{8-10}
				
				\multirow{2}{*}[0.4in]{\rotatebox[origin=c]{90}{\Bw{\Large (ii) \texttt{SpSOS} SPARKLING}}}&
				\multirow{2}{*}[0.4in]{\parbox[m]{.23\linewidth}{\includegraphics
						[trim={6cm 4cm 5.5cm 5cm},clip,width=\linewidth]
						{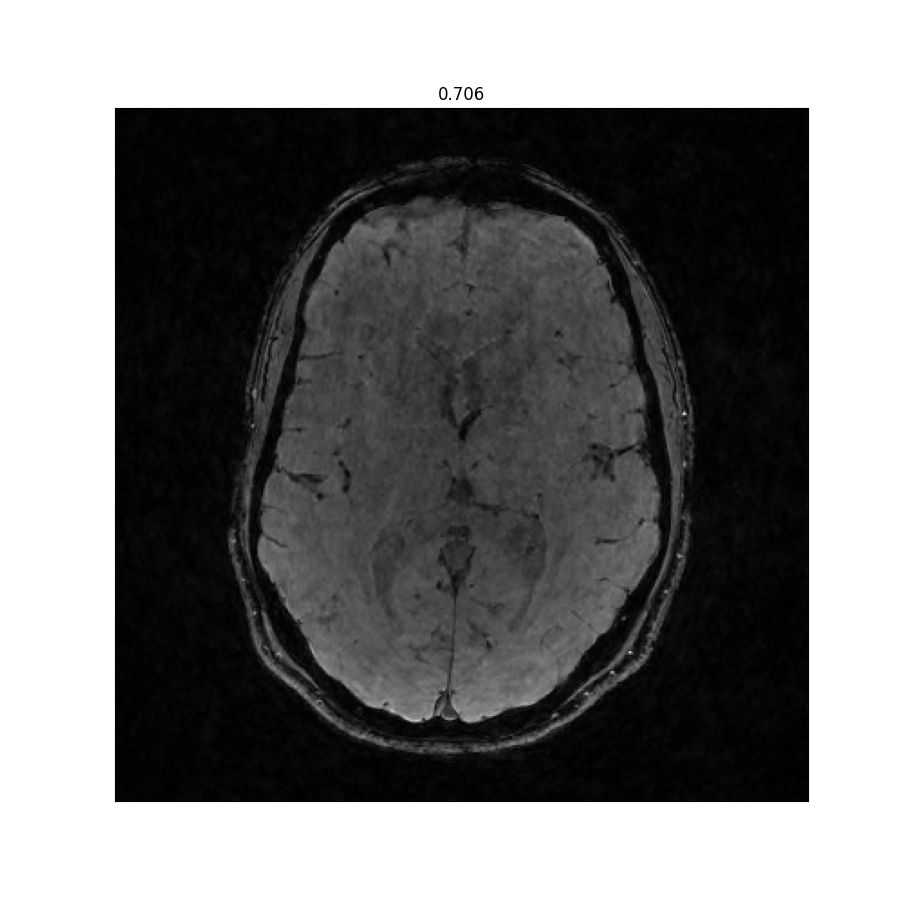}}}&
				\parbox[m]{.2\linewidth}{\includegraphics[trim={5.2cm 6.5cm 4cm 7cm},clip,width=\linewidth]{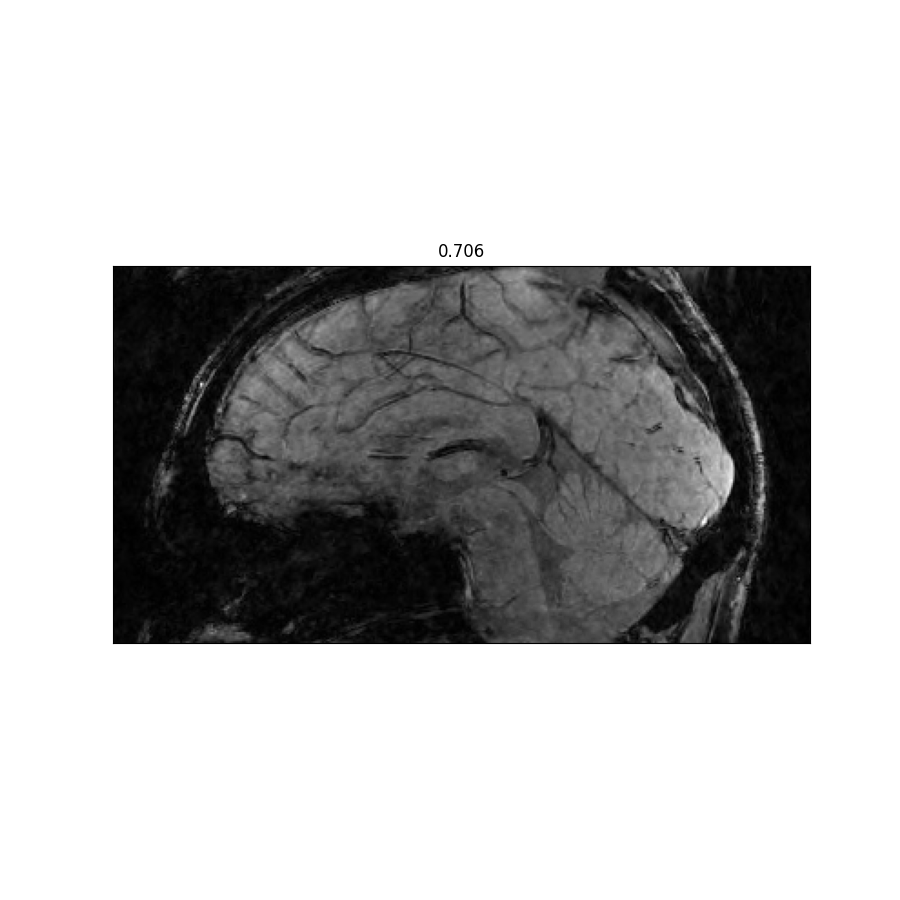}}&
				\multirow{2}{*}[0.4in]{\parbox[m]{.23\linewidth}{\includegraphics[trim={6cm 4cm 5.5cm 5cm},clip,width=\linewidth]
						{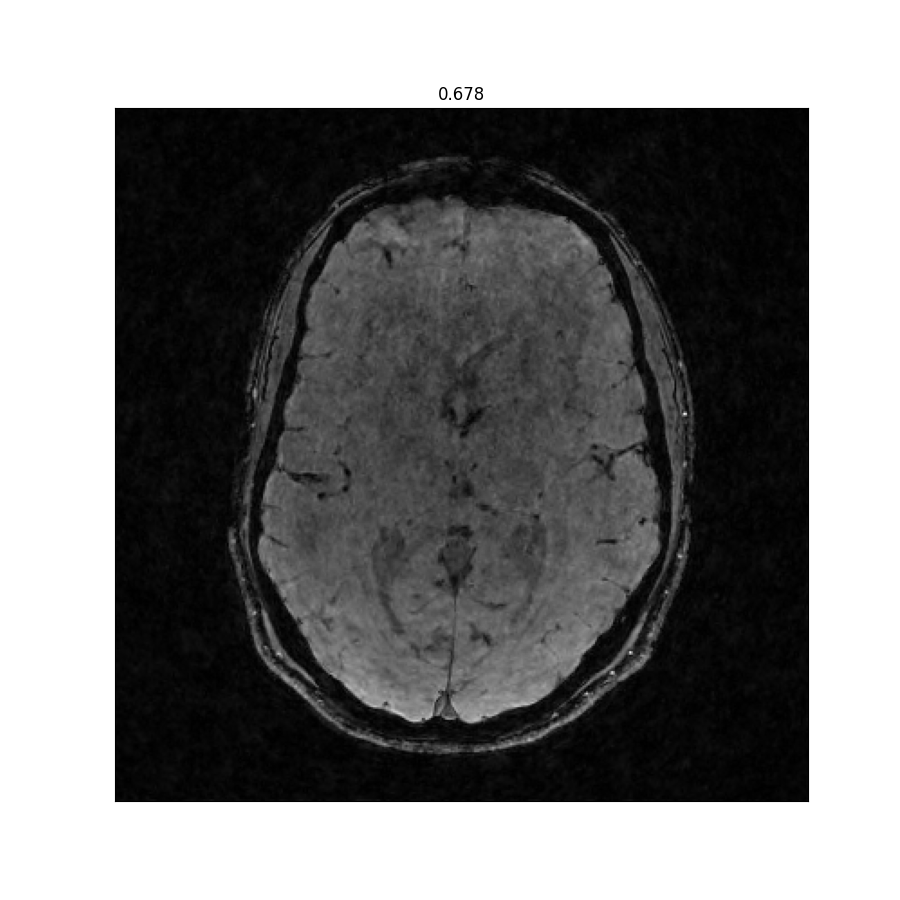}}}&
				\parbox[m]{.2\linewidth}{\includegraphics[trim={5.2cm 6.5cm 4cm 7cm},clip,width=\linewidth]{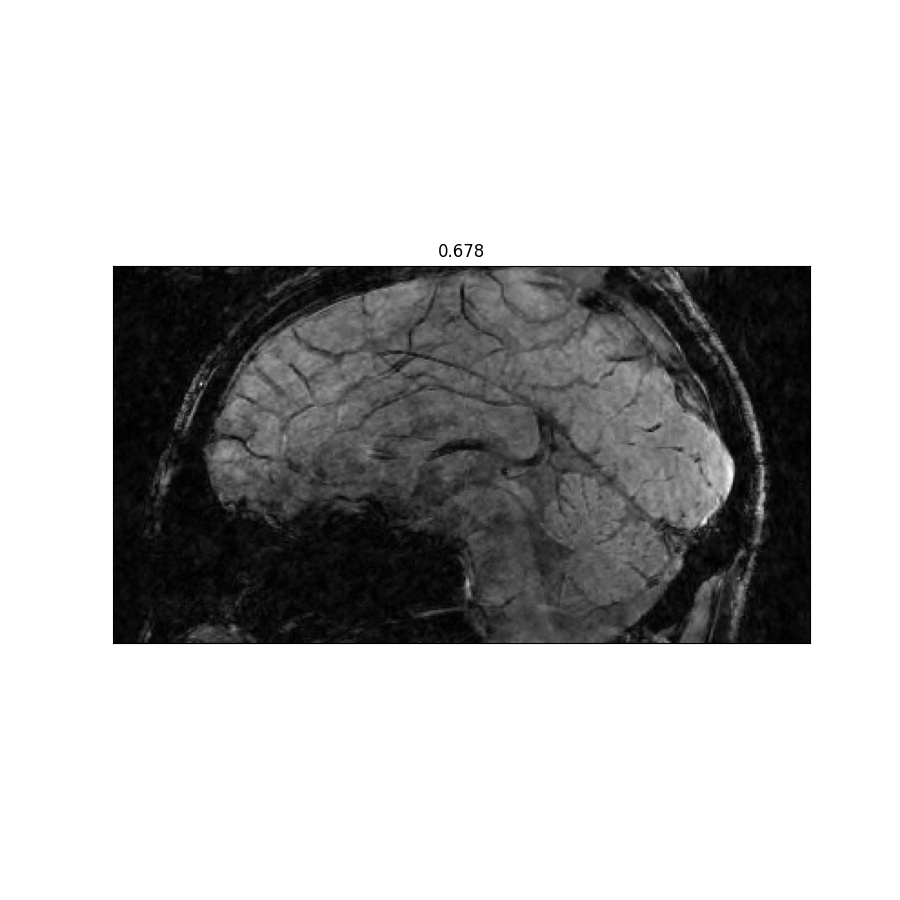}}&				\multirow{2}{*}[0.4in]{\parbox[m]{.23\linewidth}{\includegraphics[trim={6cm 4cm 5.5cm 5cm},clip,width=\linewidth]
						{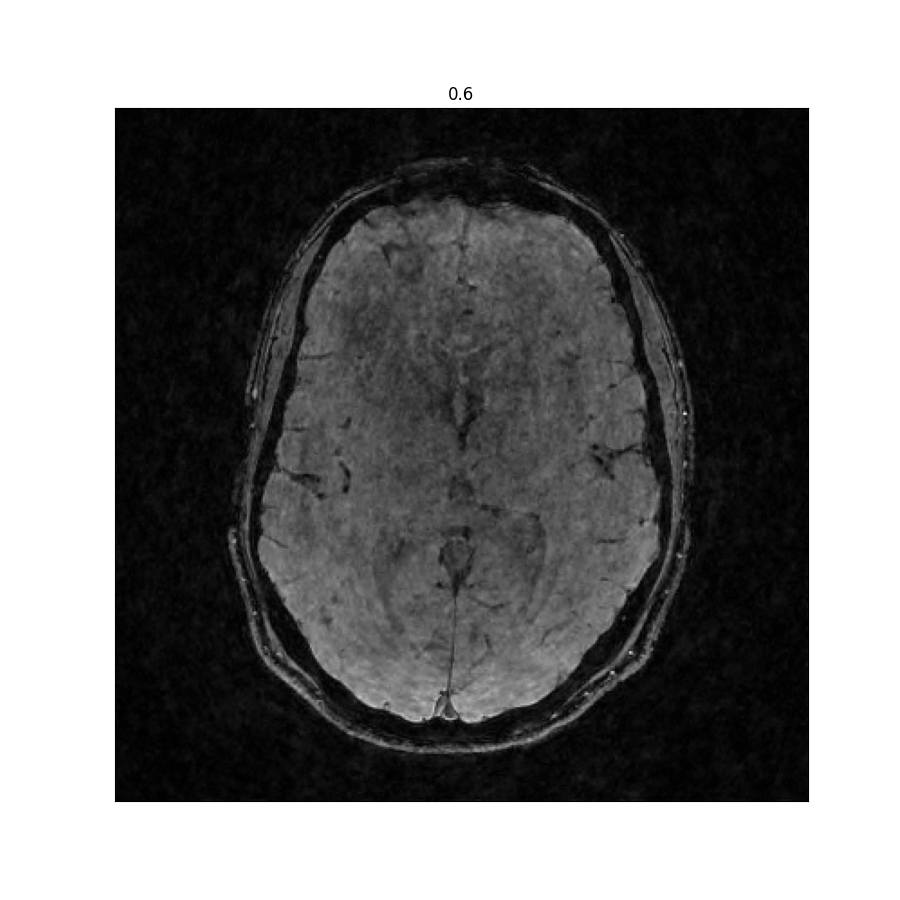}}}&
				\multicolumn{1}{c!{\color{white}\vrule width 2pt \hspace*{2mm}}}
				{\parbox[m]{.2\linewidth}{\includegraphics[trim={5.2cm 6.5cm 4cm 7cm},clip,width=\linewidth]{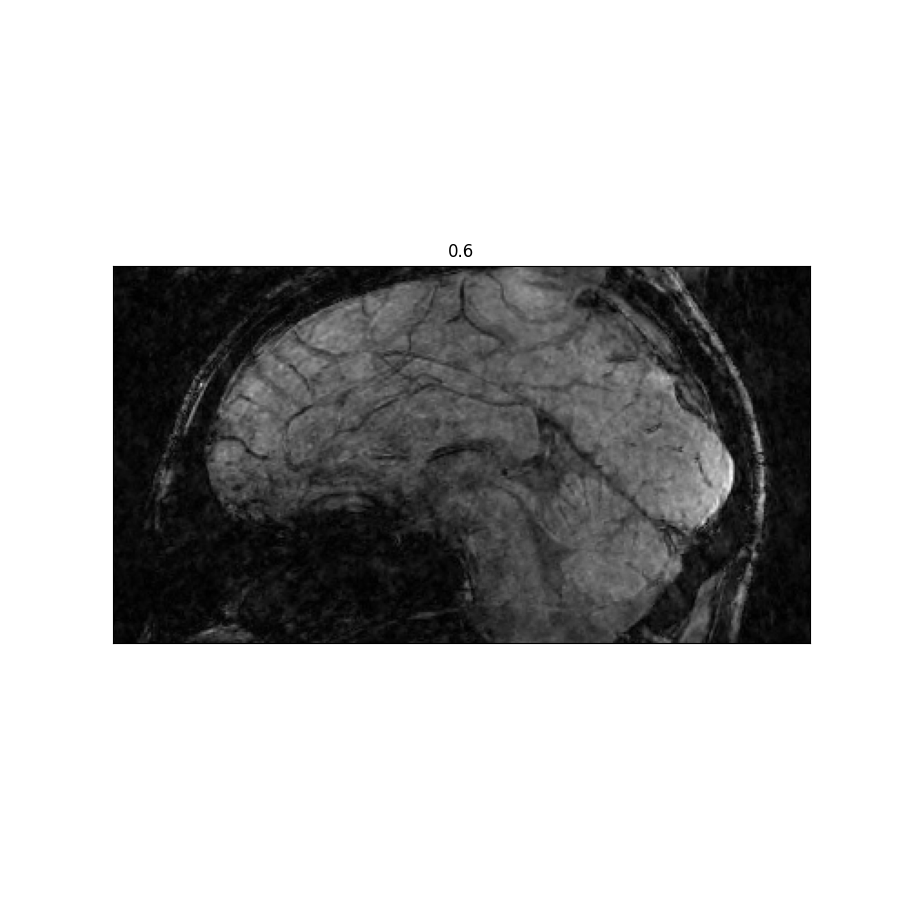}}}
				&
				\multirow{2}{*}[0.4in]{\parbox[m]{.23\linewidth}{\includegraphics[trim={6cm 4cm 5.5cm 5cm},clip,width=\linewidth]{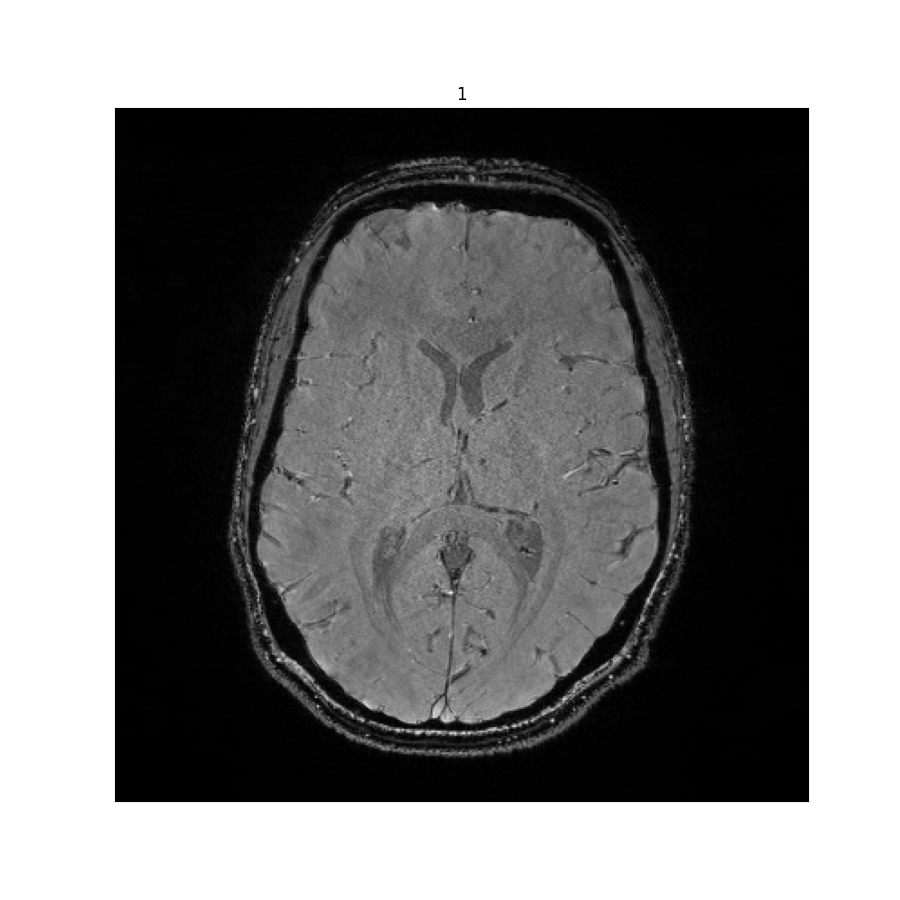}}}&
				\parbox[m]{.2\linewidth}{\includegraphics[trim={5.2cm 6.5cm 4cm 7cm},clip,width=\linewidth]{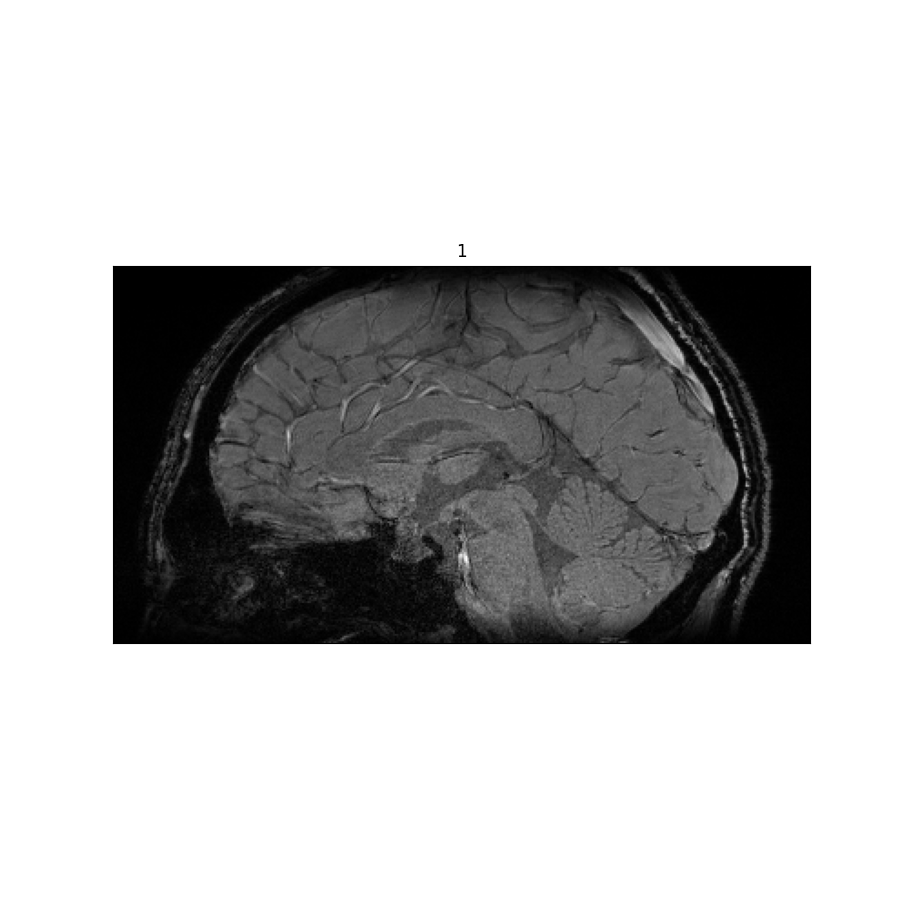}}&
				\multirow{2}{*}[0.15in]{\rotatebox[origin=c]{270}{\Bw{\thead{\Large (e) Cartesian p4 \\ \small 15min 13sec}}}}
				\\
				&&\parbox[m]{.2\linewidth}{\includegraphics[trim={6.2cm 7cm 5.5cm 6.8cm},clip,width=\linewidth]{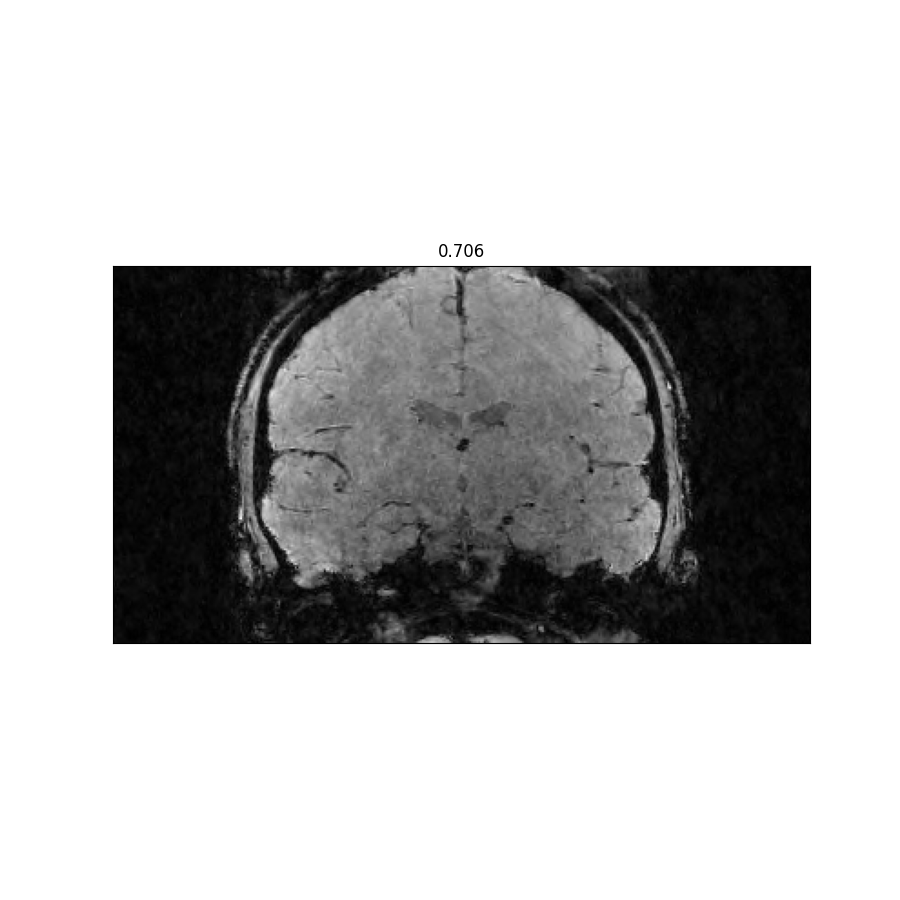}}&
				&\parbox[m]{.2\linewidth}{\includegraphics[trim={6.2cm 7cm 5.5cm 6.8cm},clip,width=\linewidth]{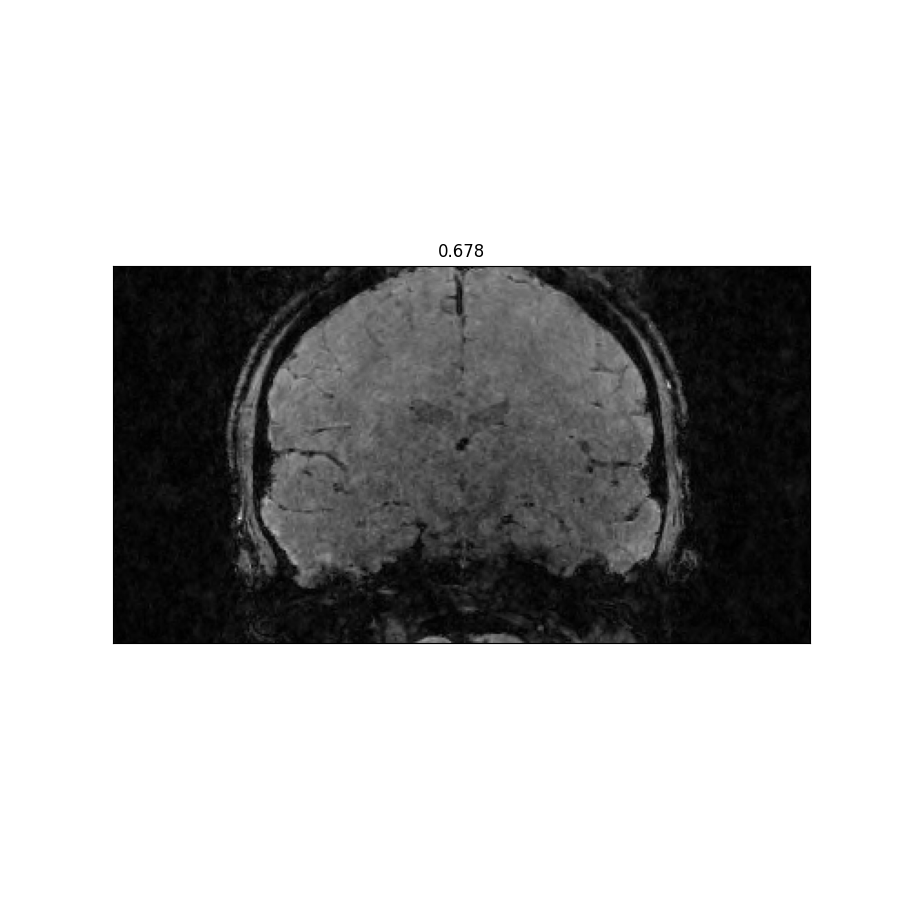}}&
				&\multicolumn{1}{c!{\color{white}\vrule width 2pt \hspace*{2mm}}}{\parbox[m]{.2\linewidth}{\includegraphics[trim={6.2cm 7cm 5.5cm 6.8cm},clip,width=\linewidth]{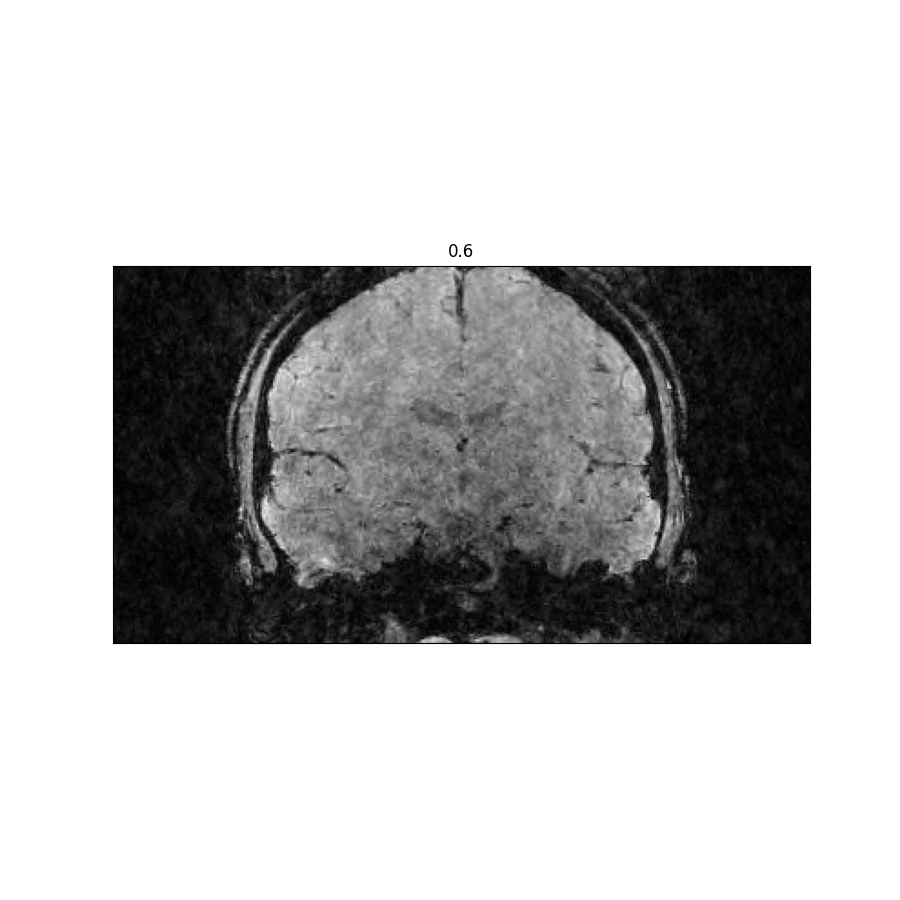}}}&
				&\parbox[m]{.2\linewidth}{\includegraphics[trim={6.2cm 7cm 5.5cm 6.8cm},clip,width=\linewidth]{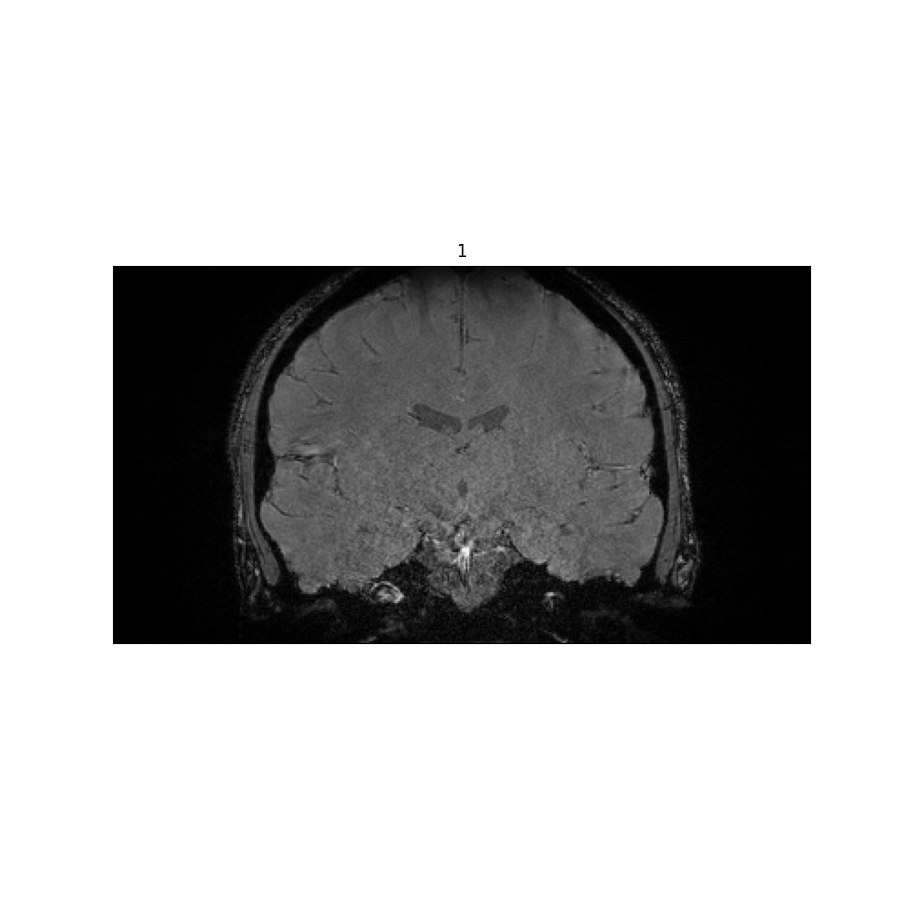}}
				\\
		\end{tabular}}
	\end{mdframed}
	\caption{Comparison of prospective results for (i) fully optimized 3D SPARKLING~(top row) and (ii) SpSOS~(bottom row) for varying acceleration factors~(from left to right, AF=10~(a), 15~(b) and 20~(c)) on in vivo human brain scans. Cartesian p4 scan (e) is provided for comparison and results for full 3D trajectory at AF 40~(d) is also presented. The scan times are reported for each AF.}
	\label{fig:prospective_invivo}
\end{figure*}

Finally, we collected prospectively accelerated in vivo data at 3T on the same individual using the same SPARKLING trajectories. We present the reconstructed images for various accelerations factors in Fig.~\ref{fig:prospective_invivo}. They clearly show superiority of the full 3D SPARKLING pattern compared to SpSOS. Image quality is well preserved for AF=10 and 15 and slightly noisy at AF=20 in full 3D strategy~(Fig.~\ref{fig:prospective_invivo}, top row), while it tends to get noisy at AF=15 and severely impaired at AF=20 for SpSOS strategy~(Fig.~\ref{fig:prospective_invivo}, bottom row). Moreover, we observed that the quality of AF=15, AF=20 and AF=40 in full 3D strategy is comparable to AF=10, AF=15 and AF=20 in SpSOS pattern respectively, allowing for an additional 2x shorter scan time. Further, we found that the full 3D SPARKLING pattern at AF=15 is comparable to GRAPPA Cartesian p4.

%and then the comparison between the Cartesian reference, the restrospective and prospective results in Fig.~\ref{fig:prospective_invivo_zoom}
 
\multirevcomment{It is important to note that the volunteer slightly moved between some scans, hence prospective image comparisons can only be carried out qualitatively. To better understand reconstruction quality, we present zoomed in visualizations for prospective result at AF=10 in Fig.~\ref{fig:prospective_invivo_zoom}.}{\ref{r:resolution_loss}} Further, for the sake of comparison between retrospective simulations and actual prospective scans, we also show the retrospective results for AF=10 with full 3D trajectory. 
We find that full 3D strategy retains better structures of the brain in the MR image than SpSOS, which is clearly visible in the cerebellum in the sagittal view. 

The comparison with retrospective image allows us to directly identify some degradation and loss of small details in prospective images. \revcomment{Potential  explanations for this effect are the $T_2$* blurring and off-resonance artifacts, which drastically drop the effective SNR obtained (see Fig.~\ref{supp-fig:psf_b0} in supplementary material). This confirms that in vivo acquisitions are more challenging.}{r:gap}

\newcommand{\proInvivo}{images/results/prospective/invivo}

\newcommand{\zoomboxx}{(0.4cm, 1.4cm) rectangle (2.2cm, 2.6cm)}
\newcommand{\zoomboxy}{(1.5cm, 0.7cm) rectangle (3cm, 1.8cm)}
\newcommand{\zoomboxz}{(0.6cm, 0.7cm) rectangle (3cm, 3.5cm)}
\begin{figure*}[h]
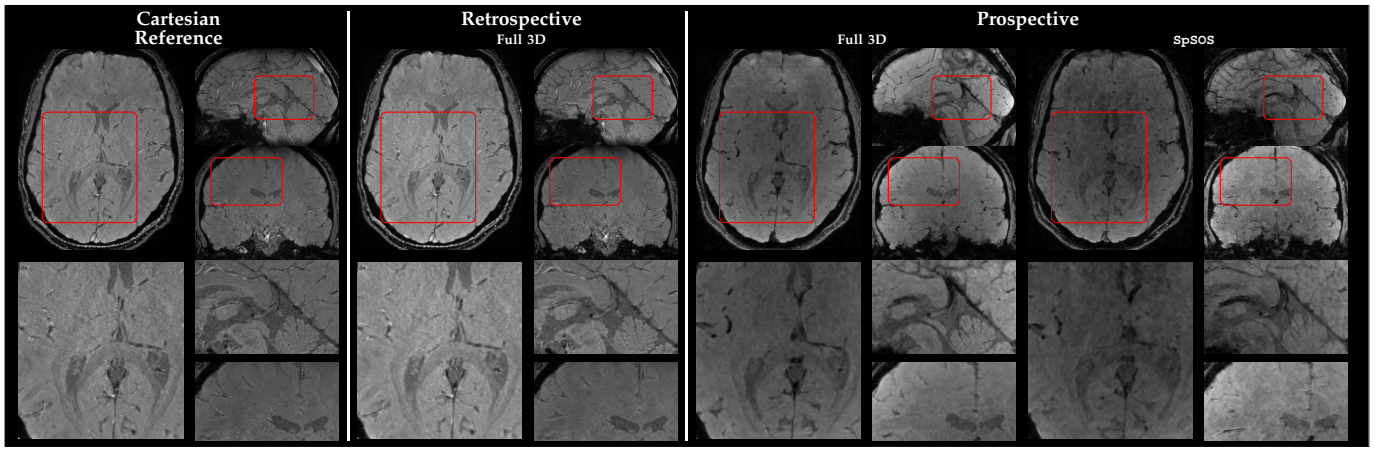

	\centering
\begin{mdframed}[innertopmargin=2pt, innerbottommargin=2pt, innerleftmargin=0pt, innerrightmargin=0pt, backgroundcolor=black, leftmargin=0cm,rightmargin=0cm,usetwoside=false]
\resizebox{\linewidth}{!}{
	\begin{tabular}{c@{\hspace*{1mm}}c@{\hspace*{1mm}}c@{\hspace*{1mm}}c@{\hspace*{1mm}}c@{\hspace*{1mm}}c@{\hspace*{1mm}}c@{\hspace*{1mm}}c@{\hspace*{1mm}}c}		
	\multicolumn{2}{c!{\color{white}\vrule width 2pt \hspace*{2mm}}}{\Bw{\Large Cartesian}}&
	\multicolumn{2}{c!{\color{white}\vrule width 2pt \hspace*{2mm}}}{\Bw{\Large Retrospective}}&
	\multicolumn{4}{c}{\Bw{\Large Prospective}}
	\\
	\multicolumn{2}{c!{\color{white}\vrule width 2pt \hspace*{2mm}}}{\Bw{\Large Reference}}&
	\multicolumn{2}{c!{\color{white}\vrule width 2pt \hspace*{2mm}}}{\Bw{Full 3D}}&
	\multicolumn{2}{c}{\Bw{Full 3D}}&
	\multicolumn{2}{c}{\Bw{\texttt{SpSOS}}}
	\\
	\multirow{2}{*}[0.4in]{\parbox[m]{.23\linewidth}{
			\begin{tikzpicture}
			\node[anchor=south west,inner sep=0] at (0,0){\includegraphics
				[trim={6cm 4cm 5.5cm 5cm},clip,width=\linewidth]{\proInvivo/cartesian_z.png}};
			\draw[red, thick, rounded corners] \zoomboxz;
			\end{tikzpicture}}}&
	\multicolumn{1}{c!{\color{white}\vrule width 2pt \hspace*{2mm}}}{\parbox[m]{.2\linewidth}{
		\begin{tikzpicture}
		\node[anchor=south west,inner sep=0] at (0,0){\includegraphics
			[trim={5.2cm 6.5cm 4cm 7cm},clip,width=\linewidth]{\proInvivo/cartesian_y.png}};
		\draw[red, thick, rounded corners] \zoomboxy;				
		\end{tikzpicture}}}&
	
	\multirow{2}{*}[0.4in]{\parbox[m]{.23\linewidth}{
	\begin{tikzpicture}
		\node[anchor=south west,inner sep=0] at (0,0){\includegraphics
		[trim={6cm 4cm 5.5cm 5cm},clip,width=\linewidth]{\retroInvivo/dim3_Nc89_z.png}};
		\draw[red, thick, rounded corners] \zoomboxz;
	\end{tikzpicture}}}&
	\multicolumn{1}{c!{\color{white}\vrule width 2pt \hspace*{2mm}}}{\parbox[m]{.2\linewidth}{
	\begin{tikzpicture}
		\node[anchor=south west,inner sep=0] at (0,0){\includegraphics
		[trim={5.2cm 6.5cm 4cm 7cm},clip,width=\linewidth]{\retroInvivo/dim3_Nc89_y.png}};
		\draw[red, thick, rounded corners] \zoomboxy;				
	\end{tikzpicture}}}&

	\multirow{2}{*}[0.4in]{\parbox[m]{.23\linewidth}{
	\begin{tikzpicture}
		\node[anchor=south west,inner sep=0] at (0,0){\includegraphics
			[trim={6cm 4cm 5.5cm 5cm},clip,width=\linewidth]{\proInvivo/FULL_SqNc89_z.png}};
		\draw[red, thick, rounded corners] \zoomboxz;
	\end{tikzpicture}}}&
	\parbox[m]{.2\linewidth}{
	\begin{tikzpicture}
		\node[anchor=south west,inner sep=0] at (0,0){\includegraphics
		[trim={5.2cm 6.5cm 4cm 7cm},clip,width=\linewidth]{\proInvivo/FULL_SqNc89_y.png}};
		\draw[red, thick, rounded corners] \zoomboxy;				
	\end{tikzpicture}}&

	\multirow{2}{*}[0.4in]{\parbox[m]{.23\linewidth}{
	\begin{tikzpicture}
		\node[anchor=south west,inner sep=0] at (0,0){\includegraphics
		[trim={6cm 4cm 5.5cm 5cm},clip,width=\linewidth]{\proInvivo/SOS_Nc38_z.png}};
		\draw[red, thick, rounded corners] \zoomboxz;
	\end{tikzpicture}}}&
	\parbox[m]{.2\linewidth}{
	\begin{tikzpicture}
		\node[anchor=south west,inner sep=0] at (0,0){\includegraphics
			[trim={5.2cm 6.5cm 4cm 7cm},clip,width=\linewidth]{\proInvivo/SOS_Nc38_y.png}};
		\draw[red, thick, rounded corners] \zoomboxy;				
	\end{tikzpicture}}&
	\\
	&\multicolumn{1}{c!{\color{white}\vrule width 2pt \hspace*{2mm}}}{\parbox[m]{.2\linewidth}{
	\begin{tikzpicture}
		\node[anchor=south west,inner sep=0] at (0,0) {\includegraphics
		[trim={6.2cm 7cm 5.5cm 7cm},clip,width=\linewidth]{\proInvivo/cartesian_x.png}};
		\draw[red, thick, rounded corners] \zoomboxx;
	\end{tikzpicture}}}&
	&\multicolumn{1}{c!{\color{white}\vrule width 2pt \hspace*{2mm}}}{\parbox[m]{.2\linewidth}{
	\begin{tikzpicture}
		\node[anchor=south west,inner sep=0] at (0,0) {\includegraphics
			[trim={6.2cm 7cm 5.5cm 7cm},clip,width=\linewidth]{\retroInvivo/dim3_Nc89_x.png}};
		\draw[red, thick, rounded corners] \zoomboxx;
	\end{tikzpicture}}} &
	&\parbox[m]{.2\linewidth}{
	\begin{tikzpicture}
		\node[anchor=south west,inner sep=0] at (0,0) {\includegraphics
			[trim={6.2cm 7cm 5.5cm 7cm},clip,width=\linewidth]{\proInvivo/FULL_SqNc89_x.png}};
		\draw[red, thick, rounded corners] \zoomboxx;
	\end{tikzpicture}}&
	&\parbox[m]{.2\linewidth}{
	\begin{tikzpicture}
		\node[anchor=south west,inner sep=0] at (0,0) {\includegraphics
			[trim={6.2cm 7cm 5.5cm 7cm},clip,width=\linewidth]{\proInvivo/SOS_Nc38_x.png}};
		\draw[red, thick, rounded corners] \zoomboxx;
	\end{tikzpicture}}
	\\[1.2cm]
	\multirow{2}{*}[0.36in]{\parbox[m]{.23\linewidth}{\includegraphics[trim={7.5cm 6cm 8.5cm 9.5cm},clip,width=\linewidth]
			{\proInvivo/cartesian_z.png}}}&
	\multicolumn{1}{c!{\color{white}\vrule width 2pt \hspace*{2mm}}}{\parbox[m]{.2\linewidth}{\includegraphics[trim={10.2cm 9cm 6cm 9.5cm},clip,width=\linewidth]{\proInvivo/cartesian_y.png}}}&
	
	\multirow{2}{*}[0.36in]{\parbox[m]{.23\linewidth}{\includegraphics[trim={7.5cm 6cm 8.5cm 9.5cm},clip,width=\linewidth]
			{\retroInvivo/dim3_Nc89_z.png}}}&
	\multicolumn{1}{c!{\color{white}\vrule width 2pt \hspace*{2mm}}}{\parbox[m]{.2\linewidth}{\includegraphics[trim={10.2cm 9cm 6cm 9.5cm},clip,width=\linewidth]{\retroInvivo/dim3_Nc89_y.png}}}&
	
	\multirow{2}{*}[0.36in]{\parbox[m]{.23\linewidth}{\includegraphics[trim={7.5cm 6cm 8.5cm 9.5cm},clip,width=\linewidth]
			{\proInvivo/FULL_SqNc89_z.png}}}&
	\parbox[m]{.2\linewidth}{\includegraphics[trim={10.2cm 9cm 6cm 9.5cm},clip,width=\linewidth]{\proInvivo/FULL_SqNc89_y.png}}&
	
	\multirow{2}{*}[0.36in]{\parbox[m]{.23\linewidth}{\includegraphics[trim={7.5cm 6cm 8.5cm 9.5cm},clip,width=\linewidth]
			{\proInvivo/SOS_Nc38_z.png}}}&
	\parbox[m]{.2\linewidth}{\includegraphics[trim={10.2cm 9cm 6cm 9.5cm},clip,width=\linewidth]
		{\proInvivo/SOS_Nc38_y.png}}
	\\[1.2cm]
	&\multicolumn{1}{c!{\color{white}\vrule width 2pt \hspace*{2mm}}}{\parbox[m]{.2\linewidth}{\includegraphics[trim={7.2cm 11.5cm 9.5cm 8cm},clip,width=\linewidth]{\proInvivo/cartesian_x.png}}}&
	&\multicolumn{1}{c!{\color{white}\vrule width 2pt \hspace*{2mm}}}{\parbox[m]{.2\linewidth}{\includegraphics[trim={7.2cm 11.5cm 9.5cm 8cm},clip,width=\linewidth]{\retroInvivo/dim3_Nc89_x.png}}}&
	&\multicolumn{1}{c!{\hspace*{1mm}}}{\parbox[m]{.2\linewidth}{\includegraphics[trim={7.2cm 11.5cm 9.5cm 8cm},clip,width=\linewidth]{\proInvivo/FULL_SqNc89_x.png}}}&
	&\multicolumn{1}{c!{\hspace*{1mm}}}{\parbox[m]{.2\linewidth}{\includegraphics[trim={7.2cm 11.5cm 9.5cm 8cm},clip,width=\linewidth]{\proInvivo/SOS_Nc38_x.png}}}
\end{tabular}}
\end{mdframed}
\caption{Comparison of prospective results for fully optimized 3D SPARKLING~(right-left) and SpSOS~(right most) with Cartesian reference~(left) and retrospective full 3D SPARKLING~(center) for AF=10 (scan time = 4min 58sec).
In each panel of the top row, axial~(left), sagittal~(top right) and coronal~(bottom right) slices are shown and a red frame is delineated in the central part of the brain for zooming purpose. Bottom row shows the magnified views with the same layout~(axial, sagittal and coronal slices in the left, top-right and bottom right insets, respectively).}
\label{fig:prospective_invivo_zoom}
\end{figure*}

\section{Discussions}
\label{sec:discussions}
One key aspect of optimized \revcomment{full}{r:title} 3D SPARKLING trajectories is that it results in a sampling pattern that enforces variable density sampling in all the 3 dimensions. We hypothesized that this allows us to efficiently under-sample the k-space acquisitions,  thus making it possible to push the acceleration factor to a larger value than what was achieved earlier, while still maintaining a good image quality. The current work actually demonstrates that at fixed acceleration factor, full 3D SPARKLING significantly outperforms the stacking strategy~\cite{Lazarus_NMRB_20} in terms of image quality. Alternatively, we show that this gain can be translated into shorter scan time by a factor of one third~(AF=15 for full 3D vs AF=10 for SpSOS) for a given image quality.

Further, the full 3D trajectory is constrained to pass through the center of k-space for each shot at echo time. This ensures that we obtain the lower frequency image content repeatedly, hence we can potentially use these trajectories for motion correction. Also, as the center of k-space is visited repeatedly at different time intervals in scan, this allows for easy adaptability of this trajectory for dynamic imaging like functional MRI. \revcomment{Such a trajectory can also be used for correcting certain artifacts causing off-resonance effects, which are due to static and dynamic B0 inhomogeneities~(heart beat, breathing). A preliminary solution has been proposed for static B0 inhomogeneities estimation and correction in~\cite{DavalFrerot_ISMRM21}}{r:gap}.

\revcomment{As the developed trajectories exploit the scanner hardware constraints nearly to the maximum, it is worth paying attention to the eddy current effects on the trajectory. To this end, we measured the trajectory with the help of a Skope field camera~\cite{DeZanche2008} and observed in Sec.~\ref{supp-sec:meas_traj} that the error between the prescribed and actual trajectories is minimal~(cf. Fig.~\ref{supp-fig:meas_traj}).}{r:off_res}

\multirevcomment{While the current SPARKLING algorithm is generic and can be applied to any imaging contrast a priori, we choose $T_2^*$-w imaging as it allows us to keep longer $T_\text{obs}$ hence enabling a full exploration of 3D k-space. In order to understand the effects of $T_2^*$ blurring and off-resonance, we simulated the PSF under these scenarios in Sec.~\ref{supp-sec:simulations}. Additionally, as we interfaced a GRE pulse sequence~(FLASH in the Siemens taxonomy) with the SPARKLING outputs, the adaptation of this algorithm to other contrasts~(e.g. $T_2$) would potentially need the development of a turbo spin echo~(TSE) sequence that is able to play arbitrary gradients. These developments are left for future work.}{\ref{r:T2star_TObs}, \ref{r:why_t2star}, \ref{r:off_res}}
%We plan to work on this in our future 

One limitation of SPARKLING is that the original optimization problem~\eqref{eq:global_minimization} is non-convex and the fact we used a locally convergent optimization algorithm to compute a minimizer. Hence the final solution heavily depends on its initialization. To overcome this issue, we introduced some perturbation~(uniform random noise in the k-space locations) and
illustrated  in Fig.~\ref{fig:add_perturbationb} that a larger perturbation results in a much better k-space coverage, allowing us to reach a better minimizer to the original optimization problem. However, there is no theoretical guarantee this approach provides a systematic better solution as the underlying optimization process remains rather disconnected from MR image reconstruction and the maximization of image quality.
%This happens mainly because  

In the same vein, another limitation of the resulting reconstructed MR images is that they heavily depend on the target sampling distribution. We obtained our results by parameterizing this distribution, thereby optimizing for its parameters using a grid search on in vivo brain data. However, these optimal parameters are not generalizable for different contrast and organs. Further, such parametrization can prevent us from using more complex target sampling densities. To overcome this limitation, ongoing work intends to couple SPARKLING with the learning of the target sampling density from the magnitude spectrum of human brain MR images~\cite{GR_EUSIPCO2021}. Akin to this work, we could also jointly optimize for the acquisition (sampling pattern) and reconstruction schemes (regularization parameters) under MR hardware and imaging contrast constraints, either in a bilevel optimization~\cite{sherry2020learning} or using deep learning approaches~\cite{weiss2019pilot,vedula20203d,ramzi2020benchmarking}. 
%,Ramzi\_ISBI2021
\revcomment{These extensions would help us to take some factors, like the anatomy and the imaging contrast, into account in the design of trajectories with perfectly matched target sampling densities for these cases. 

In this work, we did not account for the coil geometry in the design of trajectories. This would require addressing more complex target sampling densities as input which are a function of the coil sensitivities. Optimizing over richer and more adaptive families of densities is left for our future prospects. However, we note that the proposed algorithm can automatically adapt to any future progress made in this direction.}{r:other_factors}

\section{Conclusion}
In this paper, we proposed an \revcomment{optimization for full}{r:title} 3D SPARKLING k-space trajectories for accelerated high resolution 3D $T_2$*-weighted imaging and demonstrated its superiority over the previously proposed stacking strategies on phantom and in vivo human brain data at 3T.
We discussed the major computational bottlenecks that prevented us earlier from proceeding towards these full 3D trajectories. We then derived some implementations~(GPU and multi-CPU) that helped us massively accelerate the original algorithm. Our results showed that a 600\textmu m isotropic scan on human brain is achievable in 1min~16sec, whereas 3m~22sec is required to reach image quality comparable to GRAPPA-4 parallel imaging. Overall, this is a significant step forward for CS acquisitions in MRI. Future work will be devoted to the extension to 4D imaging, namely for fMRI.

\section*{Acknowledgements}
Chaithya G R was supported by the CEA NUMERICS PhD program, which received European funding from Horizon 2020 research and innovation program under the Marie Sklodowska-Curie grant agreement No 800945.
We acknowledge the financial support of the Cross-Disciplinary Program on Numerical Simulation of CEA~(SILICOSMIC project, PI: P. Ciuciu), the French Alternative Energies and Atomic Energy Commission. This work was granted access to the HPC resources of TGCC in France under the allocation 2019-GCH0424 made by GENCI. Pierre Weiss was supported by the ANR JCJC Optimization on Measures Spaces ANR-17-CE23-0013-01 and the ANR-3IA Artificial and Natural Intelligence Toulouse Institute.

\bibliographystyle{IEEEtran}
\bibliography{IEEEabrv,bibliography.bib}

\end{document}

\title{\LARGE Supplementary Material -- Optimizing full 3D SPARKLING trajectories for high-resolution $T_2$*-weighted Magnetic Resonance Imaging}
	
\author{Chaithya GR,
	Pierre~Weiss,
	Guillaume Daval-Fr\'erot,
	Aur\'elien Massire,
	Alexandre Vignaud,
	and~Philippe~Ciuciu \IEEEmembership{Senior Member, IEEE}
}
\maketitle
%\section{Supplementary Material}
%\IEEEpeerreviewmaketitle

\section{Multi-shot projection Step}
\label{app:projection}
We briefly extend the single shot projection step to $N_c$ shots. Particularly, we would like to project the complete sampling pattern $\bK = \left[ \bk_1, \bk_2, \dots, \bk_{N_c}\right]$ such that each shot $\bk_i$ complies with the hardware constraints, $\mathcal{Q}_{\alpha,\beta}$ defined in \eqref{main-eq:constraints}. From \cite{Chauffert_TMI_16}, the projection step for a single shot $\bk$ is given by:
\begin{equation*}
\bk^* = \projk{} = \arg \min_{\boldsymbol{s} \in \mathcal{Q}_{\alpha,\beta}} \frac{1}{2}\|\boldsymbol{s}-\boldsymbol{k}\|_{2}^{2}
\label{eq:cstr_projection_step}
\end{equation*}

We can now further generalize above equation to project all shots as: 
\begin{equation*}
\bK^* = \Pi_{\mathcal{Q}_{\alpha,\beta}^{N_c}}(\mathbf{K}) = \left[\projk{1}, \dots, \projk{N_c} \right]
\end{equation*}
\noindent where, we can use accelerated proximal gradient descent algorithm as described in~\cite{Chauffert_TMI_16} to carry out individual projections of each shot in parallel.

\section{Trajectory}
\subsection{Gradients and Slew Rates}
\label{sec:grads_n_slew}

\begin{figure}[htb]
	\begin{center}
		\begin{tabular}{@{\hspace*{1mm}}c@{\hspace*{1mm}}c}
			{\rotatebox[origin=c]{90}{\bf (a)}} & 
			\parbox[m]{.8\linewidth}{\includegraphics[trim={0.1cm 0cm 0cm 0cm},clip,width=\linewidth]
			{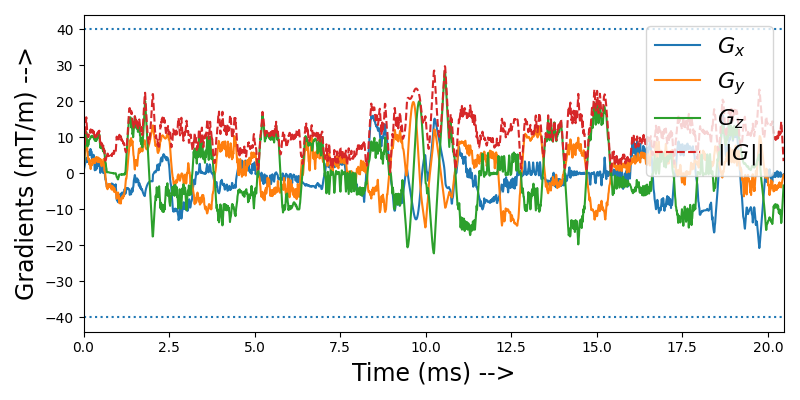}}\\
			{\rotatebox[origin=c]{90}{\bf (b)}}   &
			\parbox[m]{.8\linewidth}{\includegraphics[trim={0.1cm 0cm 0cm 0cm},clip,width=\linewidth]
			{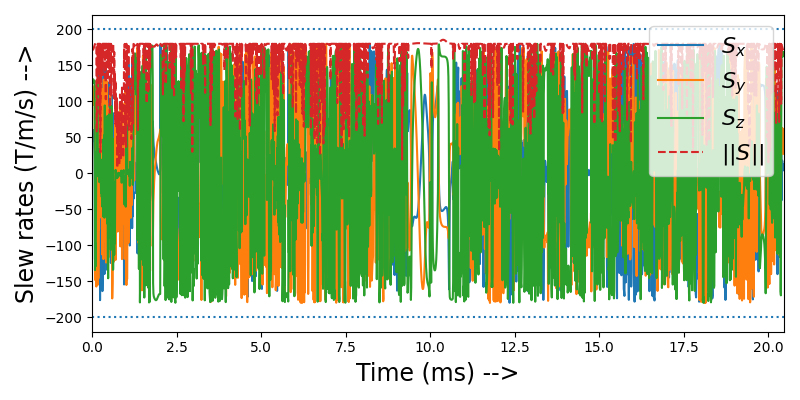}}
		\end{tabular}
		\caption{\label{fig:grads_n_slew} The {\bf (a)} gradients and {\bf (b)} slew rates for a single shot from a Full 3D SPARKLING trajectory with AF=20. We have also marked the Scanner hardware constraints ($G_\text{max} = $ 40mT/m and $S_\text{max} = $200 T/m/s) with black dotted lines.}
	\end{center}
\end{figure}

\revcomment{We present the gradients and slew rates obtained for a single shot of full 3D SPARKLING trajectory in Fig.~\ref{fig:grads_n_slew}.}{r:gradients} \revcomment{We show that the trajectory was mostly slew rate constrained, thereby making the percentage of readouts with gradient magnitude constraint active to be close to 0. The gradient waveform was never saturated but achieved its maximum nearby the center of k-space as shown in Fig.~\ref{fig:grads_n_slew}(a). 
%at a couple of time points, when the trajectory passes through 	
This is because all the trajectories pass through the center of k-space, thereby drastically increasing the sampling density of the sampling pattern. Due to this, each trajectory moves at highest velocity to achieve lesser k-space sample points here, thereby achieving the target sampling density.}{r:perc_grad_active} Note that although the scanner slew rate constraint is 200 T/m/s, the trajectories were optimized with a tighter constraint of maximum allowed slew rate of 180 T/m/s, which explains the lower slew rate limits observed in the trajectory in Fig.~\ref{fig:grads_n_slew}(b).

\subsection{Off-resonance and $T_2^*$ decay}
\label{sec:simulations}
\begin{figure}
	\includegraphics[trim={1cm 0cm 1cm 1cm},clip,width=\linewidth]{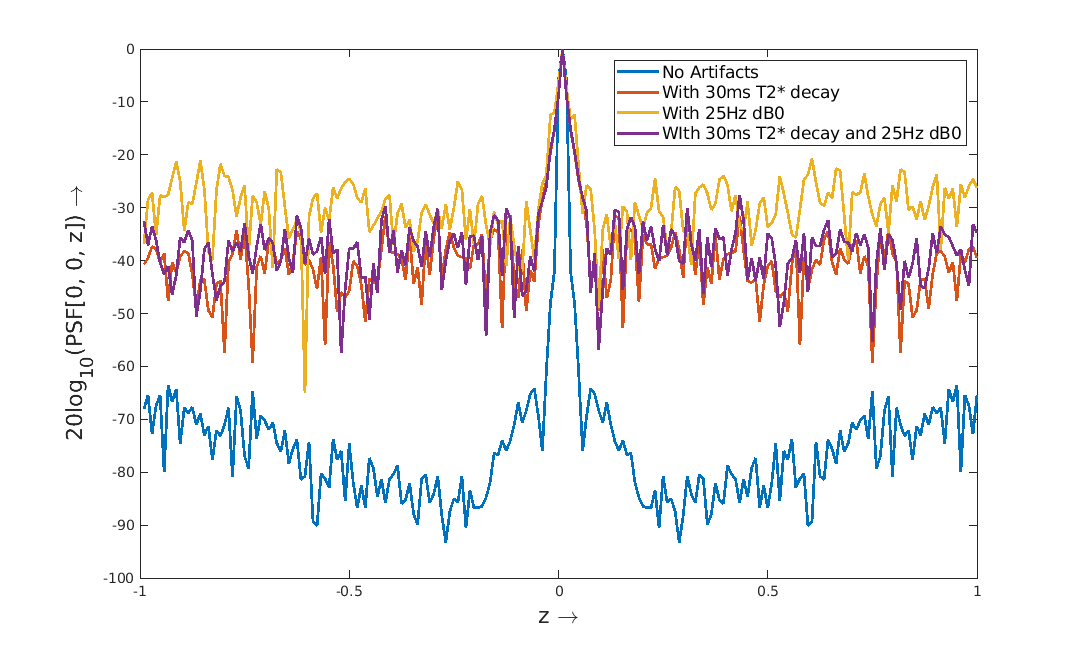}
	\caption{Simulated effects of $T_2$* decay (of 30ms) and constant $\Delta B0$ (of 25Hz) on the point spread function of the AF=10 full 3D SPARKLING trajectory. All the PSFs are normalized such that the maximum value is 1, for easier visual comparison.}
	\label{fig:psf_b0}
\end{figure}

\multirevcomment{We carried out off-resonance and $T_2^*$-decay simulations on the trajectory and analyzed the point spread functions~(PSF) in Fig.~\ref{fig:psf_b0}. We systematically added a $T_2^*$-decay by taking a constant value of $T_2^* = 30$ms, and also performed off-resonance simulations by adding a constant off-resonance~($\Delta B0$) of 25Hz. In Fig.~\ref{fig:psf_b0} we observe that the noise level in the PSF is significantly increased~(by $\simeq 25-30$~dB) when adding $\Delta B0$ and $T_2^*$ decay. Further, we see an increase in sidelobe level when adding $\Delta B0$, which however is reduced when combined with $T_2^*$ decay. We observe that the effective PSF is spread under $T_2^*$ and B0 inhomogeneities, leading to drop in effective resolution, which is studied in depth in the core paper in Sec.~\ref{main-sec:compare_res}. This study reveals some of the reasons that explain the gap in image quality between retrospective and prospective results, especially for in vivo acquisitions, where we have spatially varying $T_2^*$ and $\Delta B0$ ~(i.e. $T_2^*(\br)$ and $\Delta B0(\br)$ maps).}{\ref{r:gap}, \ref{r:simulations}, \ref{r:off_res}}

\subsection{Eddy Current and Trajectory Measurement}
\label{sec:meas_traj}
\begin{figure}[htb]
	\begin{center}
		\resizebox{\linewidth}{!}{
		\begin{tabular}{@{\hspace*{3mm}}c@{\hspace*{3mm}}c}
			{\bf (a)} & {\bf (b)}   \\
			\begin{tikzpicture}
			\node[anchor=south west,inner sep=0] at (0,0){
				\includegraphics[trim={6cm 2.2cm 4.8cm 2cm},clip,width=.5\linewidth]
				{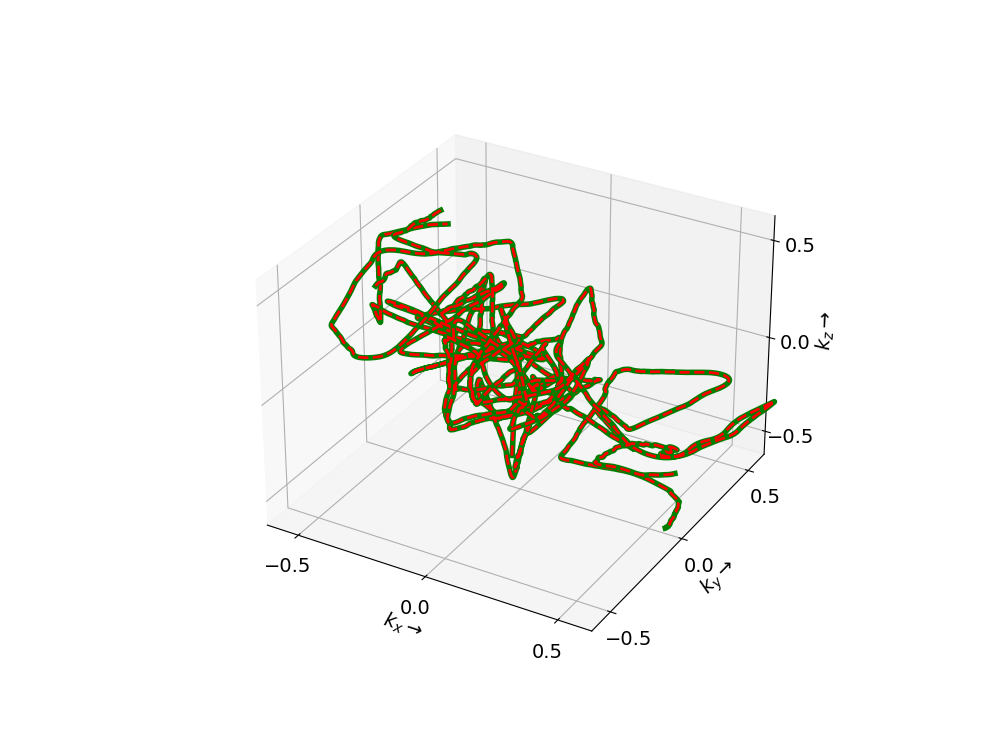}};
			\draw[blue, thick, rounded corners](1.9cm, 2.6cm) rectangle (2.4cm, 2.1cm);
			\end{tikzpicture}&
			\includegraphics[trim={6cm 2.2cm 4.8cm 2cm},clip,width=.5\linewidth]
			{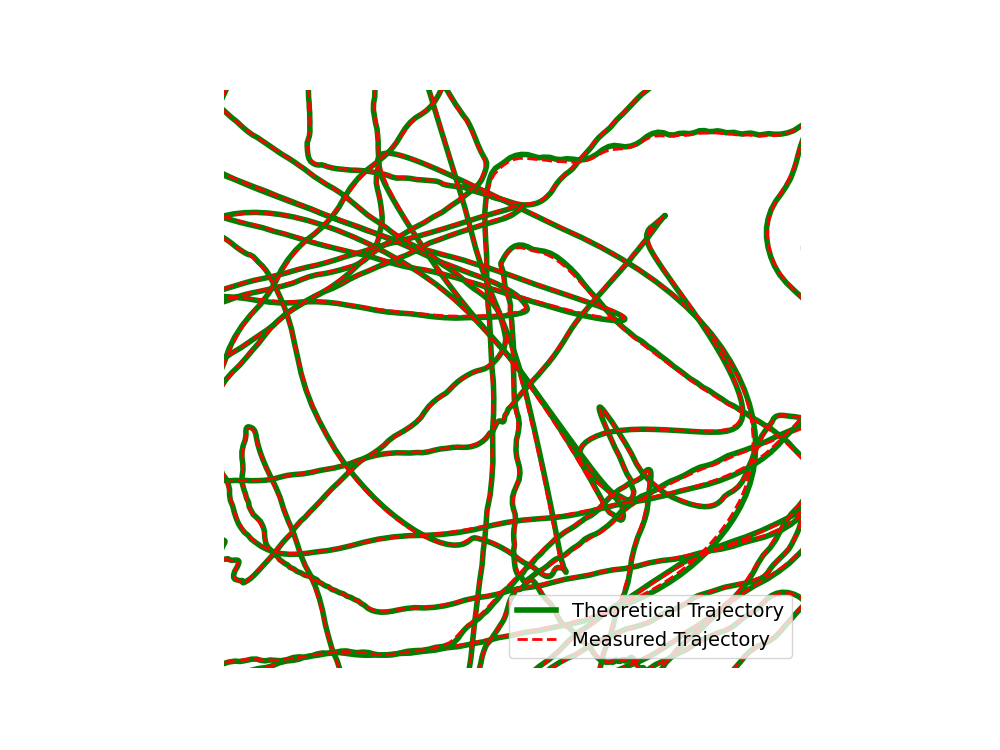}
		\end{tabular}}
		\caption{\label{fig:meas_traj}{\bf (a)} Three random shots from theoretical and measured trajectories for AF=20 full 3D SPARKLING trajectory. {\bf (b)} Zoomed in at the center of k-space.}
	\end{center}
\end{figure}

\multirevcomment{As the presented trajectory rapidly explores the k-space, it is vital to ensure that the MRI scanner is able to play the complicated gradient waveforms in Fig.~\ref{fig:grads_n_slew} with minimal errors. These errors could be induced by eddy currents and gradient imperfections. To this end, we ran the AF=20 full 3D SPARKLING trajectory on an Investigative 7T MR System~(MAGNETOM 7T, Siemens Healthcare, Erlangen, Germany) and measured the trajectory with the SKOPE dynamic field camera~\cite{DeZanche2008}. We used a 7T scanner for this study due to compatibility issues at 3T. However, we do not expect drastic changes in our results as the gradient system is the same for both scanners. We present the theoretical~(i.e. prescribed by the 3D SPARKLING algorithm) and measured trajectories for 3 random k-space shots in Fig.~\ref{fig:meas_traj}. Further, we quantitatively measured the error as to be $0.0016\pm0.0012$~(with the k-space normalized to $\Omega \in [-1, 1]^D$).}{\ref{r:gap}, \ref{r:simulations}, \ref{r:off_res}}

\section{MR image reconstruction}
\label{app:reconstruction}
The MR image reconstruction of 3D multi-channel data acquired from phased array receiver coils was carried out using a self-calibrating compressed sensing~(CS) reconstruction algorithm~\cite{ElGueddari_SAM18} in the synthesis formulation by solving for the wavelet coefficients $\mathbf{z}$ in~\eqref{synth_model}:
\begin{equation}
\label{synth_model}
\widehat{\mathbf{z}}=\underset{\mathbf{z} \in \mathbb{C}^{N_x \times N_y \times N_z}}{\operatorname{argmin}} \frac{1}{2} \sum_{\ell=1}^{L}\left\|F_{\Omega} \mathbf{S}_{\ell} \mathbf{\Psi}^* \mathbf{z} - \by_{\ell}\right\|_{2}^{2}+\lambda\|\mathbf{z}\|_{1}
\end{equation}
\noindent where the number of channels was $L=44$ and $N_x=N_y=384$ and $N_z=208$. Here the data fidelity is enforced with SENSE operators $(F_{\Omega} \mathbf{S}_{\ell})_{\ell=1}^L$, where $F_{\Omega}$ is the NUFFT operator and $\mathbf{S}_{\ell}$ is sensitivity map for $\ell^{th}$ channel estimated by density compensated adjoint of the 20 percent of acquired k-space center~(see details in~\cite{ElGueddari_SAM18}). $\lambda> 0 $ is the regularization parameter for promoting sparsity using $\ell_1$-norm regularization
in the wavelet domain $\mathbf{\Psi}$. For our reconstructions, we used Symlet 8 wavelet with 4 scales of decomposition for $\mathbf{\Psi}$. The regularization parameter $\lambda$ was grid searched between $(10^{-10}, 10^{0})$ while maximizing for the reconstruction quality using SSIM score in retrospective reconstruction. As the sampling operator was 3D non-Cartesian, the reconstruction problem was severely ill-posed with the forward operator $F_{\Omega}\mathbf{S_l}\mathbf{\Psi}^*$ having a large condition number, thereby impacting the convergence speed. In order to accelerate convergence, we preconditioned the k-space using density compensation. This translates to adding a preconditioner $D$ in the classical proximal gradient descent algorithm (here we used FISTA):

\begin{align*}
\mathbf{z}^{(k+1)} &= \text{soft}_{\lambda\tau} \left(\mathbf{z}^{(k)} \!-\! \tau  \sum_{\ell=1}^{L}  \mathbf{\Psi}\mathbf{S}_{\ell}^* F_\Omega^H D \left(  F_\Omega \mathbf{S}_{\ell} \mathbf{\Psi}^* \mathbf{z}^{(k)} \!-\! \by_{\ell}\right)\right)
\end{align*}

\noindent where $\text{soft}_{\lambda\tau}$ is the soft threshold operator and $\tau$ is the step size. The density compensators $D$ were obtained by 10 iterations of method described in \cite{pipe_dc}. The final MR image is given by $\mathbf{\widehat{x}}=  \mathbf{\Psi}^*\widehat{\mathbf{z}}\in \mathbb{C}^{N_x \times N_y \times N_z}$ as $\mathbf{\Psi}$ is a basis.

As the raw data was large (for AF=20, $p=8,388,608$ k-space points), we needed to utilize memory efficient methods to carry out the SENSE operation. For this, we implemented python wrappers for gpuNUFFT \cite{Knoll2014} which implements the NUFFT operator in CUDA and utilizes cuBLAS and cuFFT libraries to be efficient in speed and memory. The implementation of the reconstruction was completely done using \texttt{pysap-mri}\footnote{\url{https://github.com/CEA-COSMIC/pysap-mri}}~\cite{gueddari:hal-02399267}, the plugin of PySAP~\cite{farrens2019pysap} dedicated to MR image reconstruction. Despite being a 3D reconstruction problem, the computation time was just 15-30 minutes on a machine with the same hardware specifications as described earlier in Sec.~\ref{main-system_specs}.

\begin{table}[h]
\footnotesize
	\begin{center}
	\caption{SSIM scores for retrospective reconstructions on in vivo brain MR images as a function of varying density parameterizations~(cutoff $C$ and decay $D$ are varied across columns and rows, respectively). While (C=15, D=1) looks like an optimal setting, we used (C=25, D=2) as this configuration is more robust to small changes in density. Best scores appear in bold font.
	\label{fig:VD_Study_Retro}}
	\begin{tabular}{lrrrrrr}
			\toprule
			\diagbox{D}{C} &      10 &      15 &      20 &      25 &      30 &      35 \\
			\midrule
			1 &  0.976 &  \textbf{0.982} &  0.873 &  0.833 &  0.766 &  0.770 \\
			2 &  0.937 &  0.950 &  0.967 &  \textbf{0.972} &  \textbf{0.977} &  0.961 \\
			3 &  0.836 &  0.909 &  0.939 &  0.961 &  0.968 &  0.960 \\
			\bottomrule
	\end{tabular}
\end{center}
\end{table}

\newcommand{\retroPhantom}{images/results/retrospective/phantom}
\begin{figure*}[h]
	\centering
	\begin{mdframed}[innertopmargin=2pt, innerbottommargin=2pt, innerleftmargin=0pt, innerrightmargin=0pt, backgroundcolor=black, leftmargin=0cm,rightmargin=0cm,usetwoside=false]
		\resizebox{\linewidth}{!}{
			\begin{tabular}{c@{\hspace*{1mm}}c@{\hspace*{1mm}}c@{\hspace*{1mm}}c@{\hspace*{1mm}}c@{\hspace*{1mm}}c@{\hspace*{1mm}}c@{\hspace*{1mm}}c@{\hspace*{1mm}}c@{\hspace*{1mm}}c@{\hspace*{1mm}}c}
				\multicolumn{2}{c!{\color{white}\vrule width 2pt \hspace*{2mm}}}{\Bw{\Large (a) Cartesian Reference}}&&
				\multicolumn{2}{c}{\Bw{\Large (b) AF = 10}}&
				\multicolumn{2}{c}{\Bw{\Large (c) AF = 15}}&
				\multicolumn{2}{c}{\Bw{\Large (d) AF = 20}}
				\\
				\multicolumn{2}{c!{\color{white}\vrule width 2pt \hspace*{2mm}}}{}&&
				\multicolumn{2}{c}{\Bw{\small	SSIM = 0.964}}&
				\multicolumn{2}{c}{\Bw{\small	SSIM = 0.935}}&
				\multicolumn{2}{c}{\Bw{\small   SSIM = 0.923}}
				\\
				\multirow{2}{*}[0.4in]{\parbox[m]{.23\linewidth}{\includegraphics[trim={5cm 4.5cm 5cm 3.5cm},clip,width=\linewidth]{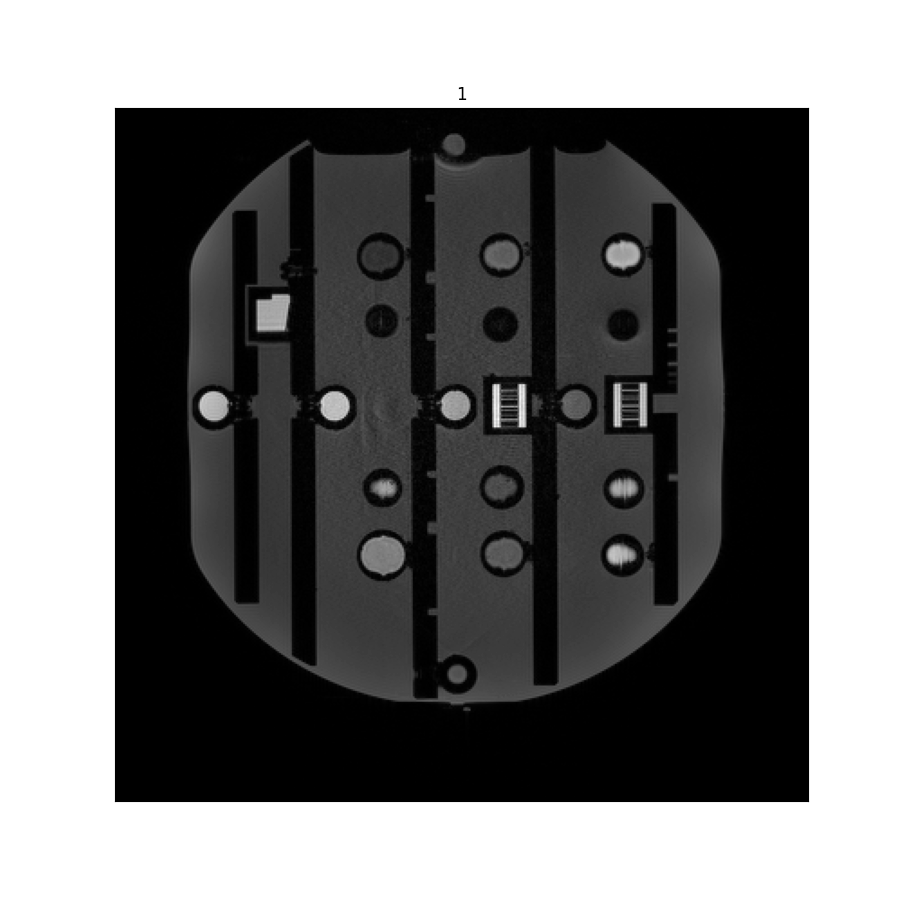}}}&
				\multicolumn{1}{c!{\color{white}\vrule width 2pt \hspace*{2mm}}}{\parbox[m]{.2\linewidth}{\includegraphics[trim={3cm 6.5cm 4.2cm 7cm},clip,width=\linewidth]{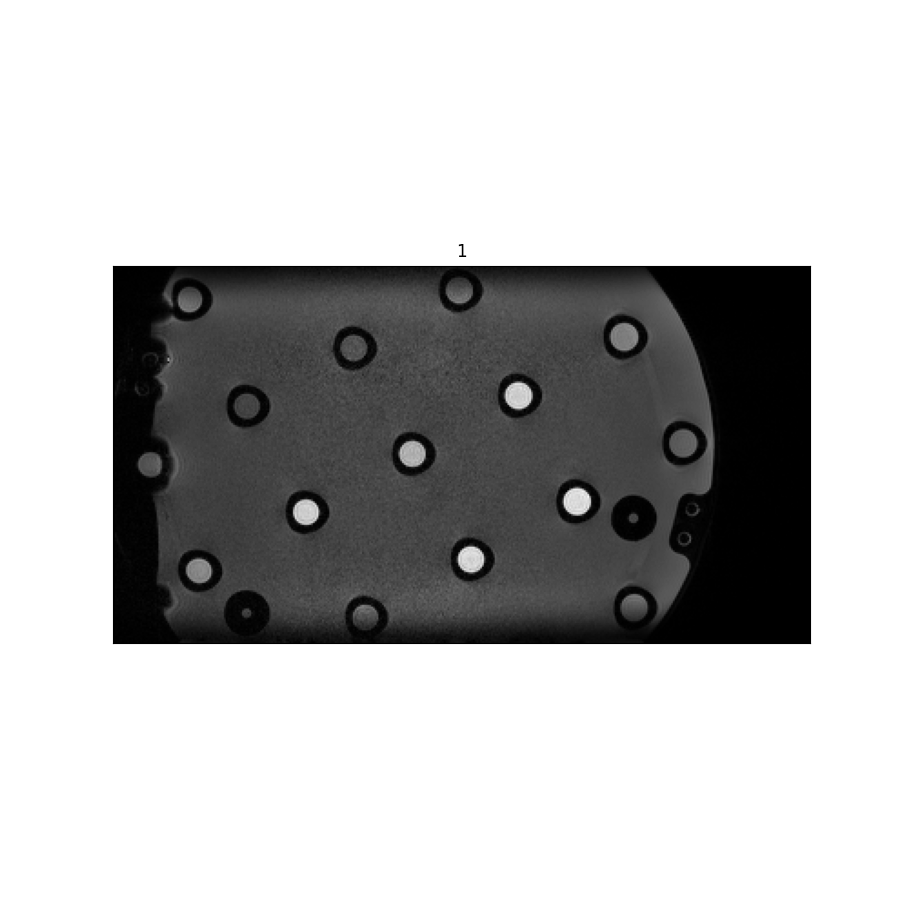}}}&
				
				\multirow{2}{*}[0.55in]{\rotatebox[origin=c]{90}{\Bw{\Large (i) Full 3D SPARKLING}}}&
				\multirow{2}{*}[0.4in]{\parbox[m]{.23\linewidth}{
						\includegraphics[trim={5cm 4.5cm 5cm 3.5cm},clip,width=\linewidth]
						{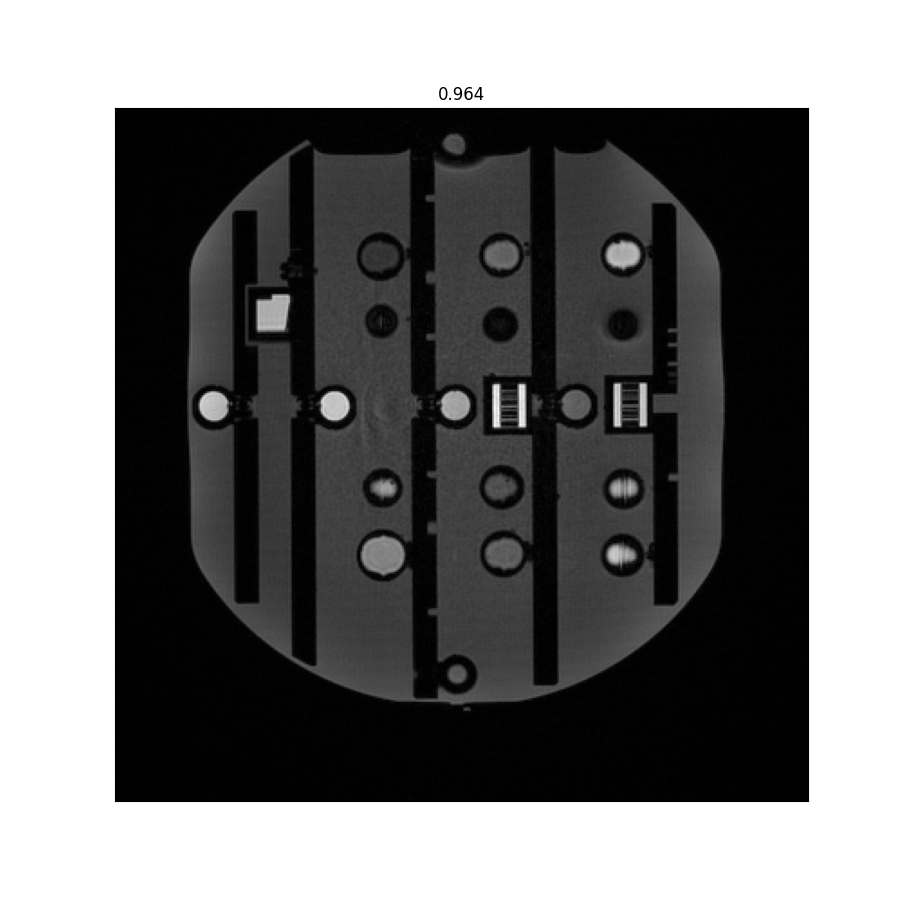}}}&
				\parbox[m]{.2\linewidth}{\includegraphics[trim={3cm 6.5cm 4.2cm 7cm},clip,width=\linewidth]{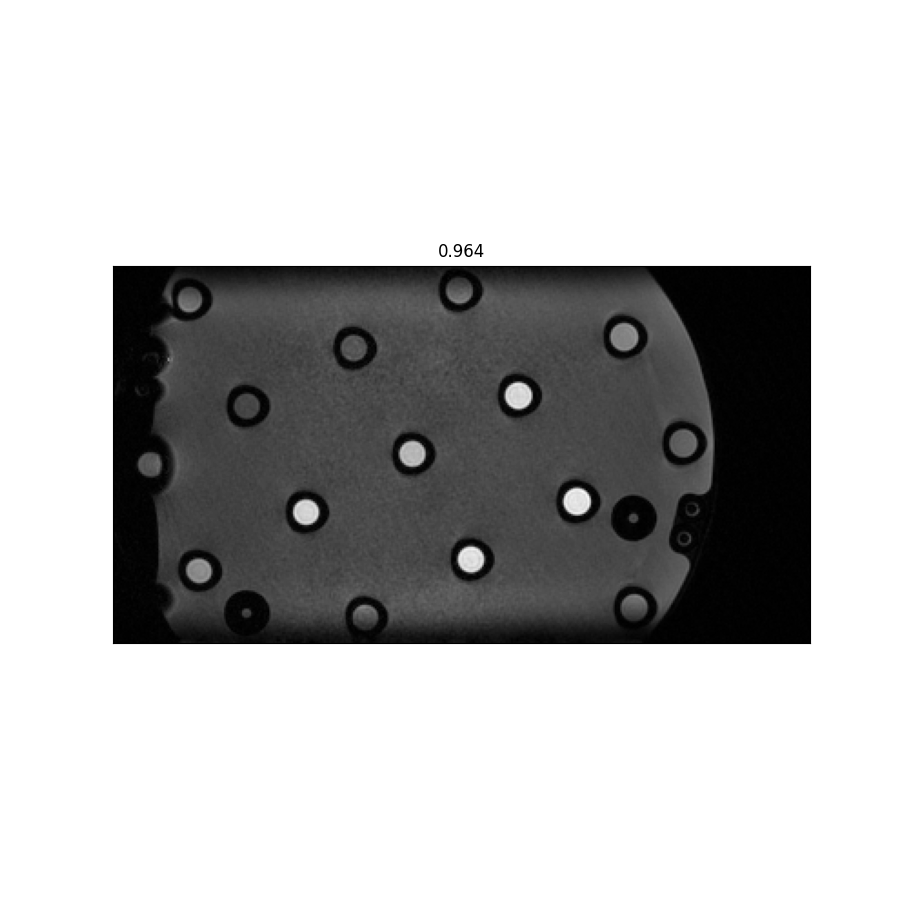}}&
				\multirow{2}{*}[0.4in]{\parbox[m]{.23\linewidth}{\includegraphics[trim={5cm 4.5cm 5cm 3.5cm},clip,width=\linewidth]
						{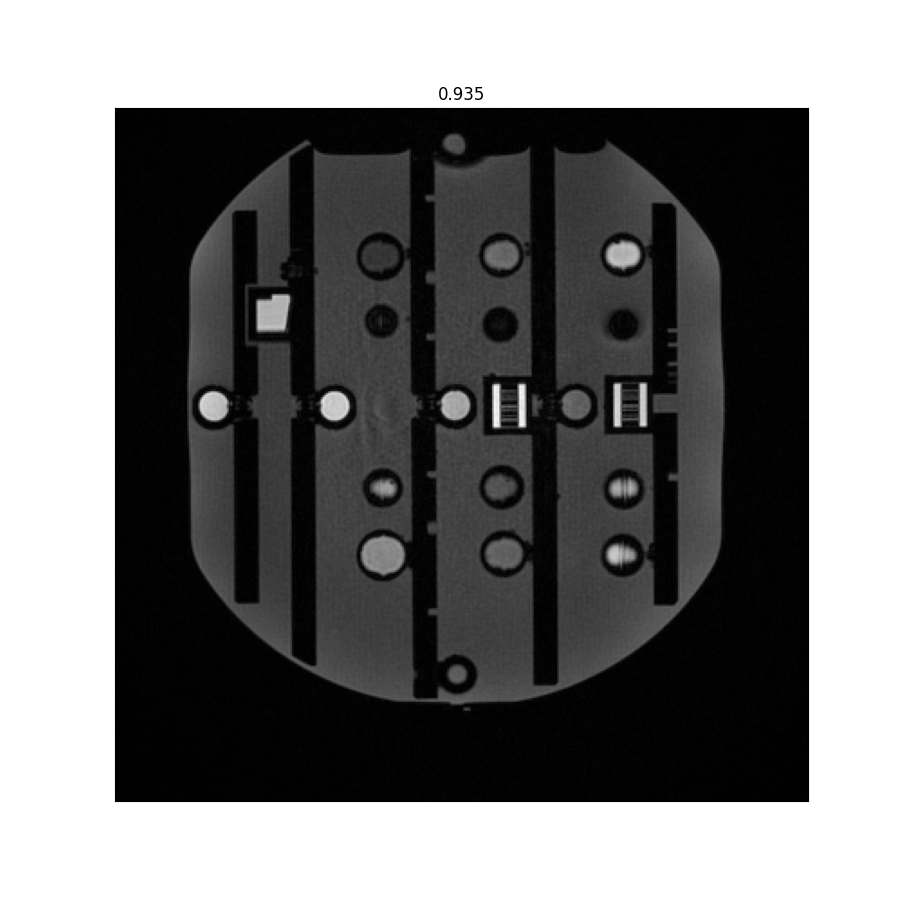}}}&
				\parbox[m]{.2\linewidth}{\includegraphics[trim={3cm 6.5cm 4.2cm 7cm},clip,width=\linewidth]	{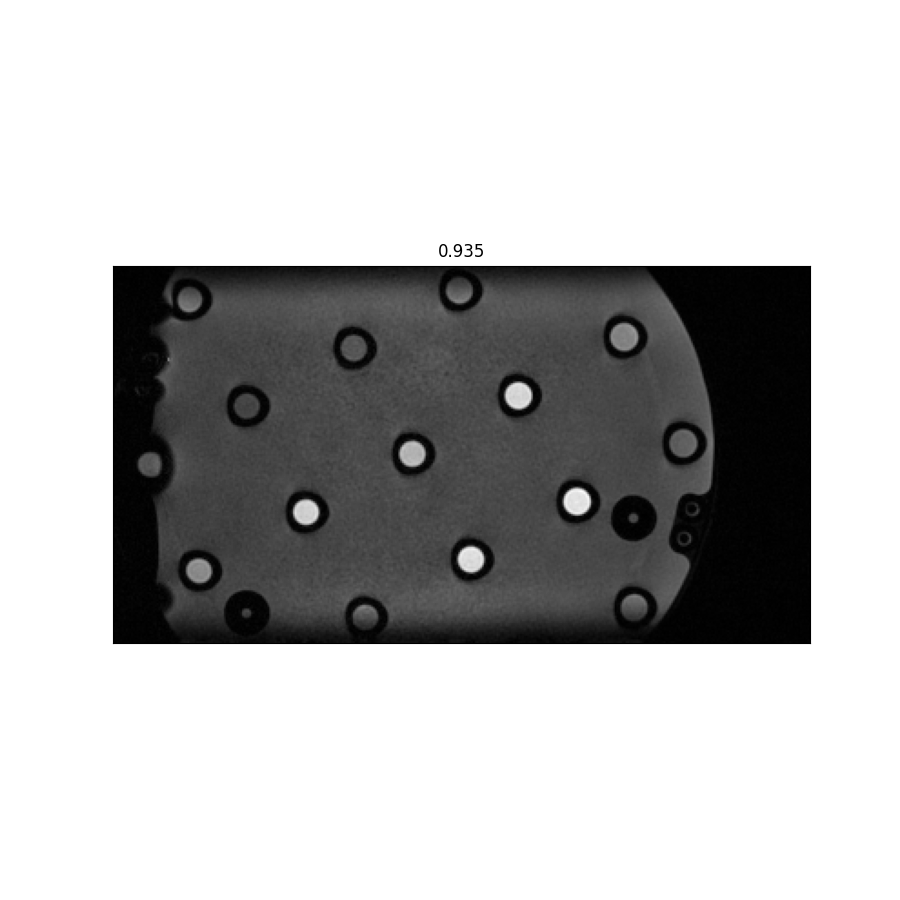}}&
				\multirow{2}{*}[0.4in]{\parbox[m]{.23\linewidth}{
						\includegraphics[trim={5cm 4.5cm 5cm 3.5cm},clip,width=\linewidth]
						{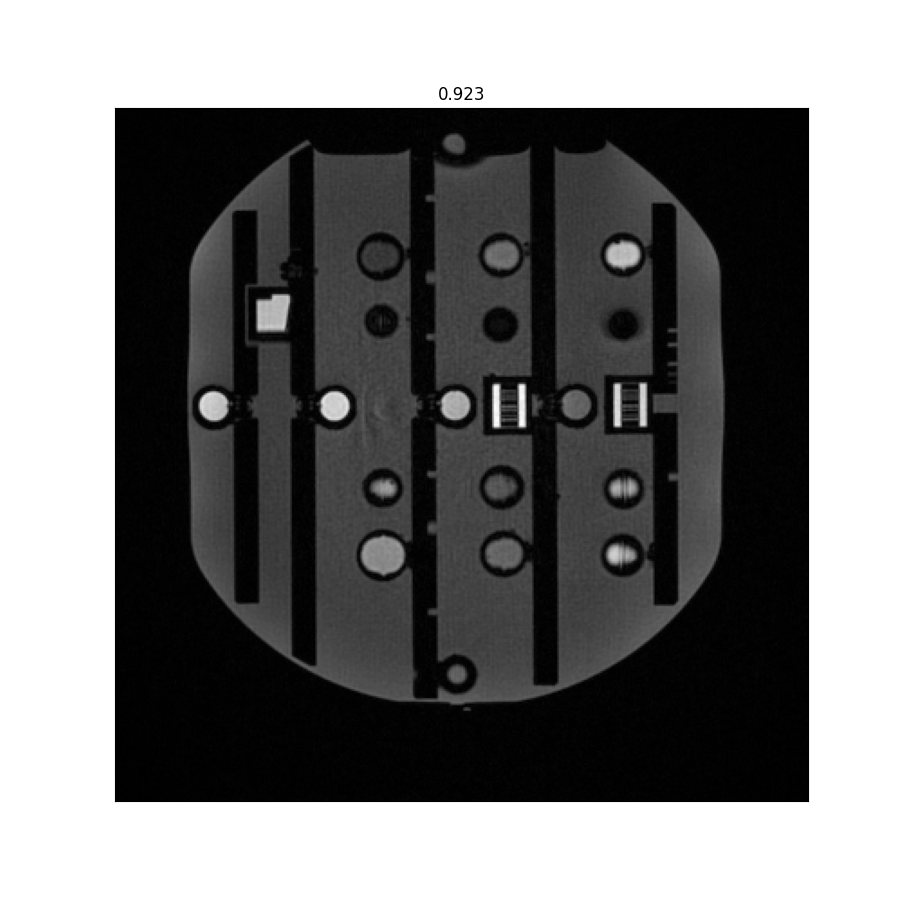}}}&
				\parbox[m]{.2\linewidth}{\includegraphics[trim={3cm 6.5cm 4.2cm 7cm},clip,width=\linewidth]{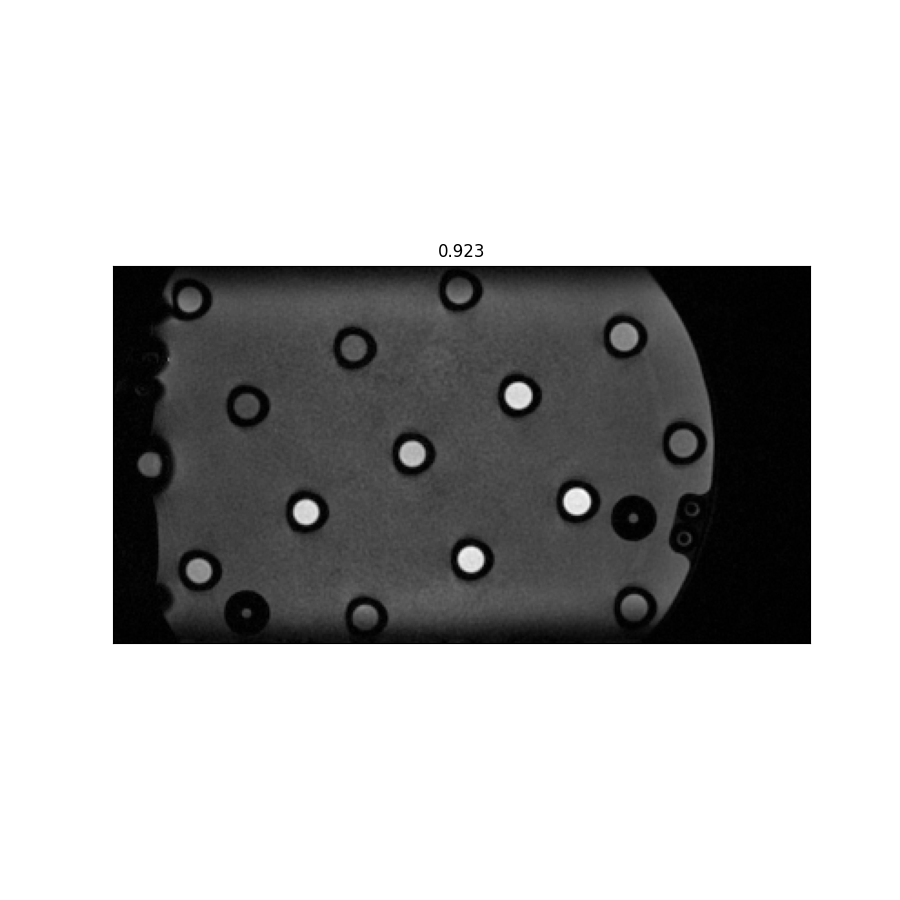}}&
				\\
				&\multicolumn{1}{c!{\color{white}\vrule width 2pt \hspace*{2mm}}}{\parbox[m]{.2\linewidth}{\includegraphics[trim={4cm 7cm 4cm 7cm},clip,width=\linewidth]{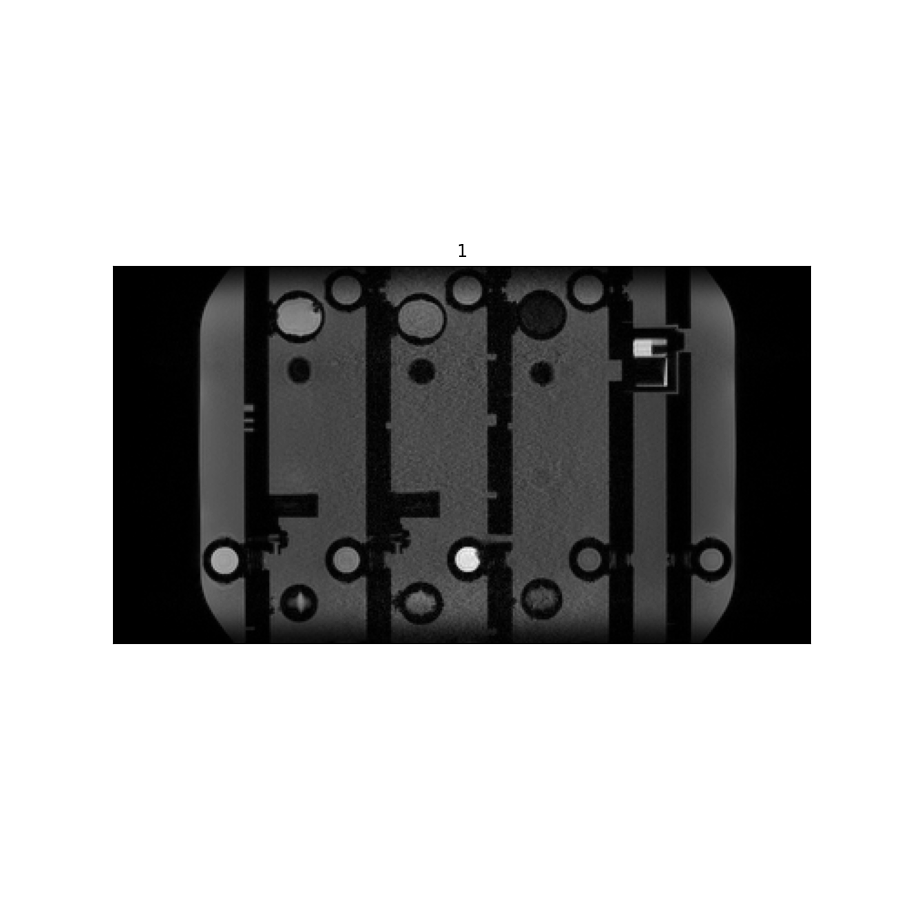}}}&
				&&\parbox[m]{.2\linewidth}{\includegraphics[trim={4cm 7cm 4cm 7cm},clip,width=\linewidth]{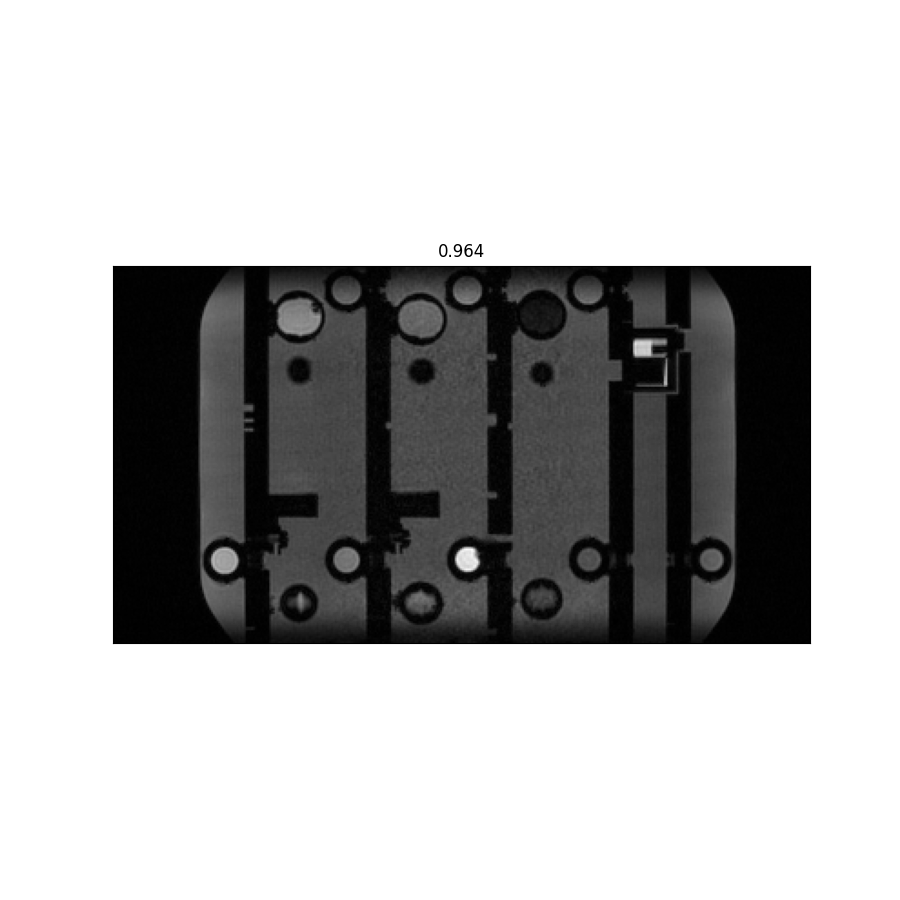}}&
				&\parbox[m]{.2\linewidth}{\includegraphics[trim={4cm 7cm 4cm 7cm},clip,width=\linewidth]{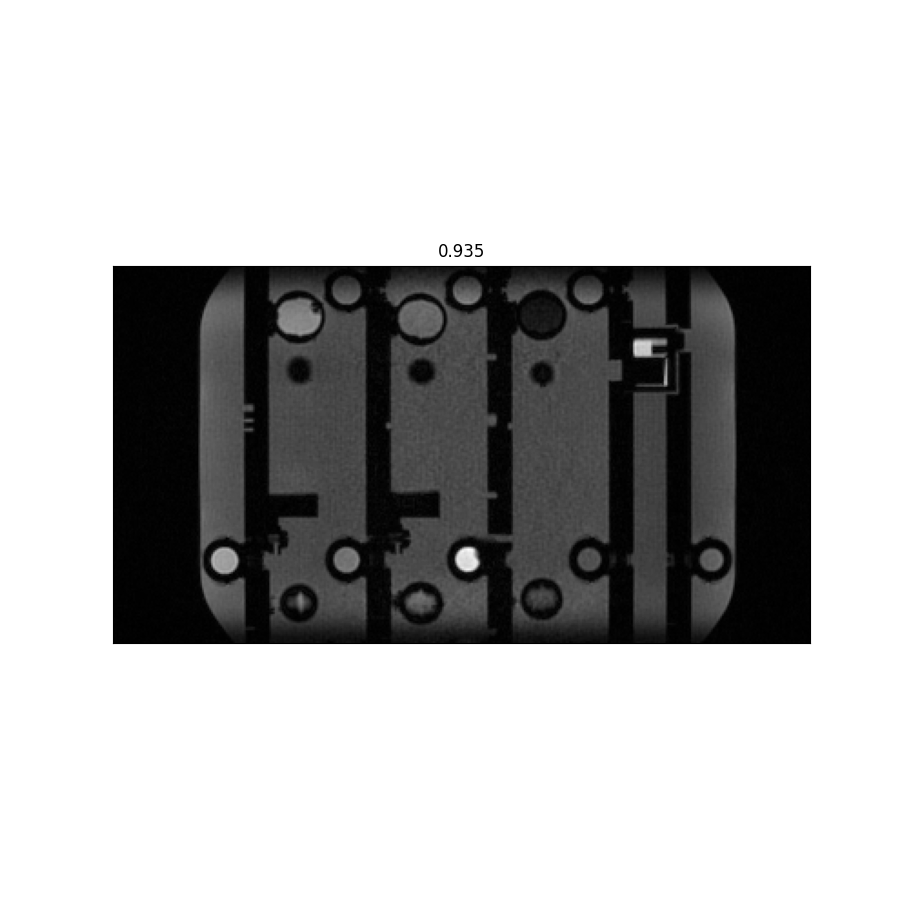}}&
				&\parbox[m]{.2\linewidth}{\includegraphics[trim={4cm 7cm 4cm 7cm},clip,width=\linewidth]{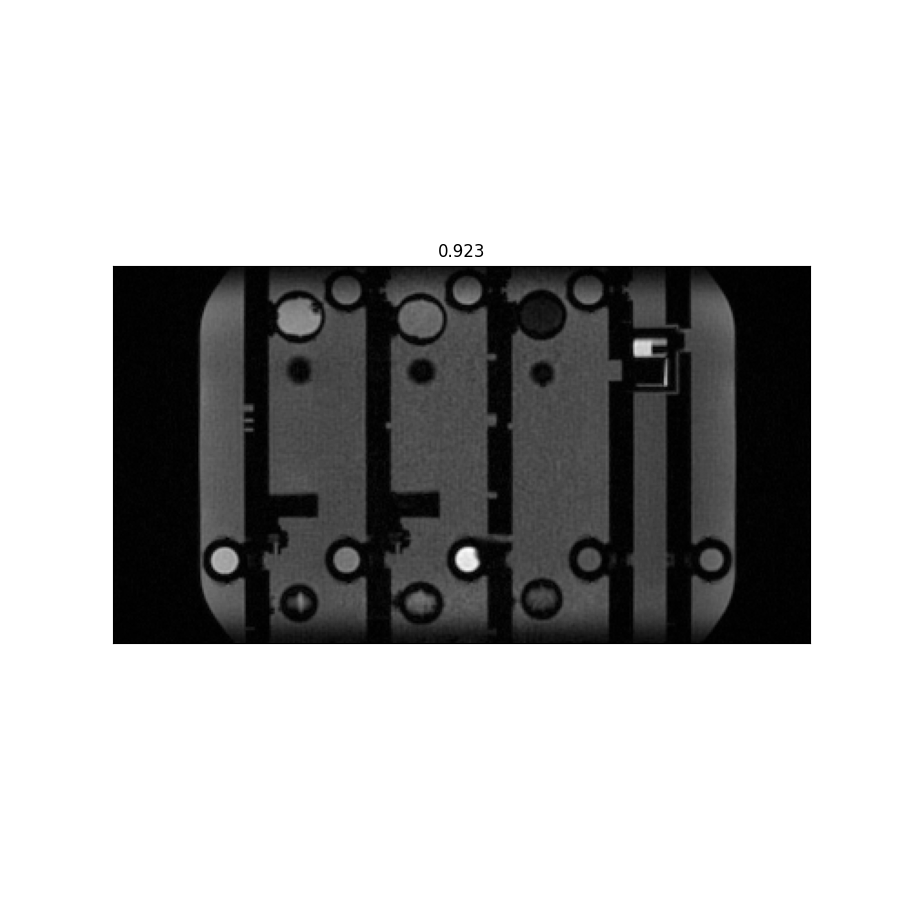}}&
				\\[1.5cm]
				\arrayrulecolor{white}\cmidrule[2pt]{1-2}
				\multicolumn{2}{c!{\color{white}\vrule width 2pt \hspace*{2mm}}}{\Bw{\thead{\Large (iii) TPI, AF = 10 \\ \small SSIM = 0.63}}}&&
				\multicolumn{2}{c}{\Bw{\small   SSIM = 0.927}}&
				\multicolumn{2}{c}{\Bw{\small	SSIM = 0.864}}&
				\multicolumn{2}{c}{\Bw{\small	SSIM = 0.737}}&
				\\
				\multirow{2}{*}[0.4in]{\parbox[m]{.23\linewidth}{\includegraphics[trim={5cm 4.5cm 5cm 3.5cm},clip,width=\linewidth]{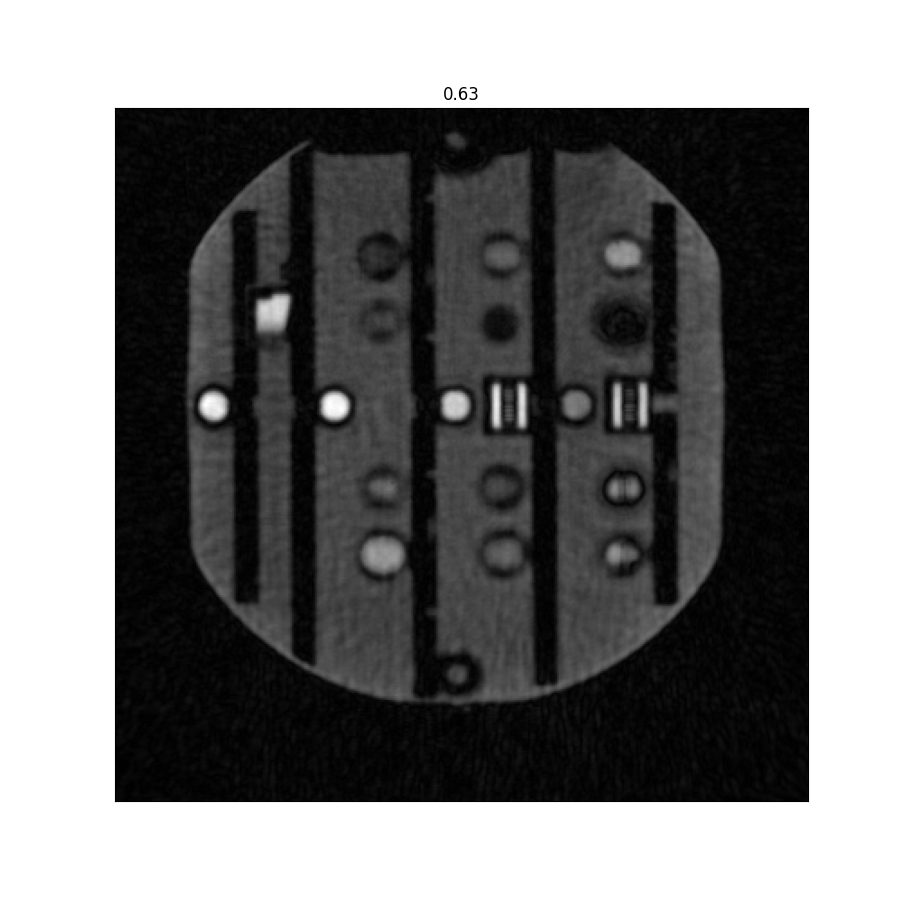}}}&
				\multicolumn{1}{c!{\color{white}\vrule width 2pt \hspace*{2mm}}}{\parbox[m]{.2\linewidth}{\includegraphics[trim={4.5cm 6.5cm 4cm 7cm},clip,width=\linewidth]{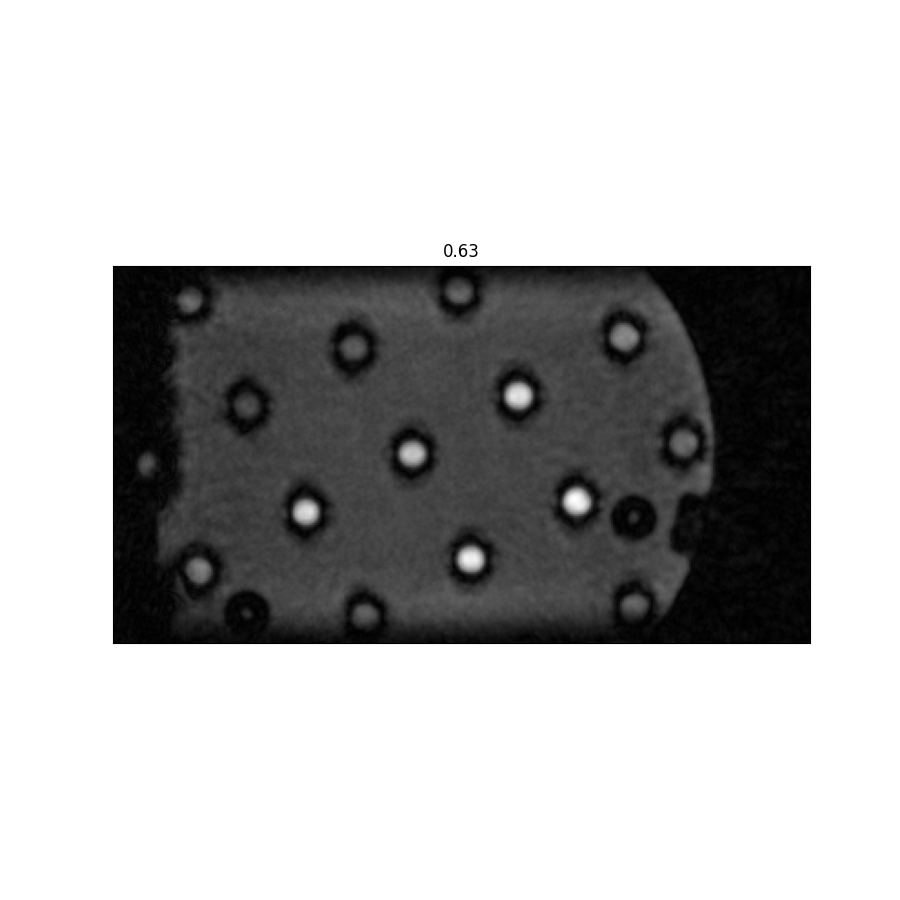}}}&
				\multirow{2}{*}[0.6in]{\rotatebox[origin=c]{90}{\Bw{\Large (ii) \texttt{SpSOS} SPARKLING}}}&
				\multirow{2}{*}[0.4in]{\parbox[m]{.23\linewidth}{\includegraphics
						[trim={5cm 4.5cm 5cm 3.5cm},clip,width=\linewidth]
						{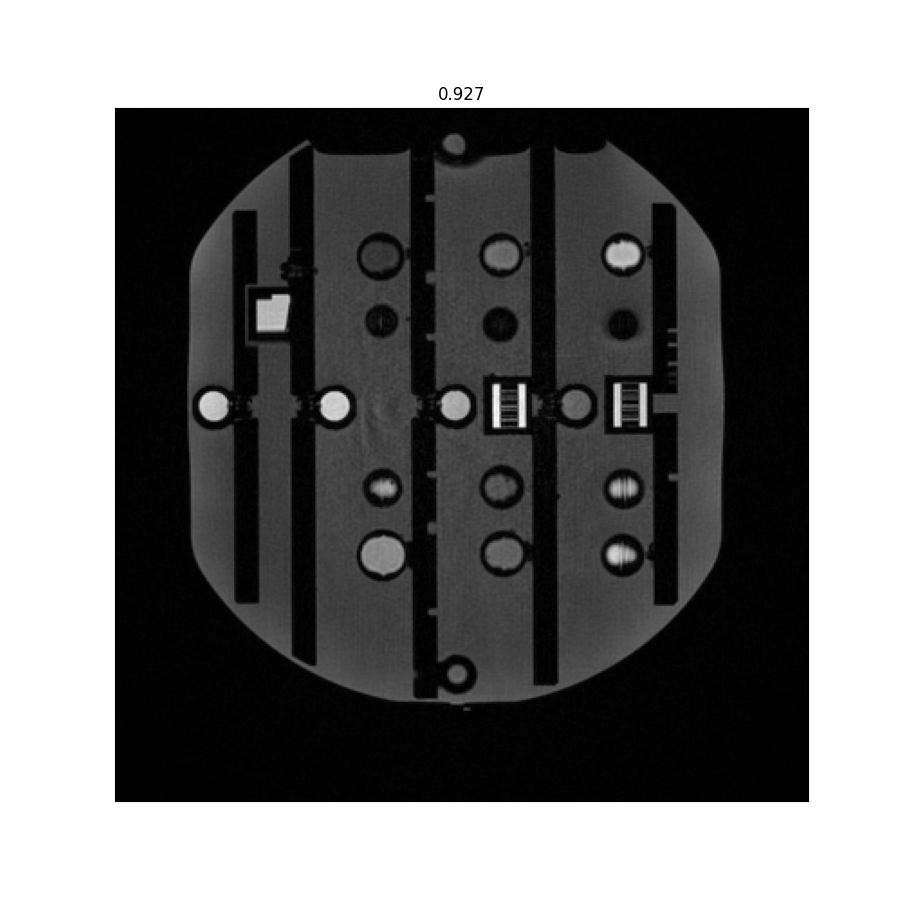}}}&
				\parbox[m]{.2\linewidth}{\includegraphics[trim={3cm 6.5cm 4.2cm 7cm},clip,width=\linewidth]{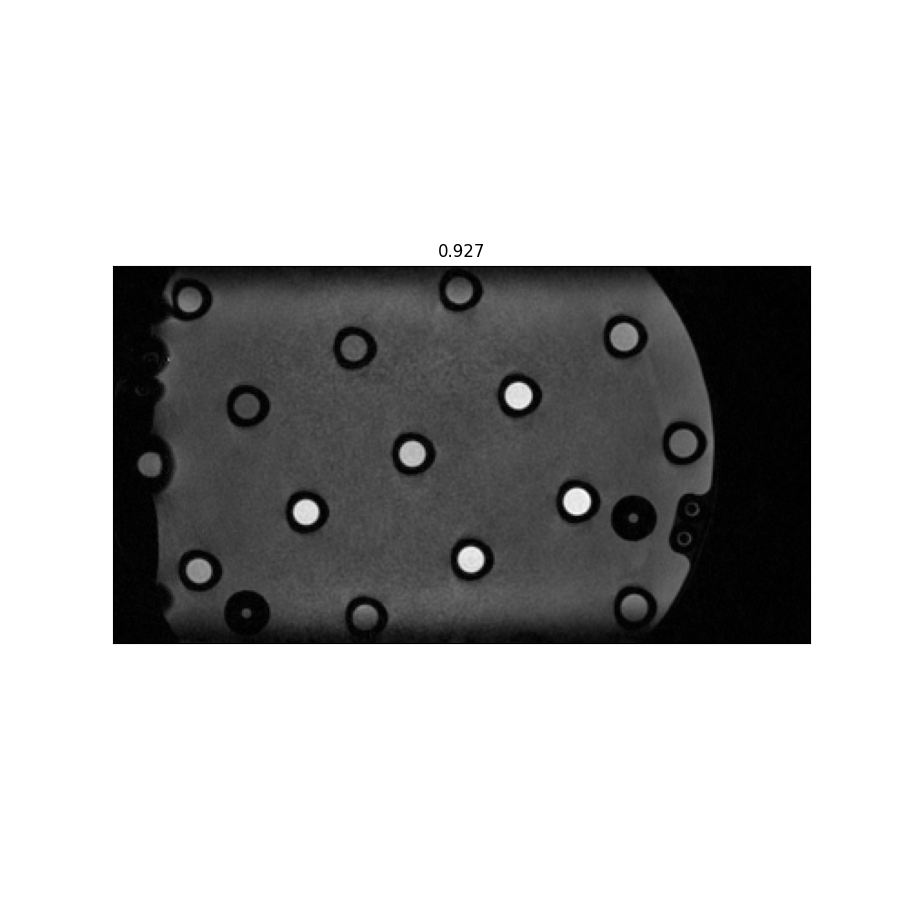}}&
				\multirow{2}{*}[0.4in]{\parbox[m]{.23\linewidth}{\includegraphics[trim={5cm 4.5cm 5cm 3.5cm},clip,width=\linewidth]
						{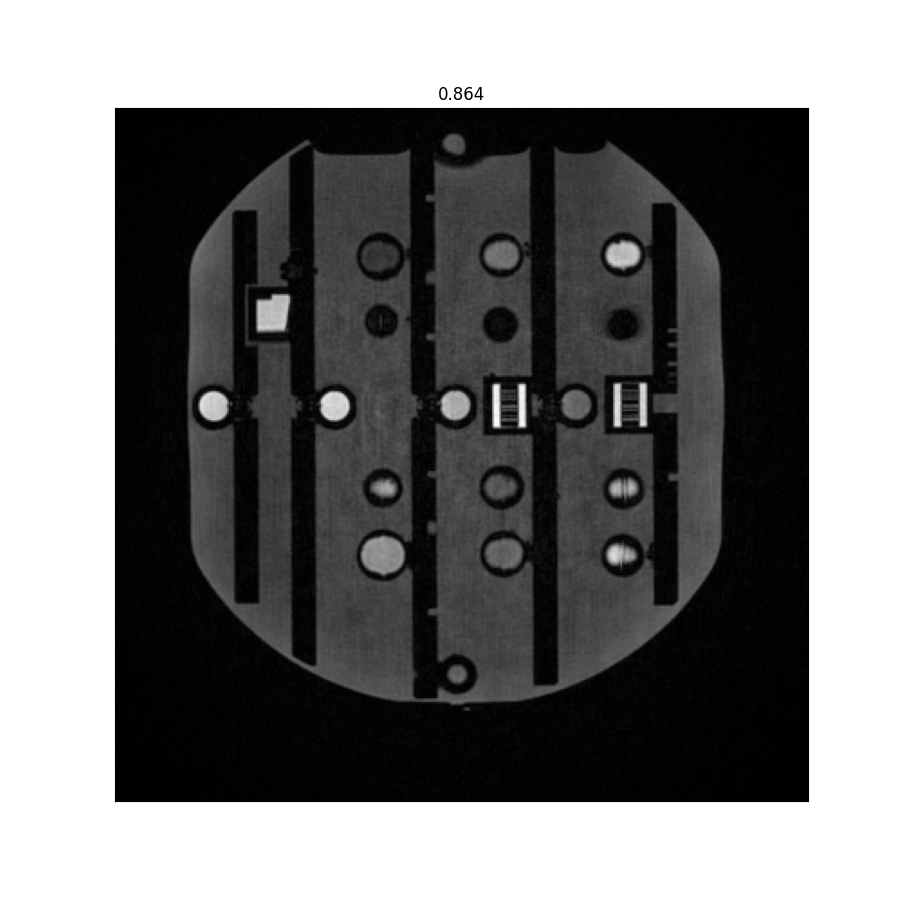}}}&
				\parbox[m]{.2\linewidth}{\includegraphics[trim={3cm 6.5cm 4.2cm 7cm},clip,width=\linewidth]{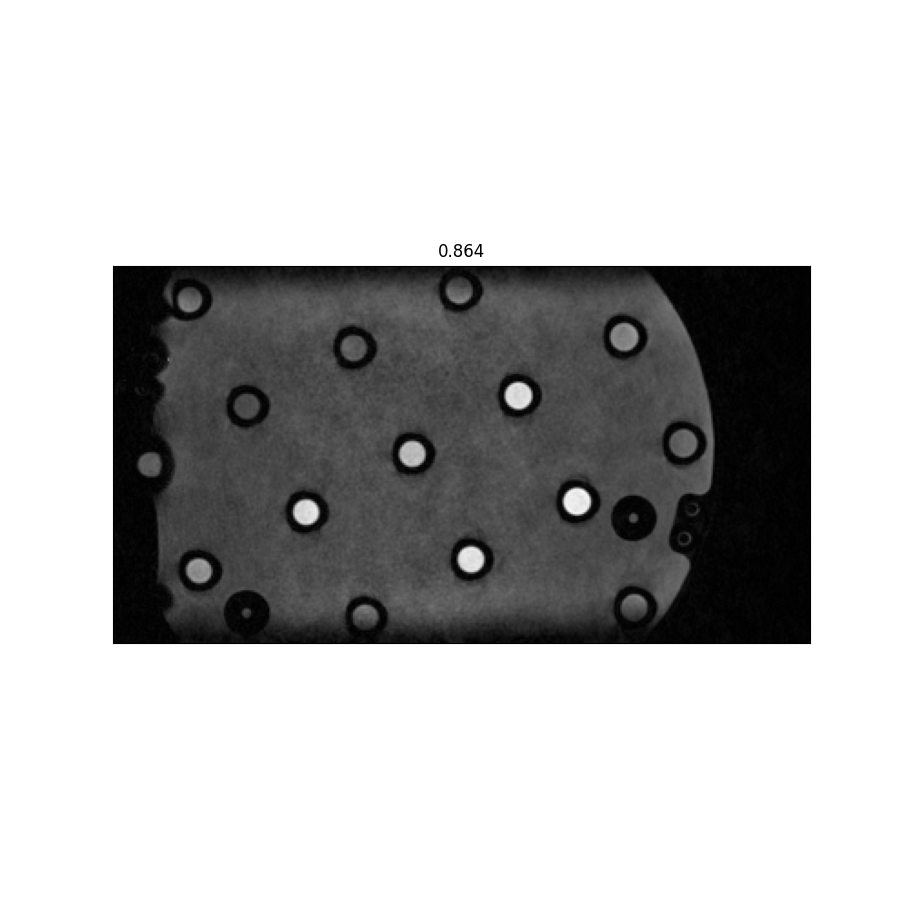}}&				\multirow{2}{*}[0.4in]{\parbox[m]{.23\linewidth}{\includegraphics[trim={5cm 4.5cm 5cm 3.5cm},clip,width=\linewidth]
						{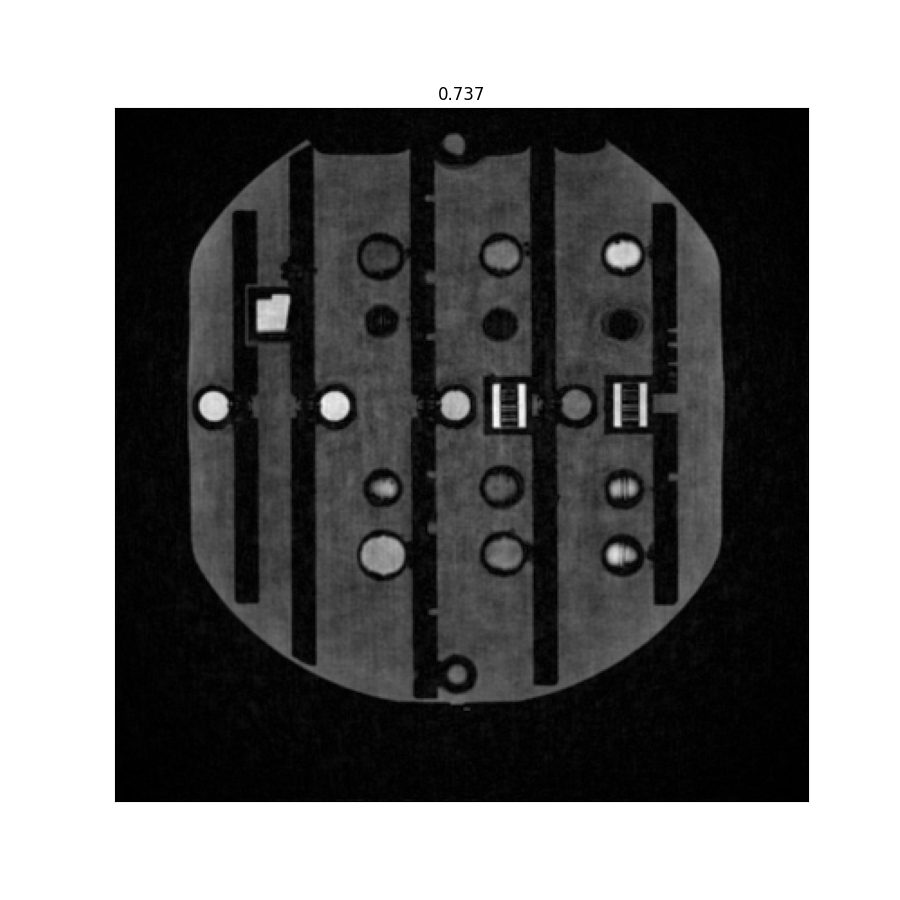}}}&
				\parbox[m]{.2\linewidth}{\includegraphics[trim={3cm 6.5cm 4.2cm 7cm},clip,width=\linewidth]{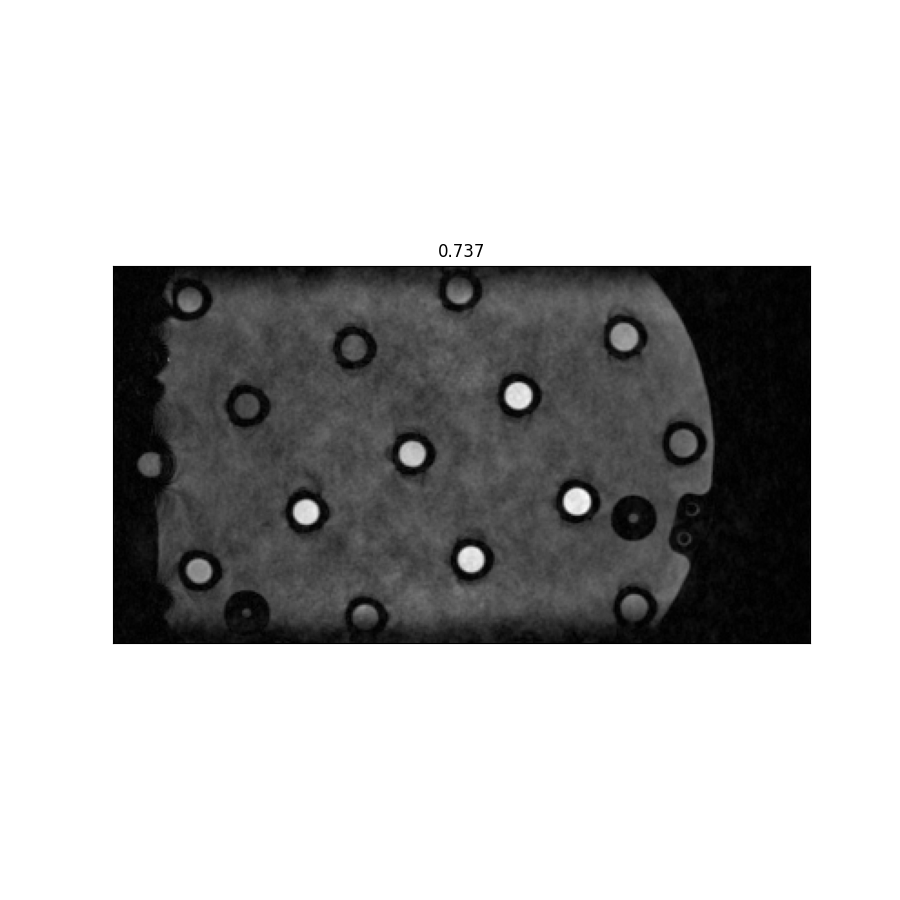}}
				
				\\
				&\multicolumn{1}{c!{\color{white}\vrule width 2pt \hspace*{2mm}}}{\parbox[m]{.2\linewidth}{\includegraphics[trim={4cm 7cm 4cm 7cm},clip,width=\linewidth]{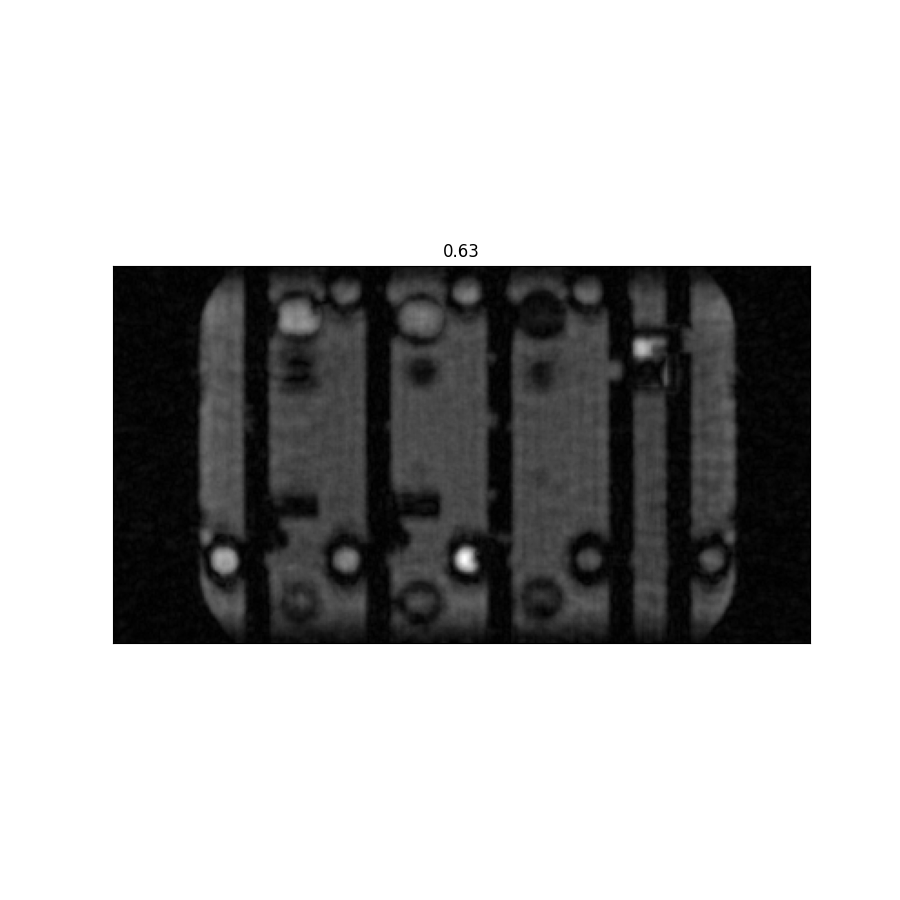}}}&
				&&\parbox[m]{.2\linewidth}{\includegraphics[trim={4cm 7cm 5.5cm 7cm},clip,width=\linewidth]{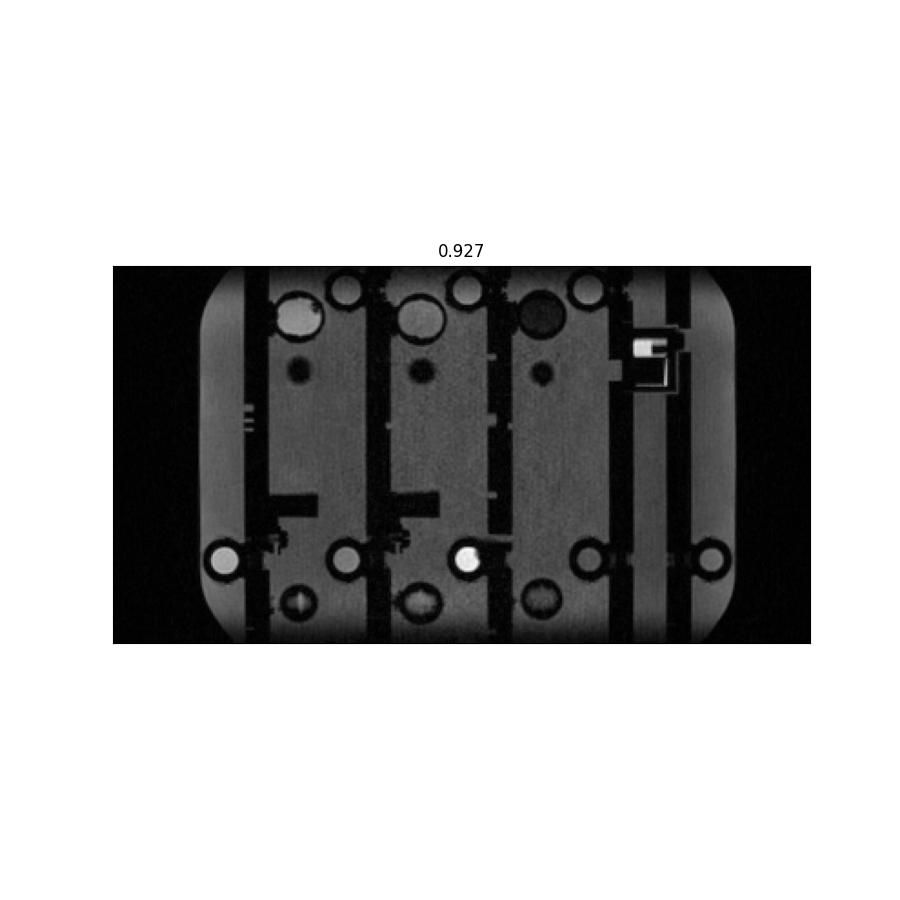}}&
				&\parbox[m]{.2\linewidth}{\includegraphics[trim={4cm 7cm 4cm 7cm},clip,width=\linewidth]{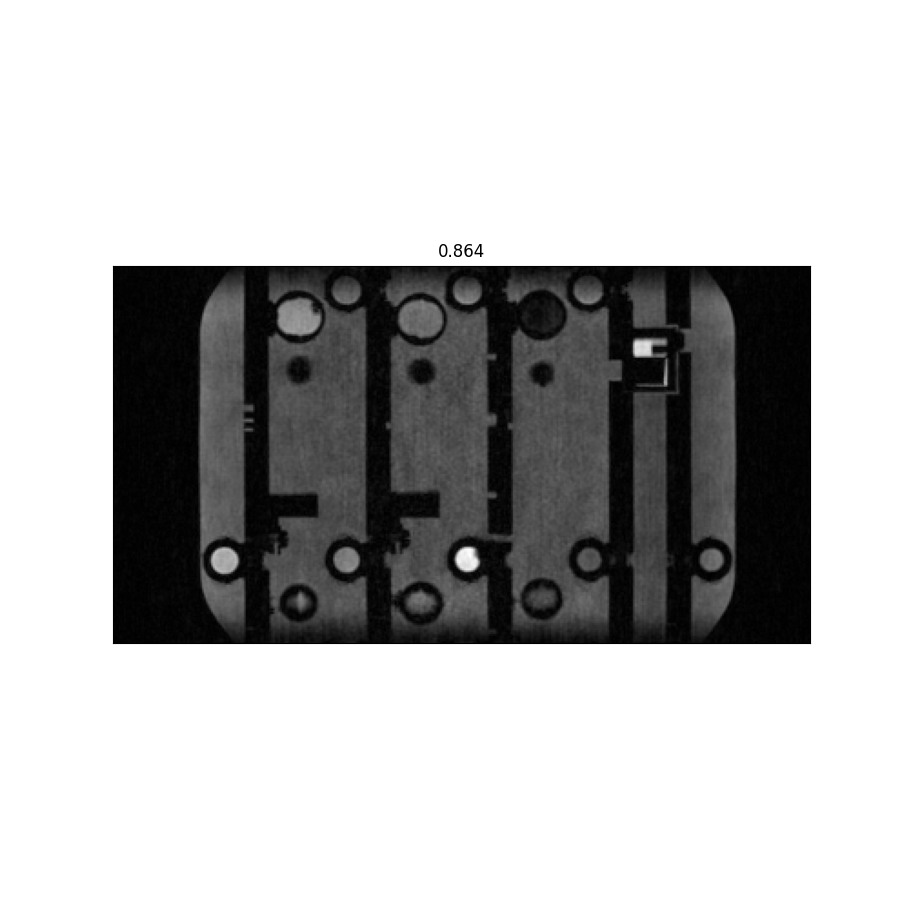}}&
				&\parbox[m]{.2\linewidth}{\includegraphics[trim={4cm 7cm 4cm 7cm},clip,width=\linewidth]{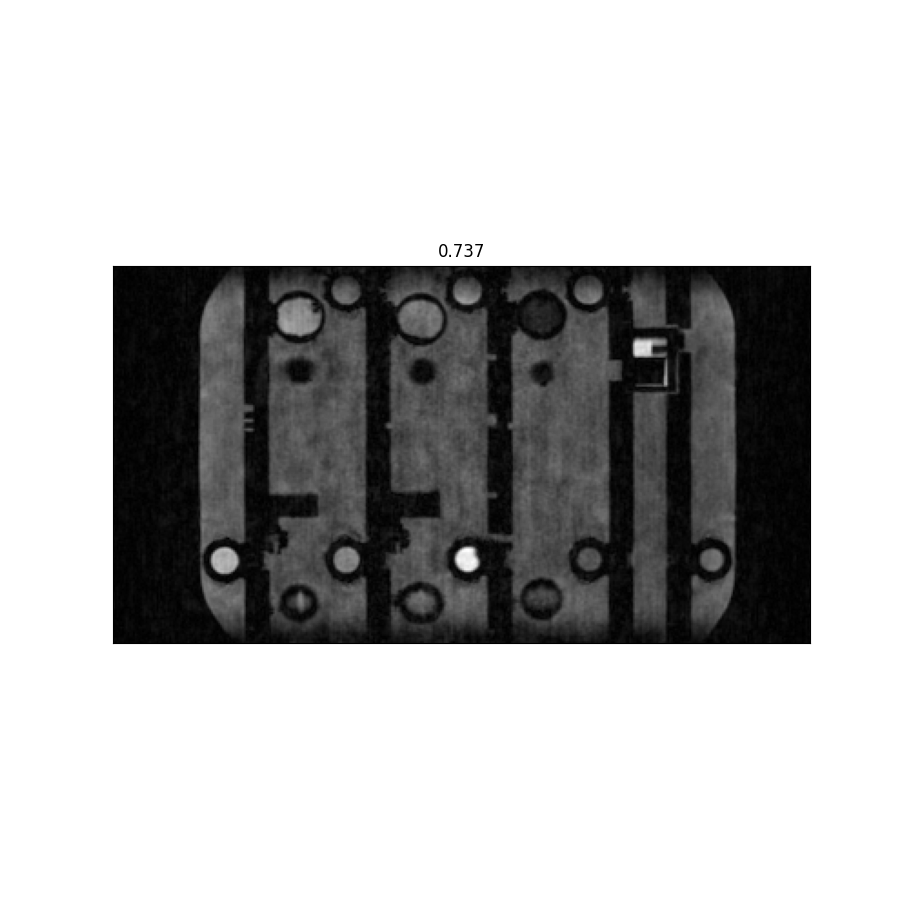}}
				
				\\
		\end{tabular}}
	\end{mdframed}
	\caption{Comparison of retrospective results for {\bf (i)} fully optimized 3D SPARKLING~(right, top row) and {\bf (ii)} SpSOS~(right, bottom row) for varying acceleration factors~(from left to right, AF=10~(b), 15~(c) and 20~(d)) on the NIST phantom. Additionally, we present the results of a retrospective study based on {\bf (iii)} TPI at AF=10 for comparison purposes with the state of the art~(left, bottom row). The Cartesian reference image~(AF=4, GRAPPA reconstructed) is shown in~(a, left top row)~\cite{Griswold02}. SSIM scores are reported for each setup. Global 3D SPARKLING gives improved results compared to the SpSOS approach which starts to get worse at AF=15 with some blurring and at AF=20 the image gets noisier. On the other hand, TPI images are extremely blurry even at AF=10.}
	\label{fig:retrospective_phantom}
	
\end{figure*}

\newcommand{\retroInvivo}{images/results/retrospective/invivo}
\begin{figure}[h]
	\begin{mdframed}[innertopmargin=2pt, innerbottommargin=2pt, innerleftmargin=0pt, innerrightmargin=0pt, backgroundcolor=black, leftmargin=0cm,rightmargin=0cm,usetwoside=false]
		\resizebox{\linewidth}{!}{
			\begin{tabular}{c@{\hspace*{1mm}}c@{\hspace*{1mm}}c@{\hspace*{1mm}}c}
				\multicolumn{2}{c}{\Bw{\footnotesize\bf (a) Cartesian Reference}} &
				\multicolumn{2}{c}{\Bw{\thead{\footnotesize\bf (b) Full 3D SPARKLING \\ \tiny SSIM = 0.964}}}
				\\
				\multirow{2}{*}[0.1in]{\parbox[m]{.23\linewidth}{
						\includegraphics[trim={6cm 4cm 5.5cm 5cm},clip,width=\linewidth]
						{\retroInvivo/cartesian_z.png}}}&
				\parbox[m]{.2\linewidth}{\includegraphics[trim={5.2cm 6.5cm 4cm 7cm},clip,width=\linewidth]{\retroInvivo/cartesian_y.png}}&
				
				\multirow{2}{*}[0.1in]{\parbox[m]{.23\linewidth}{
						\includegraphics[trim={6cm 4cm 5.5cm 5cm},clip,width=\linewidth]
						{\retroInvivo/dim3_Nc89_z.png}}}&
				\parbox[m]{.2\linewidth}{\includegraphics[trim={5.2cm 6.5cm 4cm 7cm},clip,width=\linewidth]{\retroInvivo/dim3_Nc89_y.png}}
				\\
				&\parbox[m]{.2\linewidth}{\includegraphics[trim={6.2cm 7cm 5.5cm 6.8cm},clip,width=\linewidth]{\retroInvivo/cartesian_x.png}}&
				&\parbox[m]{.2\linewidth}{\includegraphics[trim={6.2cm 7cm 5.5cm 6.8cm},clip,width=\linewidth]{\retroInvivo/dim3_Nc89_x.png}}
				\\
				\multicolumn{2}{c}{\Bw{\thead{\footnotesize\bf (c) \texttt{SpSOS} SPARKLING \\ \tiny SSIM = 0.93}}} & 
				\multicolumn{2}{c}{\Bw{\thead{\footnotesize\bf (d) TPI \\ \tiny SSIM = 0.493}}} 
				\\
				\multirow{2}{*}[0.1in]{\parbox[m]{.23\linewidth}{
						\includegraphics[trim={6cm 4cm 5.5cm 5cm},clip,width=\linewidth]
						{\retroInvivo/dim2.5_Nc38_z.png}}}&
				\parbox[m]{.2\linewidth}{\includegraphics[trim={5.2cm 6.5cm 4cm 7cm},clip,width=\linewidth]{\retroInvivo/dim2.5_Nc38_y.png}}&
				
				\multirow{2}{*}[0.1in]{\parbox[m]{.23\linewidth}{
						\includegraphics[trim={6cm 4cm 5.5cm 5cm},clip,width=\linewidth]
						{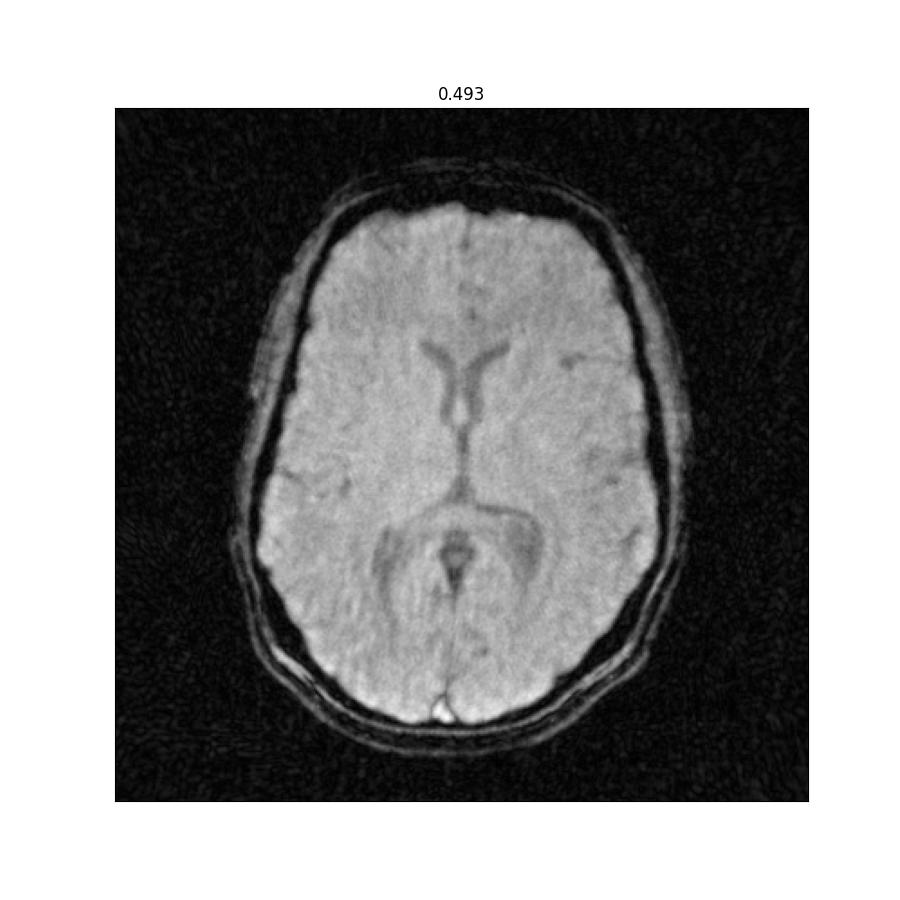}}}&
				\parbox[m]{.2\linewidth}{\includegraphics[trim={5.2cm 6.5cm 4cm 7cm},clip,width=\linewidth]{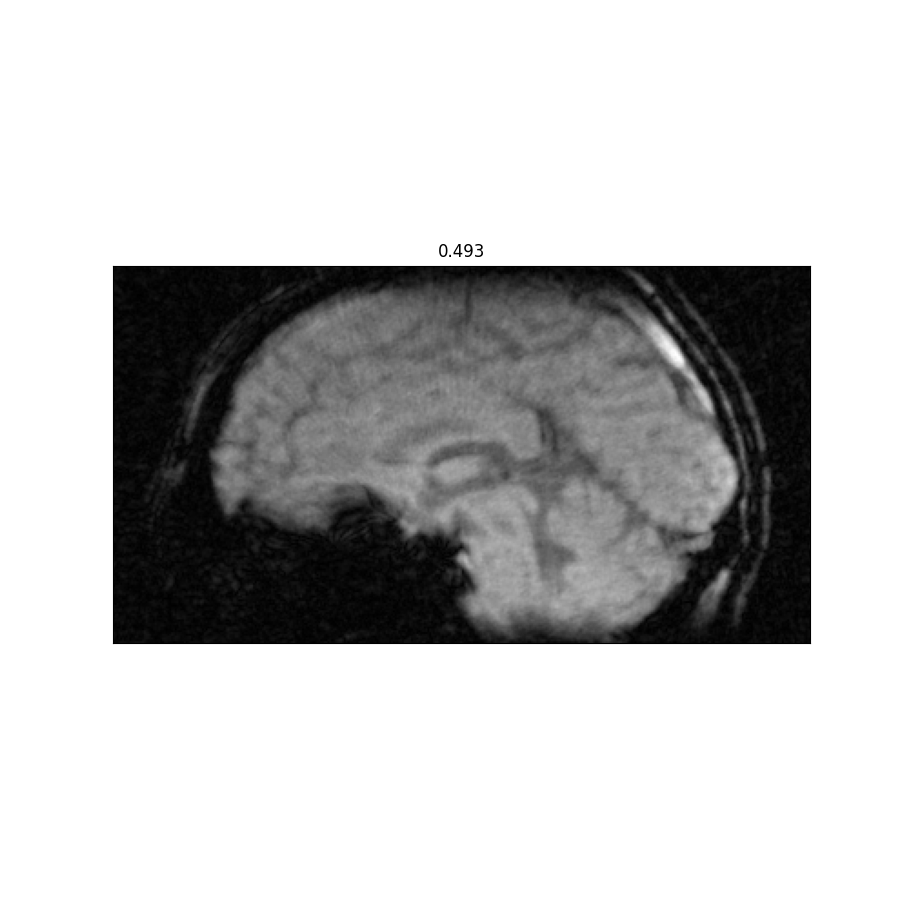}}
				\\
				&\parbox[m]{.2\linewidth}{\includegraphics[trim={6.2cm 7cm 5.5cm 6.8cm},clip,width=\linewidth]{\retroInvivo/dim2.5_Nc38_x.png}}&
				&\parbox[m]{.2\linewidth}{\includegraphics[trim={6.2cm 7cm 5.5cm 6.8cm},clip,width=\linewidth]{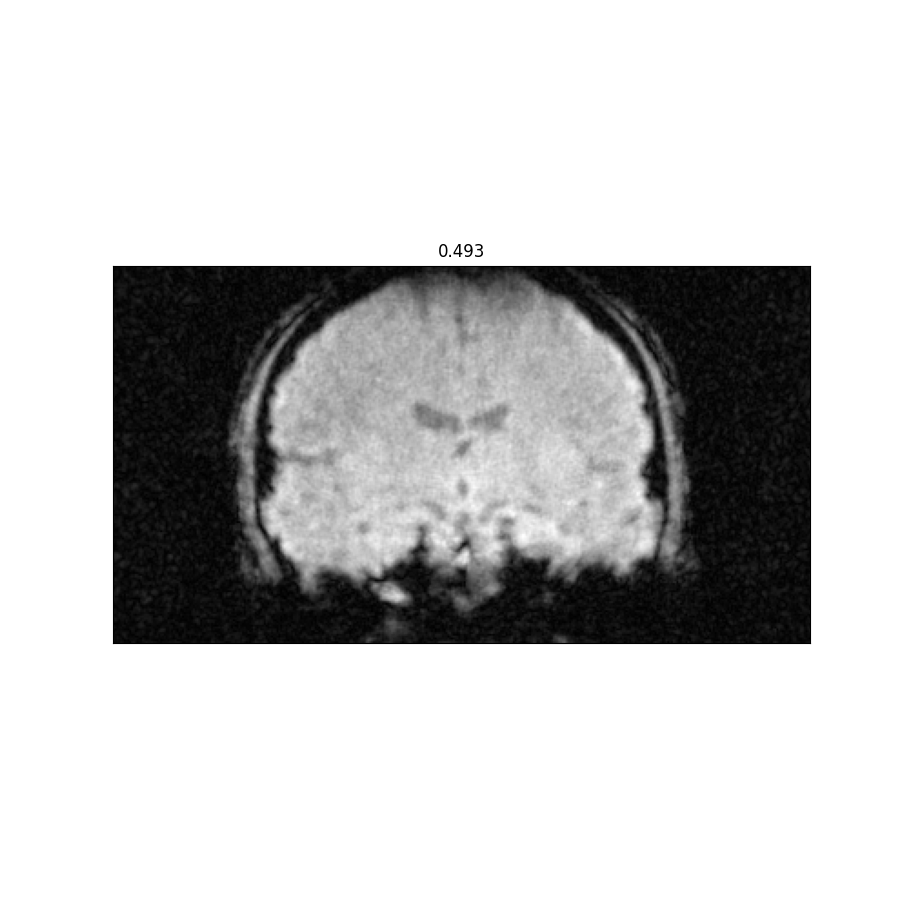}}
		\end{tabular}}
		\end{mdframed}
		\caption{\label{fig:compare_tpi_invivo} Comparing the performance of full 3D SPARKLING~(b) and SpSOS SPARKLING~(c) with twisted projection imaging (TPI)~(d) using a retrospective study at AF=10 from the Cartesian GRAPPA-4 refe\-rence~(a).}
\end{figure}

\begin{figure}[h]
	\centering
	\includegraphics[width=0.85\linewidth]{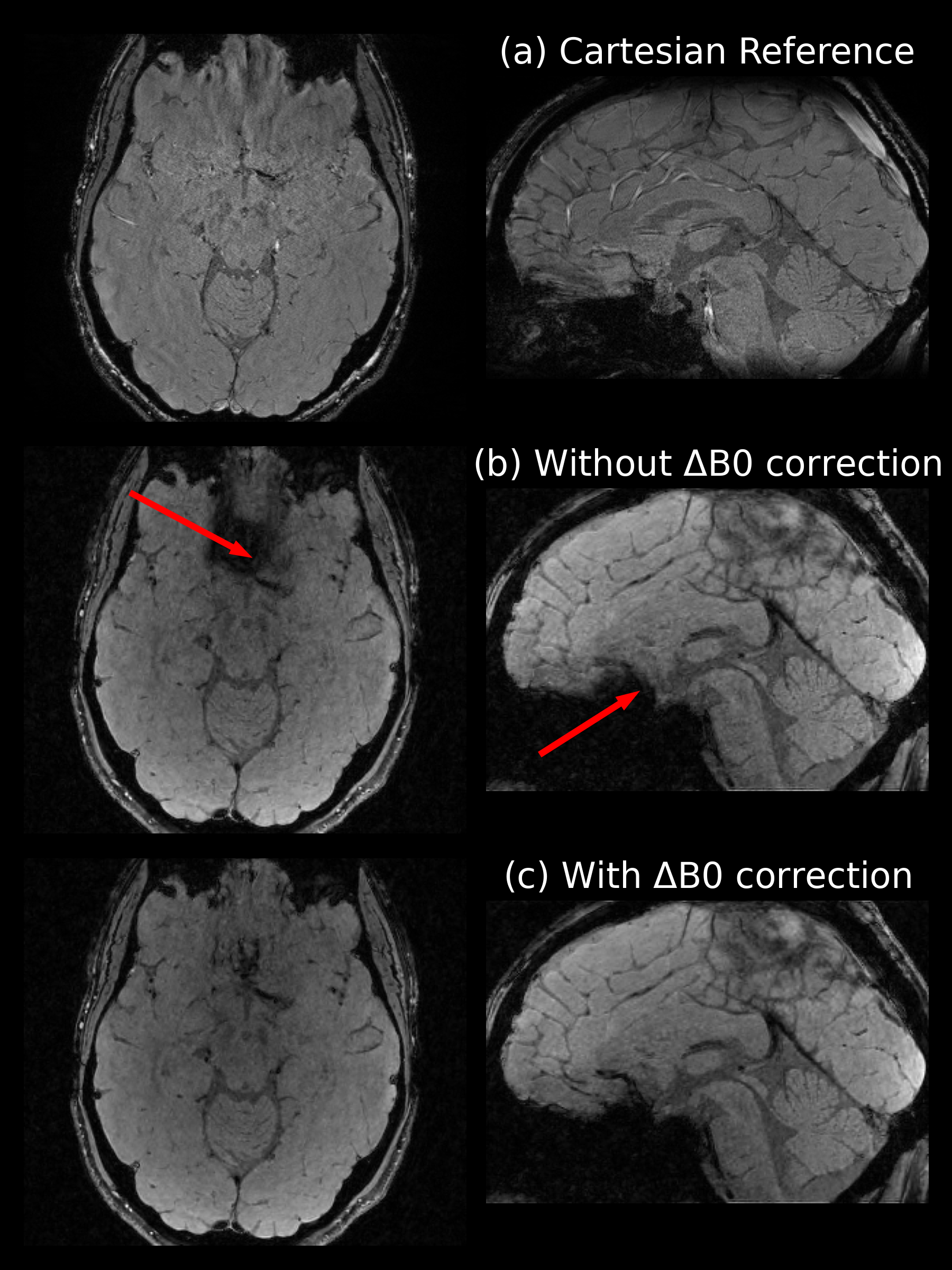}
	\caption{\label{fig:off_res_correction} Prospective reconstruction results~(axial and sagittal view only) (b) without $\Delta B0(\textbf{r})$ correction and (c) with $\Delta B0(\textbf{r})$ correction for full 3D SPARKLING trajectory at AF=10. Cartesian reference views~(a) are also shown for comparison purpose. Red arrows in (b) refer to the regions of strong $\Delta B0$ artifacts. We see that most of the MR signal in these areas is recovered in (c) using the approach proposed in~\cite{DavalFrerot_ISMRM21}.}
\end{figure} % with no additional scan time for the \$\textbackslash\{\}Delta B0\$ map